\documentclass[letterpaper, 10 pt, conference]{ieeeconf}  % Comment this line out if you need a4paper

\IEEEoverridecommandlockouts                              % This command is only needed if 
\title{\LARGE \bf
The Feedback Loop Between Recommendation Systems and\\ Reactive Users}

\author{Atefeh Mollabagher and Parinaz Naghizadeh% <-this % stops a space
\thanks{This work is supported in part by the NSF under award \#2411000.}% <-this % stops a space
\thanks{A. Mollabagher and P. Naghizadeh are with the Electrical and Computer Engineering Department, University of California, San Diego.
        {\tt \{atefeh, parinaz\}@ucsd.edu}}
}

\usepackage{amssymb} 
\usepackage{comment}
\usepackage{xcolor}
\usepackage{cite}
\usepackage{graphicx}
\usepackage[colorlinks=true]{hyperref}
\usepackage{subcaption}

\usepackage{amsmath,amsthm}
\usepackage{newtxtext,newtxmath}
\usepackage{algorithm}
\usepackage{algpseudocode}
\usepackage{forest}

\newtheorem{proposition}{Proposition}
\newtheorem{corollary}{Corollary}

\begin{document}

\maketitle
\thispagestyle{empty}
\pagestyle{empty}

%%%%%%%%%%%%%%%%%%%%%%%%%%%%%%%%%%%%%%%%%%%%%%%%%%%%%%%%%%%%%%%%%%%%%%%%%%%%%%%%
\begin{abstract}
    Recommendation systems underlie a variety of online platforms. These recommendation systems and their users form a feedback loop, wherein the former aims to maximize user engagement through personalization and the promotion of popular content, while the recommendations shape users' opinions or behaviors, potentially influencing future recommendations. These dynamics have been shown to lead to shifts in users' opinions. In this paper, we ask whether \emph{reactive} users, who are cognizant of the influence of the content they consume, can prevent such changes by actively choosing whether to engage with recommended content. We first model the feedback loop between reactive users' opinion dynamics and a recommendation system. We study these dynamics under three different policies - \emph{fixed} content consumption (a passive policy), and \emph{decreasing} or \emph{adaptive decreasing} content consumption (reactive policies). We analytically show how reactive policies can help users effectively prevent or restrict undesirable opinion shifts, while still deriving utility from consuming content on the platform. We validate and illustrate our theoretical findings through numerical experiments. 
    
\end{abstract}

%%%%%%%%%%%%%%%%%%%%%%%%%%%%%%%%%%%%%%%%%%%%%%%%%%%%%%%%%%%%%%%%%%%%%%%%%%%%%%%%

\section{Introduction}\label{sec:introduction}

Recommendation systems drive content curation and exposure in a variety of online platforms, including social networks, online shopping, streaming services, and news aggregators. Despite their benefits, it has been argued that such systems can alter users' preferences and behavior over time. For instance, it has been found that users typically follow (personalized, or even random) recommendations they receive from online shopping platforms, which in turn determines their buying/consumption choices \cite{ursu2018power}. Following recommendations can also have negative effects: exposure to content through social networks may lead to users' polarization \cite{allcott2020welfare,levy2021social} (this is also true for traditional content distribution platforms such as cable news, with users persuaded towards views promoted by the platform \cite{martin2017bias}). Similarly, the use of personalized recommendation systems can amplify users' preexisting beliefs \cite{terren2021echo}, a phenomenon known as ``echo chambers'' in the context of news/social media.  

One potential way to address the growing power of recommendation systems in shaping users' behavior is through regulating digital platforms and content moderation systems (e.g. \cite{di2022recommender}). 
In this paper, we take a complementary view and investigate the users' ability to \emph{themselves} actively respond to recommendation systems. That is, we consider \emph{reactive} users, who may sometimes intentionally ignore the recommendation made to them, to prevent substantial shifts in their innate preferences. 

We begin by proposing a model of the feedback loops between a recommendation system (platform) and \emph{reactive} users' opinion dynamics. Our model extends those of \cite{rossi2021closed,lanzetti2023impact}, which have studied the opinion dynamics of \emph{passive} users who always consume the content recommended to them. In contrast, we consider {reactive} agents following one of three policies: a baseline \emph{fixed} content consumption policy (which includes the passive agents' policy of prior works as a special case), a \emph{decreasing} content consumption policy (where users gradually cease using the platform), and an \emph{adaptive decreasing} policy (where users adjust their use of the platform adaptively so as to control their opinion drift). {In particular, our proposed model of reactive users who avoid content consumption is motivated by two factors: (i) recent large-scale surveys \cite{haupt2023recommending} showing that users indeed adopt strategic decisions when interacting with recommendation systems, and (ii) psychologically grounded models of agents' opinion dynamics \cite{curmei2022towards}, and specifically, ``hedonic adaptation'': returning to a baseline level after some time, even if there are temporary changes in one's opinion.}

We first characterize the agent's opinion under each policy (Proposition~\ref{prop:single-opinion} and Corollary~\ref{cor:compare-infinite}), and show that, unlike passive users following a \emph{fixed} policy, reactive users following the \emph{decreasing} and \emph{adaptive decreasing} policy can prevent being persuaded towards a platform's recommendation. We further account for the users' reduced utility from decreasing content consumption under the reactive policies, and identify scenarios where a user following the reactive policies can outperform the passive policy, despite the reduced content consumption (Proposition~\ref{prop:comapre-utility}). We illustrate our findings through numerical experiments in Section~\ref{sec:numerical}. 
        
\textbf{Related Work.} The closed-loop interactions between users and intelligent systems (such as recommendation systems) have been widely studied \cite{rossi2021closed, lanzetti2023impact, sinha2016deconvolving, krauth2022breaking, ou2023impact, harper2015putting, schmit2018human, haupt2023recommending, dean2024accounting}. Within the works that account for opinion changes, one line of research focuses on how recommendation systems can adjust their algorithms based on users' changing behaviour \cite{konstan2012recommender, ekstrand2011collaborative, harper2015putting, schmit2018human}. Another line of work studies how exposure to recommended content shapes users' opinions, whether it originates from recommendations or social networks \cite{rossi2021closed, lanzetti2023impact, sprenger2024control, curmei2022towards}. Among these, our work is most closely related to \cite{rossi2021closed, lanzetti2023impact}: \cite{rossi2021closed} studies how the interactions between a user and a recommendation system can lead to polarization and filter bubbles (from a microscopic view); \cite{lanzetti2023impact} studies these interactions from both macroscopic and microscopic views, showing that even if each individual's opinion is influenced by recommendations, the population's opinion distribution can remain unchanged. Our work is similar in our consideration of agents' opinion dynamics, analytically studying microscopic effects, and numerically illustrating both microscopic and macroscopic effects. 
However, we differ in our consideration of reactive agents, as opposed to the passive agents in these works. This allows us to show how reactive agents can prevent the persuasion led by recommendation systems, by intentionally adjusting their plan to consume content using a simple binary decision.
\section{Model}\label{sec:model}
\begin{figure}[t]
    \centering
    \includegraphics[scale=0.22]{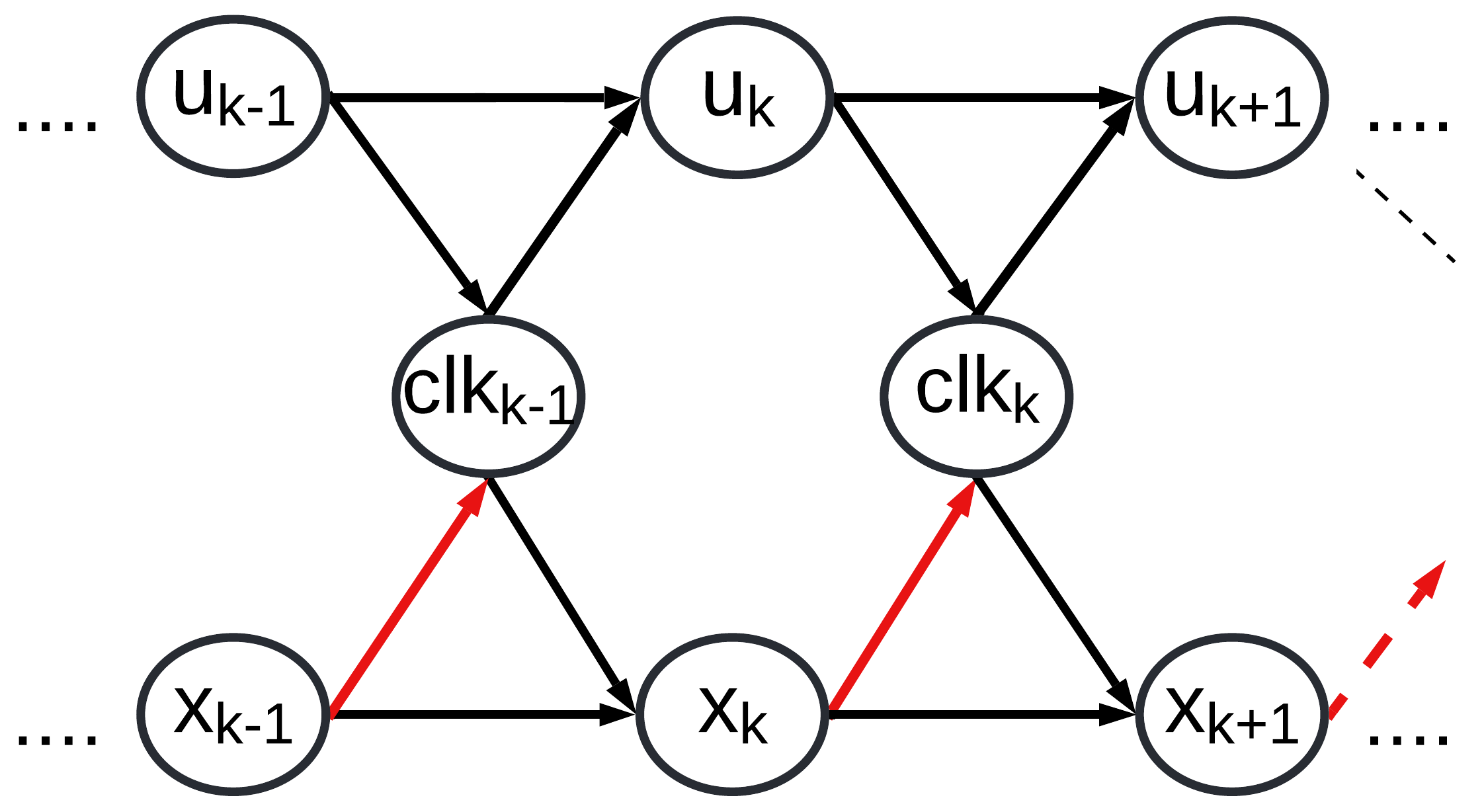}
    \caption{The feedback loop between a recommendation platform and agents. Without the red arrows, the model captures feedback loops with \emph{passive} agents (studied in prior works). Our model (with the red arrows included) captures feedback loops with \emph{reactive} agents, whose actions are informed by both the recommendation {$u_{k-1}$} and their latest opinion $x_{k-1}$.}
    \label{fig:feedbackloop}
    \vspace{-0.2in}
\end{figure}

In this section, we describe our model for capturing feedback loops between a recommendation system (platform) and (reactive) agents. Our model builds on those of \cite{lanzetti2023impact, rossi2021closed}, extending them to allow for agent responses. We present the action spaces and utilities for both the agents and the platform, and describe how they interact with each other. 
Figure \ref{fig:feedbackloop} provides an illustration. 

\paragraph{Agent's model} We consider an agent with an (evolving) \emph{opinion} $x_k\in [-1,1]$ at each time $k= 0, 1, 2, \ldots$. 
This could reflect, for instance, the agent's political viewpoints or preference for different movie genres. At each time $k$, the agent receives a recommendation  $u_{k} \in [-1,1]$ from the platform for content to consume (e.g., a news article to read, or a movie to watch). The agent decides whether to consume the content, by taking an action $\text{clk}_k \in \{0, 1\}$; 
this is a scalar showing whether the agent has clicked on the recommendation (1) or not (0). 
Accordingly, the agent's opinion will evolve. 
We model the opinion dynamics of the agent as follows: 
\begin{equation}\label{eq:user-model}
    x_{k} = 
    \begin{cases} 
    \alpha x_0 + \beta x_{k-1} + (1 - \alpha - \beta) u_{k-1}, & \text{if } \text{clk}_{k-1} = 1 \\
    \frac{\alpha}{\alpha+\beta} x_0 + \frac{\beta}{\alpha+\beta} x_{k-1}, & \text{if } \text{clk}_{k-1} = 0
    \end{cases}
\end{equation}

Here, $x_0$ is the agent's initial or \emph{innate} opinion, and $\alpha \in [0, 1]$ and $\beta \in [0, 1]$ (with $0<\alpha+\beta\leq 1$) are constants determining the relative impact of different factors in shaping the agent's future opinion. In words, when the agent clicks on a recommendation, the opinion evolves based on a convex combination of the initial opinion, the latest opinion, and the recommendation. When not clicking on the content, the opinion changes based on the initial and latest opinions.

{Specifically, note that when the agent does not click on the content, the dynamics in \eqref{eq:user-model} still allow the opinion to evolve. We will assume $\alpha\geq\beta$ throughout, given which our model allows the user to drift back towards their innate opinion $x_0$ if they stop consuming content on the platform. That is, we assume that in absence of influence from the platform, other external factors will drive the agent back to its innate opinion.}  

Clicking on a recommended content has two consequences: it influences the evolution of the agent's opinion (as detailed above), and it generates a benefit/reward for the agent (due to its entertainment value, providing knowledge, etc.). To model the latter, we define an \emph{agent reward function} $R^A: \mathbb{R}_{\geq 0} \to \mathbb{R}_{\geq 0}$, which determines the instantaneous reward for the agent at each time $k$ when evaluated at $|x_k - u_k|$ (the distance between the agent's current opinion and the consumed content). We assume $R^A$ is a non-increasing function; in words, when the difference between the recommendation and the opinion is lower (e.g., the recommended news article is more similar to the agent's political views), the user receives a higher reward. 

Accordingly, we let the agent's utility for a horizon $K$ be:
\begin{equation}\label{eq:agent-utility}
    U(h_K ^A) = \lambda \frac {1} {K} \sum_{i=0}^{K-1} \text{clk}_i R^A (|x_i - u_i|) -  (1 - \lambda) |x_K - x_0|~.
\end{equation}
Here, $h^A_K$ denotes the agent's history at time $K$, which contains the set of all agent's opinions $\{x_i\}_{i \in [0:K]}$, platform's recommendations $\{u_i\}_{i \in [0:K-1]}$, agent's actions $\{\text{clk}_i\}_{i \in [0:K-1]}$, and agent's  rewards $\{R^A(|x_i - u_i|)\}_{i \in [0:K-1]}$.

In words, the agent is interested in maintaining an opinion close to its initial opinion after $K$ time steps, while also deriving benefit from consuming content during this time, with $\lambda\in[0,1]$ capturing the relative importance of each goal.

\paragraph{Platform's model} The objective of the platform is to recommend a content that the agent chooses to consume. At each time step $k$, if the user clicks on the recommended content ($\text{clk}_k=1$), the platform collects a reward $R^P(|x_k - u_k|)$, where $R^P: \mathbb{R}_{\geq 0} \to \mathbb{R}_{\geq 0}$ is a \emph{platform reward function}. This is assumed to be a non-increasing function, which determines the platform's reward based on the difference between the recommendation and the agent's current opinion; the closer they are, the higher the reward. 

We let the platform's payoff for a horizon $K$ be the average of the rewards collected from the agent's content consumption:
\begin{equation}\label{eq:platform-utility}
    \Pi(h_K ^P) = \frac {1} {K} \sum_{i=0}^{K-1} \text{clk}_i R^P (|x_i - u_i|),
\end{equation}
where $h_K^P$ is the platform's history up to time $K$, and includes the set of all recommendations by the platform $\{u_i\}_{i \in [0:K-1]}$, the agent's content consumption actions $\{\text{clk}_i\}_{i \in [0:K-1]}$, and the platform rewards $\{R^P(|x_i - u_i|)\}_{i \in [0:K-1]}$.
\section{Platform's and Agent's Policies}\label{sec:policy}

Our goal is to use the model of Section~\ref{sec:model} to analyze the evolution of agents' opinions $x_k$ over time, contrasting passive and (different types of) reactive agents. This section outlines the platform and agent policies we consider for this analysis.  

\subsection{Platform's Policies}\label{sec:platform-policy}
We consider two platform policies: \emph{``fixed recommendation''} for analytical results, and \emph{``explore periodically''} for numerical experiments. 
In the former, the platform makes the same recommendation $u_0$ at all times, i.e., $u_k = u_0, \forall k$. This allows us to contrast our results with that of \cite{lanzetti2023impact}, which has analyzed the impacts of this policy on \emph{passive} agents. 

In the latter, \emph{``explore periodically''} policy, also considered in prior works \cite{lanzetti2023impact, rossi2021closed} with \emph{passive} agents, the platform always chooses the action which leads to the highest reward observed so far, with the exception of the exploration steps, in which the recommendation will be chosen from a probability distribution; i.e., $u_k \sim \rho \text{ for } k \in \{0, \Delta, 2\Delta, \ldots\}$. This policy resembles $\epsilon$-greedy exploration in the multi-armed bandit/reinforcement learning literatures \cite{kuleshov2014algorithms}. The recommendation at exploitation steps is selected as follows:
\begin{equation} \label{eq:platform-recommendation}
    \begin{aligned}
        u_{k+1} = \operatorname*{argmax}_{u_0, \dots, u_{k}} \Big\{& \text{clk}_0 R^P(|x_0 - u_0|), \ldots, 
        & \text{clk}_{k} R^P(|x_{k} - u_{k}|) \Big\}
    \end{aligned}
\end{equation}

\subsection{Agent's Policies}
On the agent's side, we consider three policies in both our analysis and numerical experiments: ``\emph{fixed}'', ``\emph{decreasing}'', and ``\emph{adaptive decreasing}'' clicking policies, as detailed below. 

Consider a time horizon of length $K$, divided into $n$ blocks of length $s$ each (i.e., $K=ns$). In all three policies, during the $i$-th block of time, the agent clicks on the recommended content for the first $T_i$ time steps, and refrains from clicking on the content for the subsequent $s-T_i$ time steps. The three policies will differ in how $T_i$ is selected. 

Specifically, we first consider a \emph{fixed} clicking policy (Algorithm~\ref{alg:policy-one}), where $T_i=T_0, \forall i$. If we let $T_0=s$, this will be equivalent to the ``always click'' policy of passive agents considered in prior works \cite{rossi2021closed, lanzetti2023impact}. 

\begin{algorithm}
    \hspace*{\algorithmicindent} \textbf{Input:} Total time steps $K$, divided into $n$ blocks of length $s$. Constant $T_0 \in \mathbb{N}$, $T_0 \leq s$. 
    \caption{Agent's ``Fixed'' Clicking Policy}\label{alg:policy-one}
    \begin{algorithmic}
        \For{$i \in  {0, 1, 2, \dots, n-1}$}
            \State Click for first $T_0$ time steps.
            \State Do not click for the remaining $s-T_0$ time steps.
        \EndFor
    \end{algorithmic}
\end{algorithm}
\vspace{-0.1in}

We then consider a \emph{decreasing} clicking policy (Algorithm~\ref{alg:policy-two}), where $T_i$ is reduced by a factor of $\kappa$ after each block; i.e., $T_{i+1}=\lfloor{\frac{T_i}{\kappa}}\rfloor$.\footnote{For our analysis, we assume $T_0$ and $\kappa$ are selected so that $T_i$'s are integers (e.g., $T_0$ even and $\kappa=2$, which decays the clicking rate by halving it each time). Our experiments allow for the more general case of $T_{i+1}=\lfloor{\frac{T_i}{\kappa}}\rfloor$.} Decreasing the clicking period at each block is similar to decreasing exploration rates in the multi-armed bandit/reinforcement learning literature; e.g., decaying $\epsilon_t$ in $\epsilon$-greedy policies \cite{kuleshov2014algorithms, auer2002finite}. Note also that for $\kappa=1$, this is equivalent to the \emph{fixed} clicking policy. 

\begin{algorithm}
    \hspace*{\algorithmicindent} \textbf{Input:} Total time steps $K$, divided into $n$ blocks of length $s$. Constant $T_0 \in \mathbb{N}$, $T_0 \leq s$. Clicking decrease rate $\kappa\geq 1$. 
    \caption{Agent's ``Decreasing'' Clicking Policy}\label{alg:policy-two}
    \begin{algorithmic}
        \For{$i \in  \{0, 1, 2, \dots, n-1\}$}
            \State Click for first $T_i$ time steps.
            \State Do not click for the remaining $s-T_i$ time steps.
            \State Reduce clicking times to $T_{i+1} \gets \min\{0, \lfloor\frac{T_i}{\kappa}\rfloor\}$.
        \EndFor
    \end{algorithmic}
\end{algorithm}

Finally, we also consider an \emph{adaptive decreasing} clicking policy (Algorithm~\ref{alg:policy-three}), where the agent reduces its clicking period between blocks as $T_{i+1}=T_i - \tau$, but that this decrease is only triggered if the agent finds its opinion to have deviated substantially from its innate opinion (if $|x_k-x_0|>x_{\text{drift}}$ for a given tolerance $x_{\text{drift}}>0$). 

\begin{algorithm}
    \hspace*{\algorithmicindent} \textbf{Input:} Total time steps $K$, divided into $n$ blocks of length $s$. Constant $T_0 \in \mathbb{N}$, $T_0 \leq s$. Clicking decrease number $\tau\geq 1$. Drift tolerance $x_{\text{drift}}>0$.  
    \caption{Agent's ``Adaptive Decreasing'' Clicking Policy}\label{alg:policy-three}
    \begin{algorithmic}
        \For{$i \in  \{0, 1, 2, \dots, n-1\}$}
            \State Click for first $T_i$ time steps.
            \State Do not click for remaining $s-T_i$ time steps.
            \If{$|x_{(i+1) * s} - x_0| \geq x_{\text{drift}}$}
                \State $T_{i+1} \gets \min\{0, T_i - \tau\}$
            \EndIf
        \EndFor
    \end{algorithmic}
\end{algorithm}

\begin{figure}[thpb]
    \centering
    \includegraphics[scale=0.21]{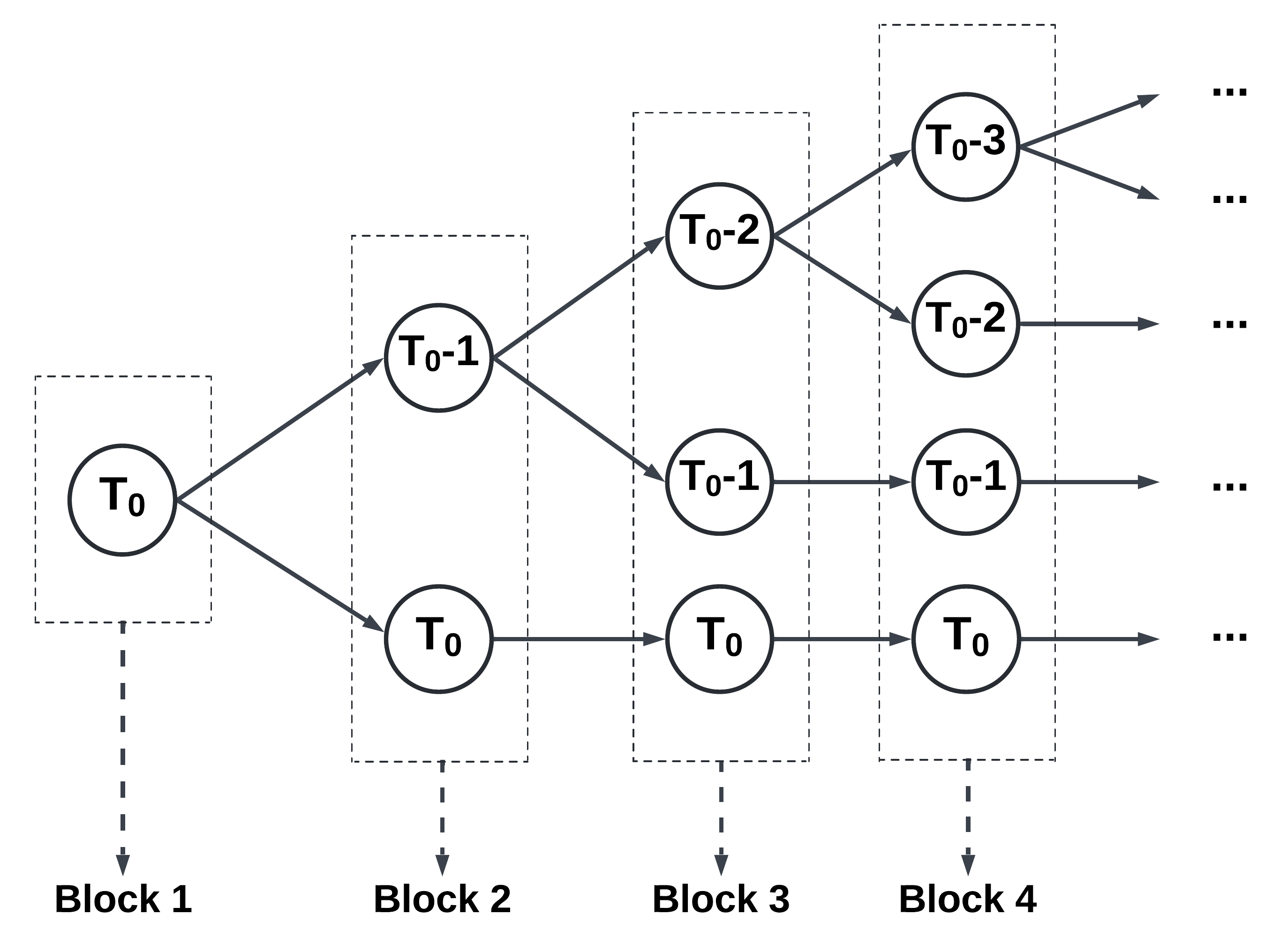}
    \caption{Possible length of clicking periods in Policy~\ref{alg:policy-three}.}
    \vspace{-0.2in}
    \label{fig:clicking-rates}
\end{figure}

Figure~\ref{fig:clicking-rates} illustrates the possible paths that the length of the clicking period can take in each block of time for the \emph{adaptive decreasing} policy. The particular realization of clicking times depends on the agent's initial opinion $x_0$ and the recommended content $u_0$, making this a ``closed-loop''  policy. 

\section{Analysis}\label{sec:analysis}

In this section, we analytically characterize how the feedback loop between the platform and reactive agents leads to the evolution of the agent's opinion $x_k$, in finite time and in the limit, when the agents follow each of the policies outlined in Section~\ref{sec:policy}. All proofs are given in Appendix~\ref{sec:appendixB}.

We start with a characterization of an agent's opinion $x_k$ at the beginning of each block of time under different policies. 

\begin{proposition}\label{prop:single-opinion}
    Consider an agent with innate opinion $x_0 \in [-1,1]$ who receives the recommendation $u_0 \in [-1,1]$. Let $x_{i}^{(p)}$ be the agent's opinion at the beginning of block $i$ when following policy $p \in \{1,2,3\}$, from Algorithms~\ref{alg:policy-one}, \ref{alg:policy-two}, and \ref{alg:policy-three}, respectively. Then, $x_i^{(p)}$ is given by
    \[x_{i}^{(p)} = (1-\Upsilon_{i}^{(p)}) x_0 + \Upsilon_{i}^{(p)} u_0 \]
    where $\Upsilon_{i}^{(p)}$ (resp. $\Gamma_{i}^{(p)}:=1-\Upsilon_{i}^{(p)}$) is the influence of the recommendation $u_0$ (resp. innate opinion $x_0$) on the agent's opinion under policy $(p)$. The exact forms of these weights are given in Appendix~\ref{app:single-opinion-prop-details}. 
\end{proposition}

In the proof, we employ a recursive approach that traces back from block $i$ to the first block, enabling us to express the opinion at the start of each block as a function of $x_0$ and $u_0$; the complexity arises from the agent's opinion dynamics, as well as varying clicking periods for \emph{decreasing} and \emph{adaptive decreasing} policies across blocks.

Proposition \ref{prop:single-opinion} shows that all policies lead to the agent's opinion evolving as a convex combination of the innate opinion $x_0$ and the recommendation $u_0$, with the difference in the weight of each factor. These weights are dependent on not only the policy $p$, but also on the block of time at which the opinion is evaluated. In particular, under the \emph{decreasing} policy (Algorithm~\ref{alg:policy-two}), the agent repeatedly decreases the clicking period by a factor of $\kappa$ (i.e., $T_{i+1}=\lfloor\frac{T_i}{\kappa}\rfloor$). Then, there exists a block $m_D$ at which $T_{m_{D}}=0$, so that the agent no longer clicks on the recommendation going forward. The blocks $i \leq m_D$ (resp. $i > m_{D}$) are then the transient (resp. steady state) blocks under this policy. Similarly, under the \emph{adaptive decreasing} policy (Algorithm~\ref{alg:policy-three}), the agent decreases the clicking period until $|x_k-x_0|<x_{\text{drift}}$ at the beginning of some block $m_{\text{AD}}$, after which the clicking time $T_i, i\geq m_{\text{AD}}$, remains fixed. This again divides the timeline of this policy into transient and steady state blocks. The detailed expressions in Appendix~\ref{app:single-opinion-prop-details} show the weights $\Upsilon_{i}^{(p)}$ of Proposition~\ref{prop:single-opinion} for both transient and steady state phases of each policy $p$.

We next use the expressions identified in Proposition~\ref{prop:single-opinion} to characterize the dependence of the weights $\Upsilon_{i}^{(p)}$ of each policy on the problem parameters. This allows us to identify when and why the recommendation of the platform has a stronger impact on the agent's opinion under each policy. 

\begin{proposition}\label{prop:compare-single-opinion} 
    Consider a \emph{fixed} policy (Algorithm~\ref{alg:policy-one}) with $T_0=s$ (i.e., the agent always clicks on the recommendation in the first block). Then: 
    \begin{enumerate}
        \item $\Upsilon^{(1)}_i$, $\Upsilon^{(2)}_i$, and $\Upsilon^{(3)}_i$ decrease in $\alpha$ (for both transient and steady-state blocks). 
        \item $\Upsilon_{i}^{(1)}$ increases in $i$, and $\Upsilon_{i}^{(2)}$ and $\Upsilon_{i}^{(3)}$ are increasing concave in $i$ for transient blocks. $\Upsilon_{i}^{(2)}$ is decreasing in $i$ for steady-state blocks.
        \item $\Upsilon_{i}^{(1)}$ increases in $T_0$. 
        \item $\Upsilon_{i}^{(2)}$ decreases in $\kappa$ for transient blocks.
        \item $\Upsilon_{i}^{(3)}$ decreases in $\tau$ for transient blocks.
    \end{enumerate}
\end{proposition}

Part (1)  of Proposition \ref{prop:compare-single-opinion} states that as $\alpha$, the importance of the innate opinion in the opinion dynamics \eqref{eq:user-model}, increases, the impact of the recommendation on the agent's opinion decreases, as intuitively expected. Part (2) considers what happens as the number of blocks increases. It states that in all policies, the influence of the recommendations on the agent's opinion grows as the number of blocks of time increases, as the agent has clicked (cumulatively) more on the recommended content when there are more blocks {(in transient blocks)}. However, in \emph{decreasing} and \emph{adaptive decreasing} policies, the rate of increase of $\Upsilon_i^{(p)}$ decreases over time. This is due to the decreasing length of clicking period at each block. In other words, although the agent still clicks on the recommendation some times, the length of clicking period decreases. Consequently, the recommendation's influence growth decreases. 
Parts (3)-(5) consider the impact of policy-specific parameters. In the \emph{fixed} policy, a higher initial clicking number $T_0$ in the first block leads the agent to click on the recommendation more frequently throughout the horizon, thereby amplifying the importance of recommendations. Conversely, as $\kappa$ (clicking decreasing rate in the \emph{decreasing} policy) and $\tau$ (clicking decrease number in the \emph{adaptive decreasing} policy) increase, the duration of clicking periods within each block shortens, resulting in a reduced emphasis on recommendations.

Propositions~\ref{prop:single-opinion} and \ref{prop:compare-single-opinion} identifies the impacts of different factors on the agents' opinions when they follow each policy for potentially large, but still finitely many blocks. The following Corollary of Proposition~\ref{prop:single-opinion}, in contrast,  considers the limit of the opinions under each policy. 

\begin{corollary}\label{cor:compare-infinite}
Consider the settings of Proposition~\ref{prop:single-opinion}, and $T_0=s$ for all policies. Then, we have: 
\begin{itemize}
    \item $\lim_{i\rightarrow \infty} x^{(1)}_i = \eta x_0 +  (1 - \eta) u_0$.
    \item $\lim_{i\rightarrow \infty} x^{(2)}_i = x_0$. 
    \item If $x_0 < u_0$, $x_0 \leq \lim_{i\rightarrow \infty} x^{(3)}_i \leq \min\{1, x_0 + x_{\text{drift}}\}$. Otherwise, $\max\{-1, x_0 - x_{\text{drift}}\} \leq \lim_{i\rightarrow \infty} x^{(3)}_i \leq x_0$. 
\end{itemize}
\end{corollary}

To illustrate the intuition behind Corollary~\ref{cor:compare-infinite}, consider an agent exposed to opinion pieces, articles, or posts, on a (controversial) issue through the recommendation system within a social network. The articles can range anywhere from in favor $(1)$ to against $(-1)$ the issue. The agent is initially against the issue $(x_0 = -1)$, yet the platform repeatedly exposes the agent to articles in favor of it $(u_0 = +1)$. 
Corollary \ref{cor:compare-infinite} states that a (significant) drift from the initial opinion happens when the agent is following the \emph{fixed} policy \ref{alg:policy-one} with $T_0=s$; this is the ``always click'' policy of \emph{passive} agents considered in prior work \cite{lanzetti2023impact}. In words, when the agent is always clicking on the recommended content, its opinion will ultimately become biased towards the recommendations (with the drift depending on the problem parameters $\alpha, \beta$, as captured by $\eta:=\frac{\alpha}{1 - \beta}$). In our example, in the long-run, an agent can change its initial stance to be more in favor of the controversial issue/bill. 

An \emph{active} agent can prevent such drifts from happening: an agent following the \emph{decreasing} policy in Algorithm~\ref{alg:policy-two} who gradually decreases its use of the social network (and its content recommendation system) will have its opinion drift back to its innate opinion ($x_0$). That said, this policy may be extreme as it leads to ceasing the use of the platform altogether. The \emph{adaptive decreasing} policy compensates for this by only decreasing the interaction to the point where the agent can maintain (at most) a deviation $x_{\text{drift}}$; the agent will continue to enjoy the platform, while decreasing its interaction rate to not deviate too much from its innate viewpoints. The following Proposition formally establishes that such policy can bring the agent the highest cumulative utility among the three options.

\begin{proposition}\label{prop:comapre-utility}
    Consider the settings of Proposition~\ref{prop:single-opinion} and $T_0=s$ for all policies. Let the agent's reward function be $R^A (|x_i - u_i|) = 1$, and the horizon be $K=ns$. Then, the limit of the agent's utilities at the start of each block under each policy are given by:

    $\bullet~ \lim_{n\rightarrow \infty} U^{(1)}(h_{K} ^A) = \lambda -  (1 - \lambda) (1 - \eta)| u_0 - x_0|$.
    
    $\bullet~ \lim_{n\rightarrow \infty} U^{(2)}(h_{K} ^A) = 0$. 
    
        $\bullet~ 0\leq \lim_{n\rightarrow \infty} U^{(3)}(h_K ^A) \leq \lambda.$

        $\bullet~$ If $\lim_{n\rightarrow \infty} U^{(1)}(h_K ^A) \geq 0$, there exist $\tau$ and $x_{\text{drift}}$ such that $\lim_{n\rightarrow \infty} U^{(3)}(h_K ^A) = \lim_{n\rightarrow \infty} U^{(1)}(h_K ^A)$. There also exist instances in which $\lim_{n\rightarrow \infty} U^{(3)}(h_K ^A) > \lim_{n\rightarrow \infty} U^{(1)}(h_K ^A)$. 

\end{proposition}

Proposition \ref{prop:comapre-utility} can be used to contrast the utility that each policy can bring the agent in the long-run. Particularly, if the agent cares considerably about the potential drifts that consuming recommended content can cause in its opinion (sufficiently low $\lambda$), following the \emph{fixed} policy can result in a \emph{negative} utility for the agent. In contrast, the \emph{decreasing} policy can achieve a zero utility, remaining aligned with the innate opinion but at the expense of no clicking. The \emph{adaptive decreasing} policy can do better than this by its choice of clicking decrease parameter $\tau$ and drift tolerance $x_{\text{drift}}$. In particular, this policy can always mimic the \emph{decreasing} policy, ensuring at least zero utility by choosing a large $\tau$; smaller $\tau$ and small enough $x_{\emph{drift}}$ can then lead to positive utility, achieving a moderate clicking rate in the long-run while keeping the drift of opinion below $x_{\text{drift}}$. Finally, note that if $\lambda$ is large (i.e., the agent cares more about content consumption than drifts in its opinion), then the \emph{adaptive decreasing} policy can perform as well as the \emph{fixed} policy at best. To conclude, either the \emph{fixed} or \emph{adaptive decreasing} may be preferred depending on $\lambda$.
\section{Numerical Experiments}\label{sec:numerical}

In this section, we use numerical experiments to validate our theoretical findings of Section~\ref{sec:analysis} for the \emph{fixed recommendation} policy of the platform, and also provide numerical results for the \emph{explore periodically} policy of the platform. We consider the reward functions $R^A (|x_i - u_i|) = R^P (|x_i - u_i|) = 1 - c|x_i - u_i| ${, where $c \in [0,1]$ is a constant}. For all experiments, we let $\alpha = 0.25,  \beta = 0.2, \lambda = 0.5, T_0 = s = 8, \kappa = 2, c = 0.1$. {This parameter selection is to provide a clearer illustration;  additional experiments in Appendix~\ref{sec:appendixB} with different parameter values also confirm our findings.}

\begin{figure*}[thpb]
    \centering
    \begin{subfigure}[b]{0.24\textwidth}
        \centering
        \includegraphics[width=0.9\textwidth]{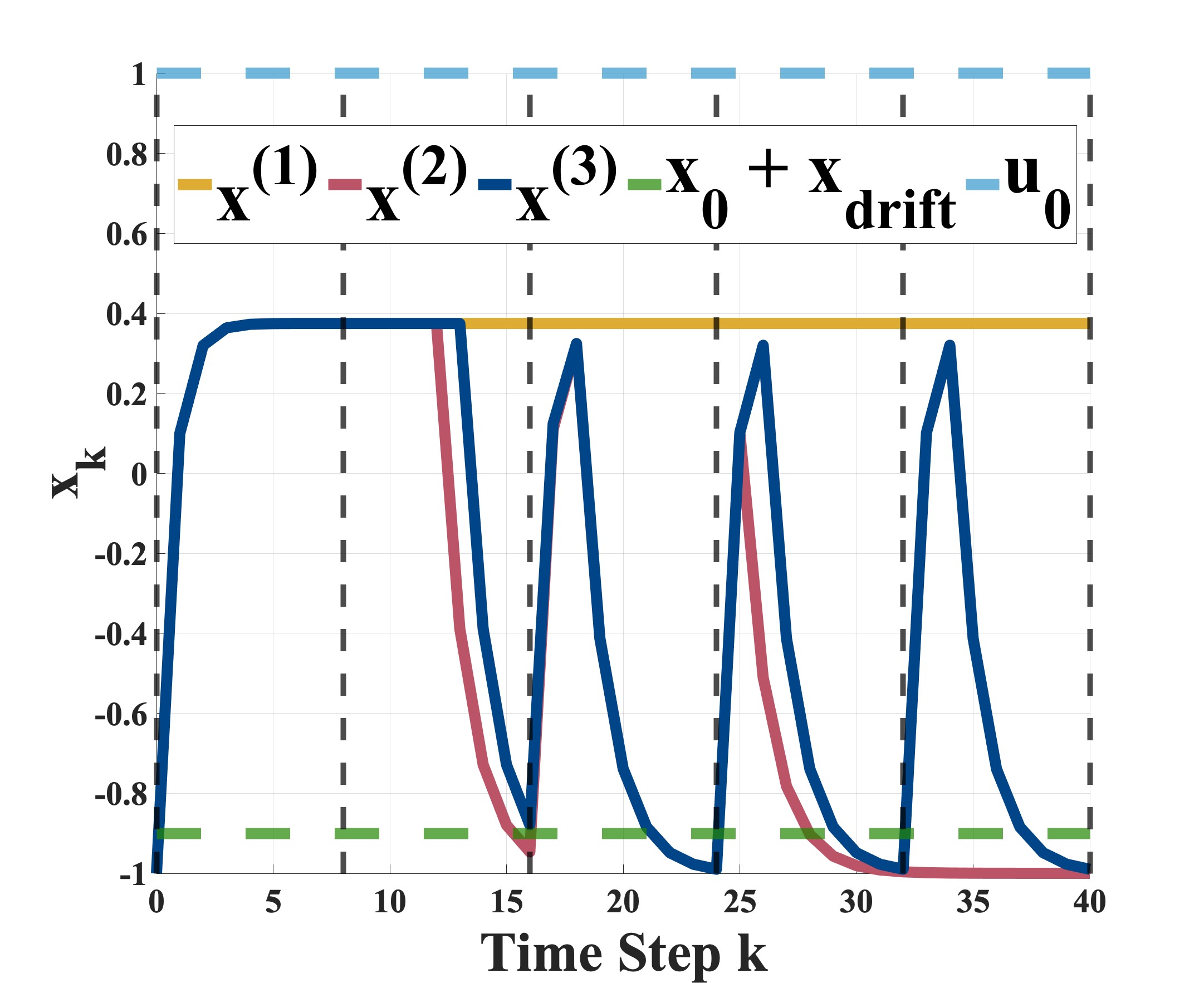}
        \caption{Agent's opinion.}
        \label{fig:1-compare-opinion}
    \end{subfigure}
    \hfill
    \begin{subfigure}[b]{0.24\textwidth}
        \centering
        \includegraphics[width=0.9\textwidth]{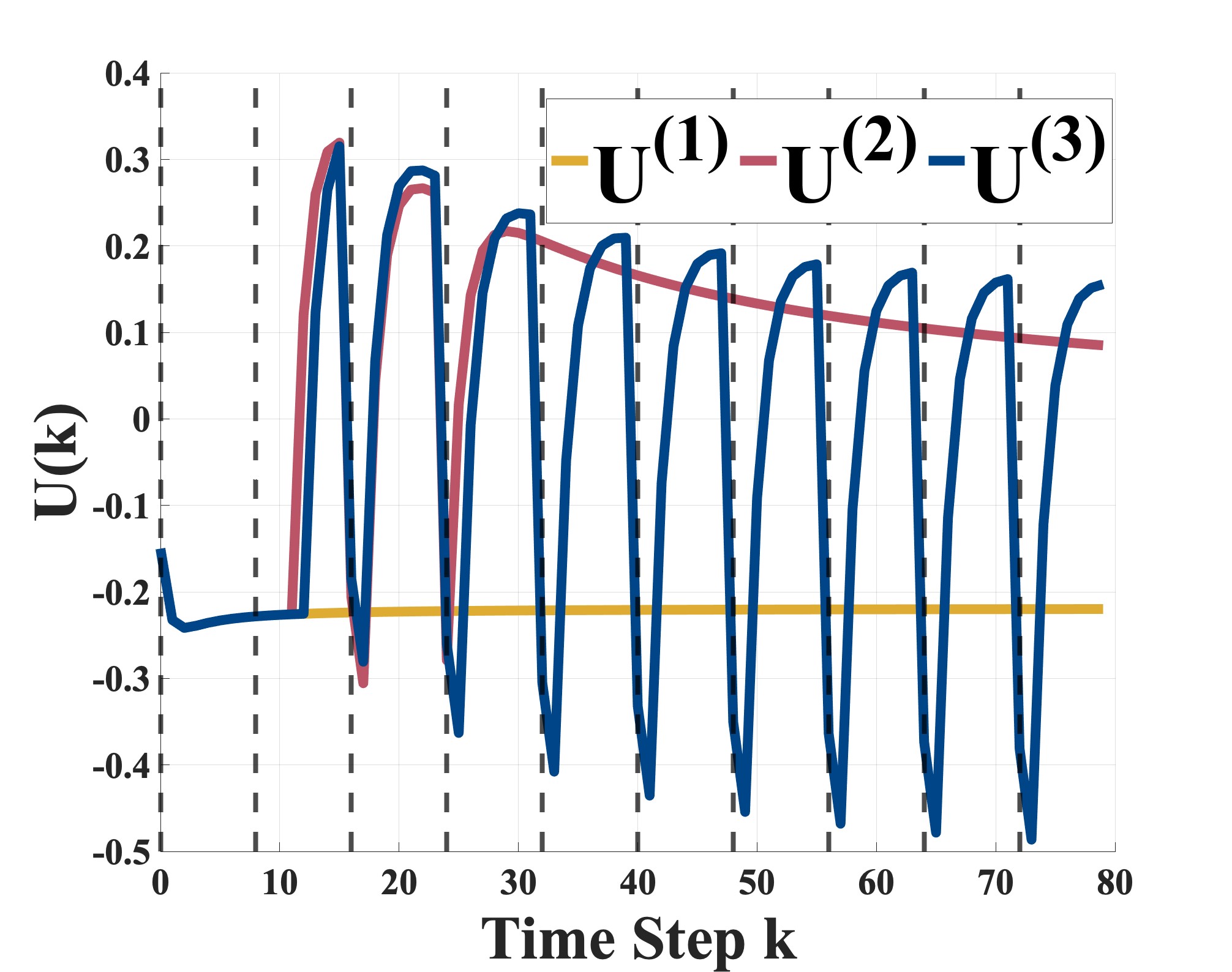}
        \caption{Agent's utility.}
        \label{fig:2-compare-agent-utility}
    \end{subfigure}
    \hfill
    \begin{subfigure}[b]{0.24\textwidth}
        \centering
        \includegraphics[width=0.9\textwidth]{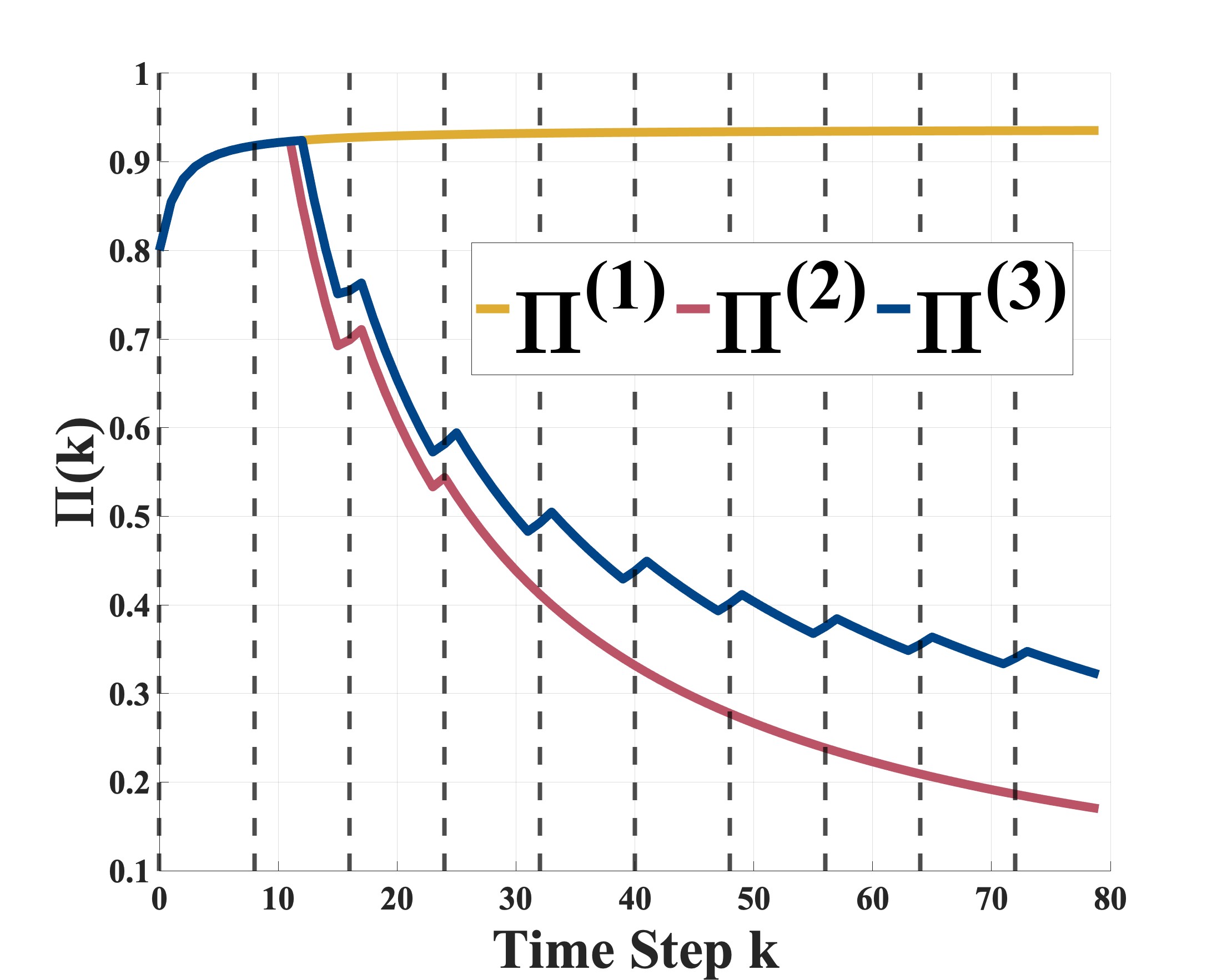}
        \caption{Platform's utility.}
        \label{fig:3-compare-platform-utility}
    \end{subfigure}
    \hfill
    \begin{subfigure}[b]{0.24\textwidth}
        \centering
        \includegraphics[width=0.9\textwidth]{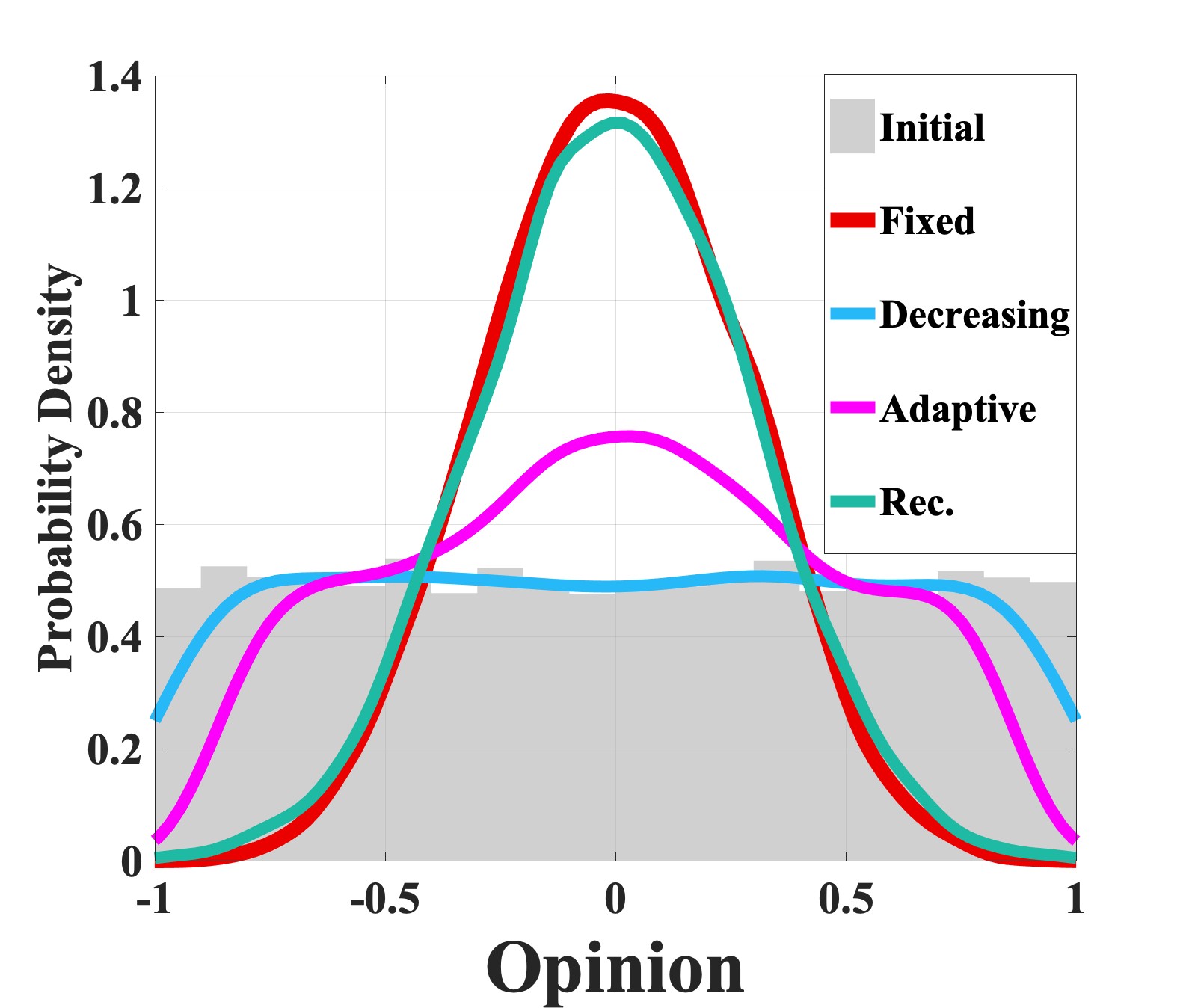}
        \caption{Final opinion distribution.}
        \label{fig:6-population}
    \end{subfigure}
    \caption{Agent's opinion and utility, platform's utility, and final opinion distribution under different agent policies and fixed platform recommendations.}
    \vspace{-0.1in}
    \label{fig:compare-all}
\end{figure*}

\paragraph{Fixed recommendations}\label{sec:experiments-fixed} Figure \ref{fig:compare-all} illustrates the evolution of an agent's opinion (\ref{fig:1-compare-opinion}), the agent's utility (\ref{fig:2-compare-agent-utility}), and the platform's utility (\ref{fig:3-compare-platform-utility}), for an agent with innate opinion $x_0 = -1$, receiving fixed recommendation $u_0 = 1$, and with $\tau = 3, x_{\text{drift}}= 0.1$. Our findings match those in Propositions~\ref{prop:single-opinion} and~\ref{prop:comapre-utility}. In particular, we see that the \emph{fixed} (passive) policy results in the maximum drift in the agent's opinion (and lowest agent utility due to the relatively high desire to prevent opinion drifts), while the agent's opinion under the \emph{decreasing} policy reverts to the innate opinion (though the agent accrues 0 utility long-term). In contrast to these, the \emph{adaptive decreasing} policy allows the opinion to oscillate, ending each block no more than $x_{\text{drift}}$ away from the innate opinion (though the opinion can drift further within the block), and yielding the highest utility among the three policies. We also note that, as intuitively expected, the platform experiences its highest (lowest) payoff under the agent's \emph{fixed} (\emph{decreasing}) policy, as the agent consumes the most (least) content on the platform. 

\paragraph{Macroscopic changes} Figure \ref{fig:6-population} illustrates the impact of each policy on the \emph{distribution of the long-run opinion} in a population of agents. Specifically, we consider a population with innate opinions $x_0$ uniformly distributed between $-1$ and $1$, and a fixed (but distinct) recommendation $u_0$ drawn from a zero-mean Gaussian distribution for each agent. Let $x_{\text{drift}}=0.4$. We observe that the final distribution under the \emph{fixed} policy closely resembles the recommendation distribution, whereas the \emph{decreasing} policy retains the innate opinion distribution. Finally, the \emph{adaptive decreasing} policy results in a distribution that is closer to the innate opinion than to the recommendation. 

\begin{figure*}[thpb]
    \centering
    \begin{subfigure}[b]{0.3\textwidth}
        \centering
        \includegraphics[width=0.8\textwidth]{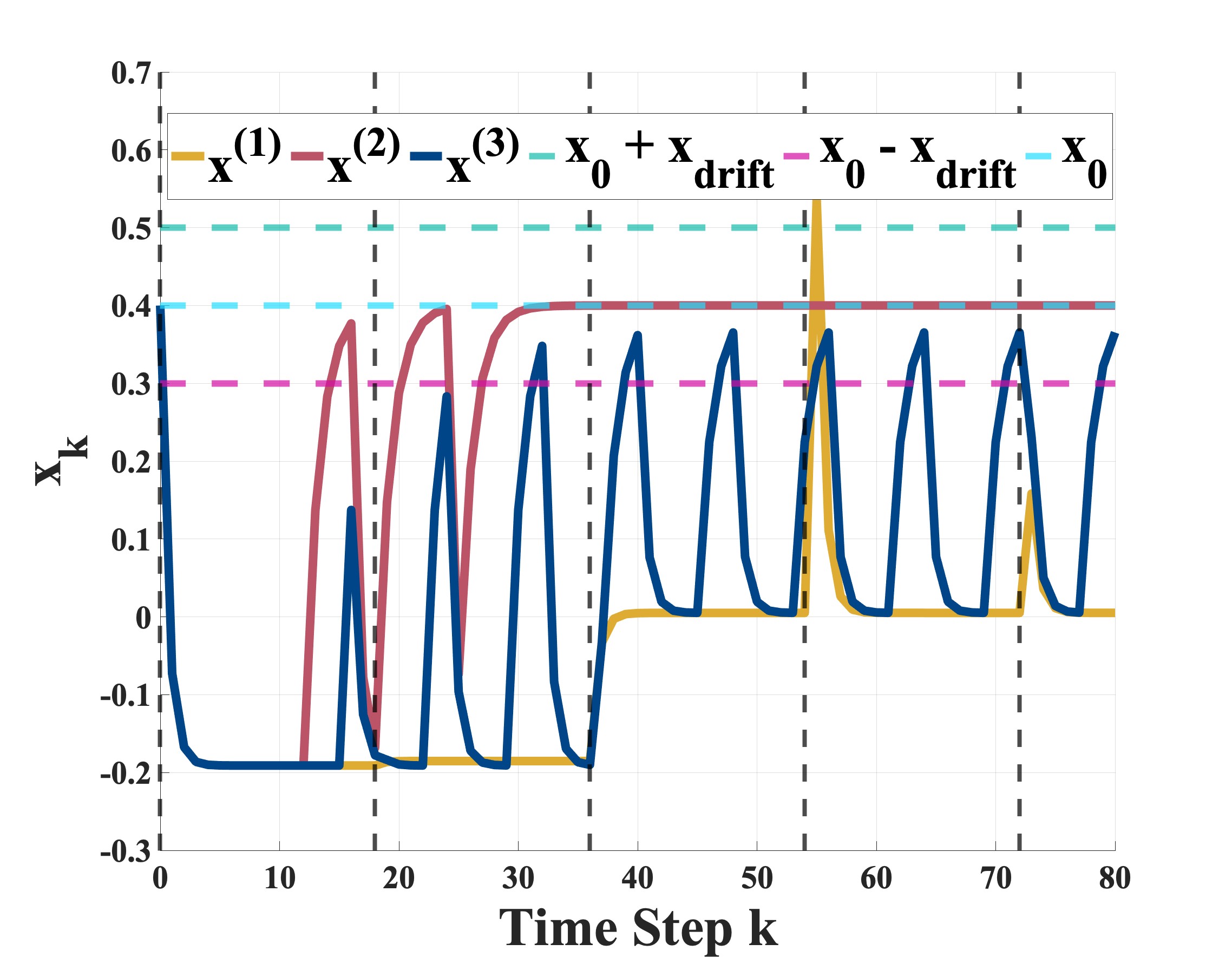}
        \caption{Agent's opinion.}
        \label{fig:4-compare-opinion-delta}
    \end{subfigure}
    \hfill
    \begin{subfigure}[b]{0.3\textwidth}
        \centering
        \includegraphics[width=0.8\textwidth]{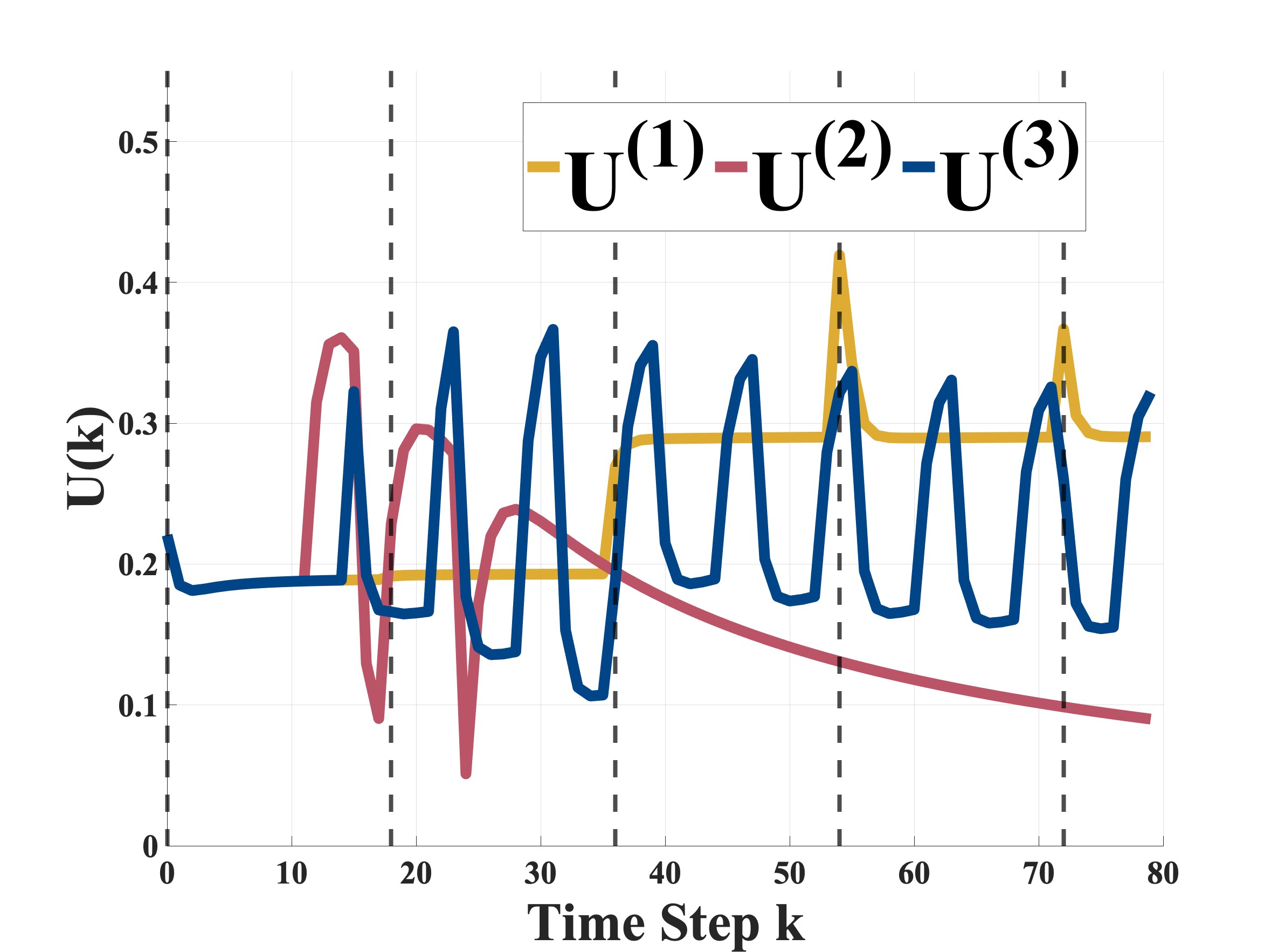}
        \caption{Agent's utility.}
        \label{fig:5-compare-agent-utility-delta}
    \end{subfigure}
    \hfill
    \begin{subfigure}[b]{0.3\textwidth}
        \centering
        \includegraphics[width=0.8\textwidth]{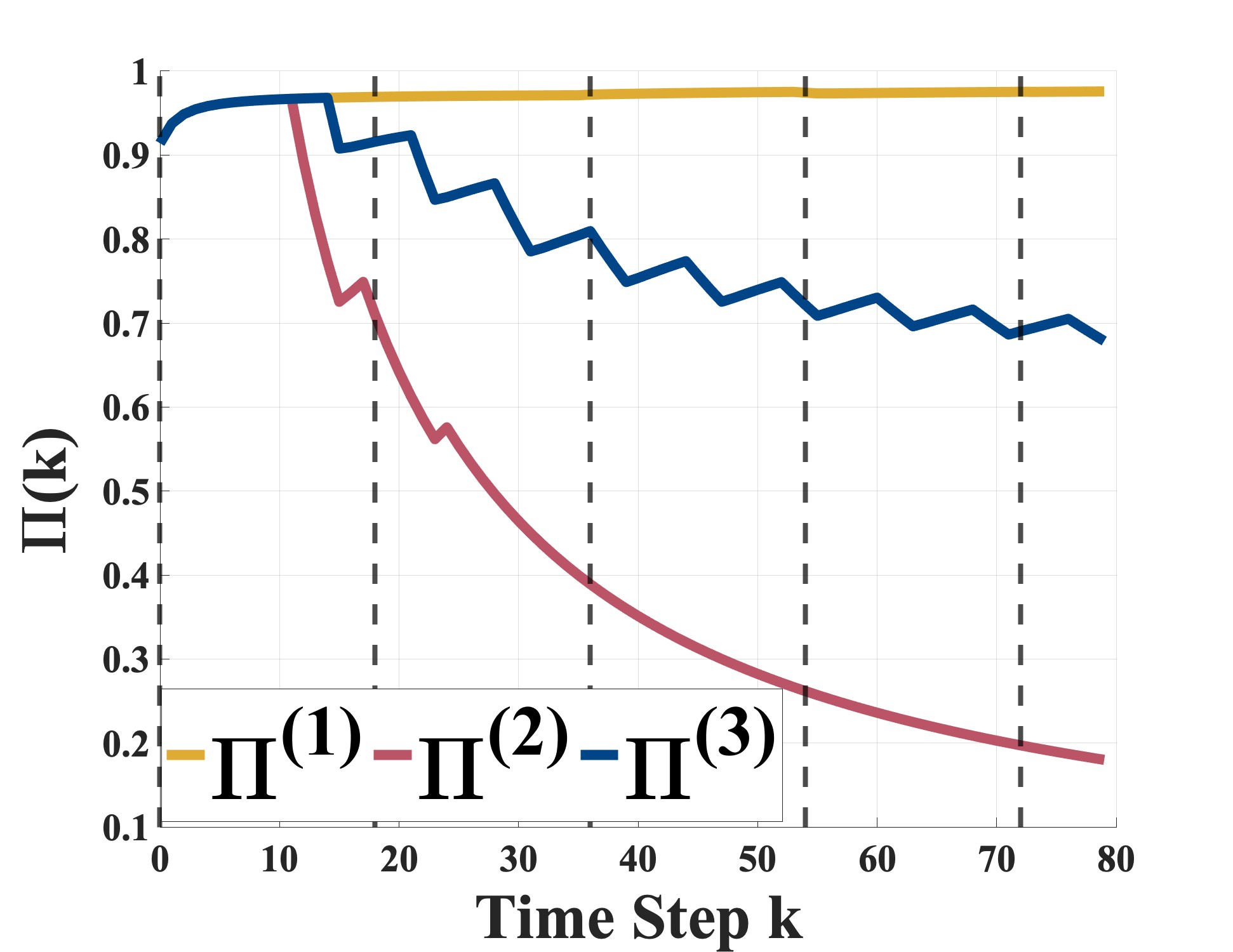}
        \caption{Platform's utility.}
        \label{fig:7-compare-platform-utility-delta}
    \end{subfigure}
    \caption{Agent's opinion and utility, and platform's utility, under different agent policies and varying platform recommendations.}
    \vspace{-0.1in}
    \label{fig:compare-all-delta}
\end{figure*}

\paragraph{Exploring recommendations} 
We next move beyond our analytical findings and allow the platform to follow the \emph{explore periodically} policy detailed in Section~\ref{sec:platform-policy}. Figure \ref{fig:compare-all-delta} {is about} this policy with $\Delta=18$ (picking a $u_k$ uniformly at random when exploring every 18 time steps), and with $\tau=1, x_{\text{drift}}=0.1$. We see that after an exploring recommendation, agents following the \emph{fixed} and \emph{adaptive decreasing} policies, due to their non-zero clicking rates, experience sudden drifts in their opinions in response to the new recommendation. For the \emph{adaptive decreasing} policy, this has further triggered a decrease in the clicking period. 
That said, in the long-run, the platform will converge to one recommendation, and the agent will resume similar behavior to those identified for fixed-recommendations under both policies. For the \emph{decreasing} policy, as the agent reached the steady state before an exploration step by the platform, there is no opinion drift.

\section{Conclusion}\label{sec:conclusions}

We modeled the opinion dynamics of \emph{reactive} agents in a recommendation system, who can decide whether or not to consume recommended content. We analytically and numerically showed that reactive agents can avoid being persuaded towards a platform's recommendation by adjusting their content consumption decisions. Main directions of future research include the analytical studies of macroscopic effects (opinion distributions), closed-loop policies by the platform (i.e., a platform that adjusts its recommendations when users' opinions shift), and other (stochastic) policies by the agents.

\appendix
\subsection{Detailed Expressions of Weights in Proposition~\ref{prop:single-opinion}}\label{app:single-opinion-prop-details}

We will use the following additional shorthand notation: $Z:= \alpha + \beta$, $B:=\frac{\beta}{\alpha + \beta}$, and $\eta:=\frac{\alpha}{1 - \beta}$.

    \noindent$\bullet$ For the \emph{fixed} policy in Algorithm~\ref{alg:policy-one}: 
    \begin{equation}\label{eq:single-opinion-policy-one}
        \begin{aligned}
            \Upsilon_{i}^{(1)} &= \tfrac{1 - (B^{s} Z^{T_0})^i}{1 - B^{s} Z^{T_0}} (1 - \eta) B^{s-T_0} (1 - \beta^{T_0})
        \end{aligned}
    \end{equation}

    \noindent$\bullet$ For the \emph{decreasing} policy in Algorithm~\ref{alg:policy-two}, at transient blocks:
    \begin{equation}\label{eq:single-opinion-policy-two}
        \begin{aligned}
            \Upsilon_{i}^{(2)} & =(1 - \eta) Z^{\frac{-T_0 \kappa^{1-i}}{\kappa - 1}} (\sum_{j=0}^{i-1} B^{(j+1)s-\frac{T_0}{\kappa^{i-j-1}}} Z^{\frac{T_0 \kappa^{1-i+j}}{\kappa - 1}} \\
            &\quad \times (1 - \beta^{\frac{T_0}{\kappa^{i-j-1}}}))
        \end{aligned}
    \end{equation}

    For steady state blocks, following $m_D$ transient blocks:
    \begin{equation}\label{eq:single-opinion-policy-two-0}
        \begin{aligned}
            \Upsilon^{(2)}_i &= B^{(i-m_D)s} \Upsilon_{m_D}^{(2)}
        \end{aligned}
    \end{equation}
    where $\Upsilon_{m_D}^{(2)}$ is found based on \eqref{eq:single-opinion-policy-two}. 

    \noindent$\bullet$ For the transient blocks of the \emph{adaptive decreasing} policy in Algorithm~\ref{alg:policy-three}:
    \begin{equation} \label{eq:single-opinion-policy-three}
        \begin{aligned}
            \Upsilon_{i}^{(3)} &= (1-\eta) (\sum_{j=0}^{i-1} B^{(j+1)s-T_0+(i-1-j)\tau} Z^{j(T_0-\tau i)+\tau\frac{j^2+j}{2}}\\
            &\quad \times (1-\beta^{T_0-(i-1-j)\tau}))
        \end{aligned}
    \end{equation}

    For the steady state blocks, following $m_{{AD}}$ transient blocks:
    \begin{equation}\label{eq:single-opinion-policy-three-fixed}
        \begin{aligned}
            \Upsilon_i^{(3)} & = \Upsilon_{i-m_{AD}}^{(1)} +  (B^{s}Z^{T_0-(m_{AD} - 1)\tau})^{i-m_{AD}} \Upsilon_{m_{AD}}^{(3)}
        \end{aligned}
    \end{equation}
    where $\Upsilon_{i-m_{AD}}^{(1)}$ and $\Upsilon_{m_{AD}}^{(3)}$ come from \eqref{eq:single-opinion-policy-one} and \eqref{eq:single-opinion-policy-three}, respectively.
\bibliographystyle{IEEEtran}
\bibliography{references}
\clearpage
\onecolumn

\begin{appendices}
    \renewcommand{\thesection}{B}
    \section{Second appendix}\label{sec:appendixB}
    \subsection{Proof of Proposition~\ref{prop:single-opinion}}\label{app:single-opinion}
    \begin{proof}\label{proof:single-opinion-policy-one}
    Let the agent's innate opinion be $x_0 \in [-1,1]$, and let the agent receive the same recommendation $u_0 \in [-1,1]$ at any time steps.

        \begin{enumerate}
            \item  We start to find the opinion at the start of each block based on a function of $x_0$ and $u_0$. To reach this goal, we use a recursive approach to write the opinion at the start of each block based on the opinion at the start of the previous block. Consequently, the agent's opinion can be expressed at the beginning of block $i$, under \emph{fixed} policy based on a function of $x_0$ and $u_0$, for any $i \geq 1, i \in \mathbb{N}$. For time step $k=is$, we have:
            \begin{align*}
                x_{k} &= \frac{\alpha}{\alpha + \beta} x_0 + \frac{\beta}{\alpha + \beta} x_{k-1} = (1-B) (1 + B) x_0 + B^2 x_{k-2} = (1-B) \sum_{q=0}^{s-T_0-1} B^q x_0 + B^{s-T_0} x_{k-s+T_0}\\
                &= (1-B) \sum_{q=0}^{s-T_0-1} B^q x_0 + B^{s-T_0} \Bigg( \alpha x_0 + \beta x_{k-s+T_0-1} + (1-Z)u_0 \Bigg)\\
                &= (1-B) \sum_{q=0}^{s-T_0-1} B^q x_0 + B^{s-T_0} \Bigg( \alpha (1 + \beta) x_0 + (1-Z) (1 + \beta) u_0 + \beta^2 x_{k-s+T_0-2} \Bigg)\\
                &= (1-B) \sum_{q=0}^{s-T_0-1} B^q x_0 + B^{s-T_0} \Bigg( \alpha (1 + \beta + \beta^2) x_0 + (1-Z) (1 + \beta + \beta^2) u_0 + \beta^3 x_{k-s+T_0-3} \Bigg)\\
                &= (1-B) \sum_{q=0}^{s-T_0-1} B^q x_0 + B^{s-T_0} \left[\alpha \sum_{q=0}^{T_0-1} \beta^q x_0 + (1-Z) \sum_{q=0}^{T_0-1} \beta^q u_0 \right] + \beta^{T_0} B^{s-T_0} x_{k-s}
            \end{align*}
            Now, we should find the opinion $x_{k-s}$ based on the previous block. We can continue this recursive approach to just express the opinion purely as a function of $x_0$ and $u_0$.
            \begin{align*}
                x_{k} &= (1-B) \sum_{q=0}^{s-T_0-1} B^q x_0 + B^{s-T_0} \left[ \alpha \sum_{q=0}^{T_0-1} \beta^q x_0 + (1-Z) \sum_{q=0}^{T_0-1} \beta^q u_0 \right] + B^{s-T_0} \beta^{T_0} \left[ (1-B) \sum_{q=0}^{s-T_0-1} B^q x_0 + B^{s-T_0} x_{k-2s+T_0} \right]\\
                &= (1-B) \sum_{q=0}^{s-T_0-1} B^q x_0 + B^{s-T_0} \left[ \alpha \sum_{q=0}^{T_0-1} \beta^q x_0 + (1-Z) \sum_{q=0}^{T_0-1} \beta^q u_0 \right] + B^{s-T_0} \beta^{T_0} (1-B) \sum_{q=0}^{s-T_0-1} B^q x_0\\
                &\quad + B^{2s-2T_0} \beta^{T_0} \left[ \alpha \sum_{q=0}^{T_0-1} \beta^q x_0 + (1-Z) \sum_{q=0}^{T_0-1} \beta^q u_0 \right] + B^{2s-2T_0} \beta^{2T_0} x_{k-2s}\\
                &= (1-B) \sum_{q=0}^{s-T_0-1} B^q x_0 + B^{s-T_0} \left[ \alpha \sum_{q=0}^{T_0-1} \beta^q x_0 + (1-Z) \sum_{q=0}^{T_0-1} \beta^q u_0 \right] + B^{s-T_0} \beta^{T_0} (1-B) \sum_{q=0}^{s-T_0-1} B^q x_0\\
                &\quad + B^{2s-2T_0} \beta^{T_0} \left[ \alpha \sum_{q=0}^{T_0-1} \beta^q x_0 + (1-Z) \sum_{q=0}^{T_0-1} \beta^q u_0 \right] + B^{2s-2T_0} \beta^{2T_0} (1-B) \sum_{q=0}^{s-T_0-1} B^q x_0 + B^{3s-3T_0} \beta^{2T_0} x_{k-3s+T_0}\\
                &= (1-B) \sum_{q=0}^{s-T_0-1} B^q x_0 + B^{s-T_0} \left[ \alpha \sum_{q=0}^{T_0-1} \beta^q x_0 + (1-Z) \sum_{q=0}^{T_0-1} \beta^q u_0 \right] + B^{s-T_0} \beta^{T_0} (1-B) \sum_{q=0}^{s-T_0-1} B^q x_0\\
                &\quad + B^{2s-2T_0} \beta^{T_0} \left[ \alpha \sum_{q=0}^{T_0-1} \beta^q x_0 + (1-Z) \sum_{q=0}^{T_0-1} \beta^q u_0 \right] + B^{2s-2T_0} \beta^{2T_0} (1-B) \sum_{q=0}^{s-T_0-1} B^q x_0\\
                &\quad + B^{3s-3T_0} \beta^{2T_0} \left[ \alpha \sum_{q=0}^{T_0-1} \beta^q x_0 + (1-Z) \sum_{q=0}^{T_0-1} \beta^q u_0 \right] + B^{3s-3T_0} \beta^{3T_0} x_{k-3s}\\
                &= (1-B) \Bigg(\sum_{q=0}^{s-T_0-1} B^q\Bigg) \sum_{j=0}^{n-1} (B^{s-T_0} \beta^{T_0})^j x_0 + B^{s-T_0} \left[ \alpha \sum_{q=0}^{T_0-1} \beta^q x_0 + (1 - Z) \sum_{q=0}^{T_0-1} \beta^q u_0 \right] \sum_{j=0}^{n-1} (B^{s-T_0} \beta^{T_0})^j + B^{n(s-T_0)} \beta^{nT_0} x_0
            \end{align*}
        
            As we know both B and $\beta$ are in $[0,1]$, so, we can reformulate the above equation as the sum of a geometric series.
        
            \begin{align*}
                x_k &= (1-B) \left(\frac{1 - B^{s-T_0}}{1 - B} \right) \left(\frac{1 - (B^{s-T_0} \beta^{T_0})^i}{1 - B^{s-T_0} \beta^{T_0}}\right) x_0 + B^{s-T_0} \left[ \alpha \frac{1 - \beta^{T_0}}{1 - \beta} x_0 + (1 - Z) \frac{1 - \beta^{T_0}}{1 - \beta} u_0 \right] \left(\frac{1 - (B^{s-T_0} \beta^{T_0})^i}{1 - B^{s-T_0} \beta^{T_0}}\right)\\
                &\quad + B^{is - iT_0} \beta^{iT_0} x_0\\\\
                &= \left[ \frac{1 - (B^{s-T_0} \beta^{T_0})^i}{1 - B^{s-T_0} \beta^{T_0}} \left(1 - B^{s-T_0} + \alpha B^{s-T_0} \frac{1 - \beta^{T_0}}{1 - \beta} \right) + B^{i(s-T_0)} \beta^{iT_0} \right] x_0 + \left[ \frac{1 - (B^{s-T_0} \beta^{T_0})^i}{1 - B^{s-T_0} \beta^{T_0}} (1 - Z) B^{s-T_0} \frac{1 - \beta^{T_0}}{1 - \beta} \right] u_0\\\\
                &= \left[ \frac{1 - (B^{s} Z^{T_0})^i}{1 - B^{s} Z^{T_0}} \left(1 - B^{s-T_0} + \alpha B^{s-T_0} \frac{1 - \beta^{T_0}}{1 - \beta} \right) + B^{is} Z^{iT_0} \right] x_0 + \left[ \frac{1 - (B^{s} Z^{T_0})^i}{1 - B^{s} Z^{T_0}} (1 - Z) B^{s-T_0} \frac{1 - \beta^{T_0}}{1 - \beta} \right] u_0\\
                &= \left[ \frac{1 - (B^{s} Z^{T_0})^i}{1 - B^{s} Z^{T_0}} \left(1 - B^{s-T_0} + \eta B^{s-T_0} (1 - \beta^{T_0}) \right) + B^{is} Z^{iT_0} \right] x_0 + \left[ \frac{1 - (B^{s} Z^{T_0})^i}{1 - B^{s} Z^{T_0}} (1 - \eta) B^{s-T_0} (1 - \beta^{T_0}) \right] u_0
            \end{align*}
    
            Then, for any $i \geq 1, i \in \mathbb{N}$, we can use $\Gamma_{i}^{(1)}$ to show the influence of the innate opinion ($x_0$) on the agent's opinion at the start of block $i$ under \emph{fixed} policy. Similarly, $\Upsilon_{i}^{(1)}$ demonstrates the impact of the recommendation ($u_0$) on the agent's opinion at the start of block $i$ under \emph{fixed} policy.
            \[x_{i}^{(1)} = \Gamma_{i}^{(1)} x_0 + \Upsilon_{i}^{(1)} u_0 \]
            Furthermore, we can show that the sum of these two influences equals one ($\Gamma_{i}^{(1)} + \Upsilon_{i}^{(1)} = 1$). This implies that the agent's opinion at the start of block $i$ is affected by a convex combination of the innate opinion and the recommendation (meaning that the weights assigned to these influences sum to 1).

            \begin{align*}
                \Gamma_{i}^{(1)} + \Upsilon_{i}^{(1)} &= \left[ \frac{1 - (B^{s} Z^{T_0})^i}{1 - B^{s} Z^{T_0}} \left(1 - B^{s-T_0} + \eta B^{s-T_0} (1 - \beta^{T_0}) \right) + B^{is} Z^{iT_0} \right] + \left[ \frac{1 - (B^{s} Z^{T_0})^i}{1 - B^{s} Z^{T_0}} (1 - \eta) B^{s-T_0} (1 - \beta^{T_0}) \right]\\
                &= \left[ \frac{1 - (B^{s} Z^{T_0})^i}{1 - B^{s} Z^{T_0}} \left(1 - B^{s-T_0} \right) + B^{is} Z^{iT_0} \right] + \left[ \frac{1 - (B^{s} Z^{T_0})^i}{1 - B^{s} Z^{T_0}}  B^{s-T_0} (1 - \beta^{T_0}) \right]\\
                &\quad + \frac{1 - (B^{s} Z^{T_0})^i}{1 - B^{s} Z^{T_0}} \eta B^{s-T_0} (1 - \beta^{T_0}) (1-1)\\
                &= \left[ \frac{1 - (B^{s} Z^{T_0})^i}{1 - B^{s} Z^{T_0}} \left(1 - B^{s-T_0} \right) + B^{is} Z^{iT_0} \right] + \left[ \frac{1 - (B^{s} Z^{T_0})^i}{1 - B^{s} Z^{T_0}}  B^{s-T_0} (1 - \beta^{T_0}) \right]\\
                &= \left[ \frac{1 - (B^{s} Z^{T_0})^i}{1 - B^{s} Z^{T_0}} - \frac{1 - (B^{s} Z^{T_0})^i}{1 - B^{s} Z^{T_0}} B^{s-T_0} + (B^{s} Z^{T_0})^i \right] + \left[ \frac{1 - (B^{s} Z^{T_0})^i}{1 - B^{s} Z^{T_0}}  B^{s-T_0}  - \frac{1 - (B^{s} Z^{T_0})^i}{1 - B^{s} Z^{T_0}}  B^{s-T_0} \beta^{T_0} \right]\\
                &=  \frac{1 - (B^{s} Z^{T_0})^i}{1 - B^{s} Z^{T_0}} + (B^{s} Z^{T_0})^i - \frac{1 - (B^{s} Z^{T_0})^i}{1 - B^{s} Z^{T_0}}  B^{s-T_0} \beta^{T_0}\\
                &=  \frac{1 - (B^{s} Z^{T_0})^i + (B^{s} Z^{T_0})^i - (B^{s} Z^{T_0})^i B^{s} Z^{T_0} - B^{s-T_0} \beta^{T_0} + (B^{s} Z^{T_0})^i B^{s-T_0} \beta^{T_0}}{1 - B^{s} Z^{T_0}}\\
                &=  \frac{1 - (B^{s} Z^{T_0})^i B^{s} Z^{T_0} - B^{s-T_0} \beta^{T_0} + (B^{s} Z^{T_0})^i B^{s} Z^{T_0}}{1 - B^{s} Z^{T_0}}\\
                &=  \frac{1 - B^{s-T_0} \beta^{T_0}}{1 - B^{s} Z^{T_0}} =  \frac{1 - B^{s} Z^{T_0}}{1 - B^{s} Z^{T_0}} = 1
            \end{align*}
            
            As a result, the agent's opinion at the start of block $i$ is a convex combination of $x_0$ and $u_0$ (for any $i \in [1,2,\dots,n-1]$).
            \[x_{i}^{(1)} = (1-\Upsilon_{i}^{(1)}) x_0 + \Upsilon_{i}^{(1)} u_0 \]

            \item Next, we consider an agent following a \emph{decreasing} policy. Using the same approach as before, we express the agent's opinion at the beginning of block $i$ as a function of $x_0$ and $u_0$. However, we first assume that block $i$ is in transient blocks, for any $i \in [1,2,\dots,m_D]$. From block 1 to block $m_D$, the agent has some clicking periods, but after that the agent does not click anymore. Let $k=is$, then we have:

            \begin{align*}
                x_k &= \frac{\alpha}{\alpha+\beta} x_{0} + \frac{\beta}{\alpha+\beta} x_{k-1} = (1-B) (1 + B) x_0 + B^2 x_{k-2} = (1-B) \sum_{q=0}^{s-\frac{T_0}{\kappa^{i-1}}-1} B^q x_0 + B^{s-\frac{T_0}{\kappa^{i-1}}} x_{k - s + \frac{T_0}{\kappa^{i-1}}} \\
                &= (1-B) \sum_{q=0}^{s-\frac{T_0}{\kappa^{i-1}}-1} B^q x_0 + B^{s-\frac{T_0}{\kappa^{i-1}}} \left(\alpha x_0 + (1 - \alpha - \beta) u_0 + \beta x_{k - s + \frac{T_0}{\kappa^{i-1}} - 1} \right)\\
                &= (1-B) \sum_{q=0}^{s-\frac{T_0}{\kappa^{i-1}}-1} B^q x_0 + B^{s-\frac{T_0}{\kappa^{i-1}}} \left(\alpha (1 + \beta) x_0 + (1 - Z) (1 + \beta) u_0 + \beta^2 x_{k - s + \frac{T_0}{\kappa^{i-1}} - 2} \right)\\
                &= (1-B) \sum_{q=0}^{s-\frac{T_0}{\kappa^{i-1}}-1} B^q x_0 + B^{s-\frac{T_0}{\kappa^{i-1}}} \left(\alpha (1 + \beta + \beta^2) x_0 + (1 - Z) (1 + \beta + \beta^2) u_0 + \beta^3 x_{k - s + \frac{T_0}{\kappa^{i-1}} - 3} \right)\\
                &= (1-B) \sum_{q=0}^{s-\frac{T_0}{\kappa^{i-1}}-1} B^q x_0 + B^{s-\frac{T_0}{\kappa^{i-1}}} \left(\alpha \sum_{q=0}^{\frac{T_0}{\kappa^{i-1}}-1} \beta^q x_0 + (1 - Z) \sum_{q=0}^{\frac{T_0}{\kappa^{i-1}}-1} \beta^q u_0 \right) + B^{s-\frac{T_0}{\kappa^{i-1}}} \beta^{\frac{T_0}{\kappa^{i-1}}} x_{k - s}
            \end{align*}
            The previous equation is the opinion at the start of block $i$ as a function of the innate opinion, recommendation, and the opinion at the start of the previous block. By applying the recursive approach, we derive an equation for the agent's opinion at the start of block $i$, which is only a function of $x_0$ and $u_0$, under \emph{decreasing} policy in transient states.

            \begin{align*}
                x_k &= (1-B) \sum_{q=0}^{s-\frac{T_0}{\kappa^{i-1}}-1} B^q x_0 + B^{s-\frac{T_0}{\kappa^{i-1}}} \left(\alpha \sum_{q=0}^{\frac{T_0}{\kappa^{i-1}}-1} \beta^q x_0 + (1 - Z) \sum_{q=0}^{\frac{T_0}{\kappa^{i-1}}-1} \beta^q u_0\right)\\
                &\quad + B^{s-\frac{T_0}{\kappa^{i-1}}} \beta^{\frac{T_0}{\kappa^{i-1}}} \left( (1-B) \sum_{q=0}^{s-\frac{T_0}{\kappa^{i-2}}-1} B^q x_0 + B^{s-\frac{T_0}{\kappa^{i-2}}} x_{k - 2s + \frac{T_0}{\kappa^{i-2}}} \right)\\
                &= (1-B) \sum_{q=0}^{s-\frac{T_0}{\kappa^{i-1}}-1} B^q x_0 + B^{s-\frac{T_0}{\kappa^{i-1}}} \left(\alpha \sum_{q=0}^{\frac{T_0}{\kappa^{i-1}}-1} \beta^q x_0 + (1 - Z) \sum_{q=0}^{\frac{T_0}{\kappa^{i-1}}-1} \beta^q u_0\right) + B^{s-\frac{T_0}{\kappa^{i-1}}} \beta^{\frac{T_0}{\kappa^{i-1}}} \left( (1-B) \sum_{q=0}^{s-\frac{T_0}{\kappa^{i-2}}-1} B^q x_0 \right)\\
                & \quad + B^{2s-\frac{T_0}{\kappa^{i-1}}-\frac{T_0}{\kappa^{i-2}}} \beta^{\frac{T_0}{\kappa^{i-1}}} \left(\alpha \sum_{q=0}^{\frac{T_0}{\kappa^{i-2}}-1} \beta^q x_0 + (1 - Z) \sum_{q=0}^{\frac{T_0}{\kappa^{i-2}}-1} \beta^q u_0\right) + B^{2s-\frac{T_0}{\kappa^{i-1}}-\frac{T_0}{\kappa^{i-2}}} \beta^{\frac{T_0}{\kappa^{i-1}} + \frac{T_0}{\kappa^{i-2}}} x_{k-2s}\\
                &= (1-B) \sum_{q=0}^{s-\frac{T_0}{\kappa^{i-1}}-1} B^q x_0 + B^{s-\frac{T_0}{\kappa^{i-1}}} \left(\alpha \sum_{q=0}^{\frac{T_0}{\kappa^{i-1}}-1} \beta^q x_0 + (1 - Z) \sum_{q=0}^{\frac{T_0}{\kappa^{i-1}}-1} \beta^q u_0\right) + B^{s-\frac{T_0}{\kappa^{i-1}}} \beta^{\frac{T_0}{\kappa^{i-1}}} \left( (1-B) \sum_{q=0}^{s-\frac{T_0}{\kappa^{i-2}}-1} B^q x_0 \right) \\
                & \quad + B^{2s-\frac{T_0}{\kappa^{i-1}}-\frac{T_0}{\kappa^{i-2}}} \beta^{\frac{T_0}{\kappa^{i-1}}} \left(\alpha \sum_{q=0}^{\frac{T_0}{\kappa^{i-2}}-1} \beta^q x_0 + (1 - Z) \sum_{q=0}^{\frac{T_0}{\kappa^{i-2}}-1} \beta^q u_0\right)\\
                &\quad +B^{2s-\frac{T_0}{\kappa^{i-1}}-\frac{T_0}{\kappa^{i-2}}} \beta^{\frac{T_0}{\kappa^{i-1}} + \frac{T_0}{\kappa^{i-2}}} \left((1-B) \sum_{q=0}^{s-\frac{T_0}{\kappa^{i-3}}-1} B^q x_0 + B^{s-\frac{T_0}{\kappa^{i-3}}} x_{k - 3s + \frac{T_0}{\kappa^{i-3}}} \right)
            \end{align*}
             \begin{align*}
                x_k &= (1-B) \sum_{q=0}^{s-\frac{T_0}{\kappa^{i-1}}-1} B^q x_0 + B^{s-\frac{T_0}{\kappa^{i-1}}} \left(\alpha \sum_{q=0}^{\frac{T_0}{\kappa^{i-1}}-1} \beta^q x_0 + (1 - Z) \sum_{q=0}^{\frac{T_0}{\kappa^{i-1}}-1} \beta^q u_0\right) + B^{s-\frac{T_0}{\kappa^{i-1}}} \beta^{\frac{T_0}{\kappa^{i-1}}} \left( (1-B) \sum_{q=0}^{s-\frac{T_0}{\kappa^{i-2}}-1} B^q x_0 \right)\\
                & \quad + B^{2s-\frac{T_0}{\kappa^{i-1}}-\frac{T_0}{\kappa^{i-2}}} \beta^{\frac{T_0}{\kappa^{i-1}}} \left(\alpha \sum_{q=0}^{\frac{T_0}{\kappa^{i-2}}-1} \beta^q x_0 + (1 - Z) \sum_{q=0}^{\frac{T_0}{\kappa^{i-2}}-1} \beta^q u_0\right) + B^{2s-\frac{T_0}{\kappa^{i-1}}-\frac{T_0}{\kappa^{i-2}}} \beta^{\frac{T_0}{\kappa^{i-1}} + \frac{T_0}{\kappa^{i-2}}} \left((1-B) \sum_{q=0}^{s-\frac{T_0}{\kappa^{i-3}}-1} B^q x_0 \right)\\
                &\quad + B^{3s-\frac{T_0}{\kappa^{i-1}}-\frac{T_0}{\kappa^{i-2}} - \frac{T_0}{\kappa^{i-3}}} \beta^{\frac{T_0}{\kappa^{i-1}} + \frac{T_0}{\kappa^{i-2}}} \left(\alpha \sum_{q=0}^{\frac{T_0}{\kappa^{i-3}}-1} \beta^q x_0 + (1 - Z) \sum_{q=0}^{\frac{T_0}{\kappa^{i-3}}-1} \beta^q u_0\right) + B^{3s-\frac{T_0}{\kappa^{i-1}}-\frac{T_0}{\kappa^{i-2}}-\frac{T_0}{\kappa^{i-3}}} \beta^{\frac{T_0}{\kappa^{i-1}} + \frac{T_0}{\kappa^{i-2}} + \frac{T_0}{\kappa^{i-3}}} x_{k-3s}\\
                &= (1-B) \left(\sum_{j=0}^{i-1} B^{js} \left(\sum_{q=0}^{s-\frac{T_0}{\kappa^{i-j-1}}-1}B^q\right) \left(\prod_{q=0}^{j-1} (B^{-1}\beta)^{\frac{T_0}{\kappa^{i-q-1}}} \right) \right) x_0 + \left(\sum_{j=0}^{i-1} B^{(j+1)s} \alpha \left(\sum_{q=0}^{\frac{T_0}{\kappa^{i-j-1}}-1}\beta^q\right) \left(\prod_{q=0}^{j} B^{-\frac{T_0}{\kappa^{i-q-1}}} \right) \left(\prod_{q=0}^{j-1} \beta^{\frac{T_0}{\kappa^{i-q-1}}} \right) \right) x_0\\
                &\quad + \left(\sum_{j=0}^{i-1} B^{(j+1)s} (1 - Z) \left(\sum_{q=0}^{\frac{T_0}{\kappa^{i-j-1}}-1}\beta^q\right) \left(\prod_{q=0}^{j} B^{-\frac{T_0}{\kappa^{i-q-1}}} \right) \left(\prod_{q=0}^{j-1} \beta^{\frac{T_0}{\kappa^{i-q-1}}} \right) \right) u_0 + B^k \left(\prod_{q=0}^{i-1} (B^{-1}\beta)^{\frac{T_0}{\kappa^{i-q-1}}} \right) x_0
            \end{align*}
    
            Since $B$ and $\beta$ are within the range $[0,1]$, we can reformulate the above equation as a sum of a geometric series.

            \begin{align*}
                x_k &= (1-B) \left(\sum_{j=0}^{i-1} B^{js} \frac{1 - B^{s - \frac{T_0}{\kappa^{i-j-1}}}}{1-B} (B^{-1}\beta)^{T_0 \kappa^{1-i}\frac{\kappa^j - 1}{\kappa - 1}} \right) x_0 + \left(\sum_{j=0}^{i-1} B^{(j+1)s} \alpha \frac{1 - \beta^{\frac{T_0}{\kappa^{i-j-1}}}}{1-\beta} B^{-T_0 \kappa^{1-i}\frac{\kappa^{j+1} - 1}{\kappa - 1}} \beta^{T_0 \kappa^{1-i}\frac{\kappa^j - 1}{\kappa - 1}} \right) x_0\\
                &\quad + \left(\sum_{j=0}^{i-1} B^{(j+1)s} (1 - Z) \frac{1 - \beta^{\frac{T_0}{\kappa^{i-j-1}}}}{1-\beta} B^{-T_0 \kappa^{1-i}\frac{\kappa^{j+1} - 1}{\kappa - 1}} \beta^{T_0 \kappa^{1-i}\frac{\kappa^j - 1}{\kappa - 1}} \right) u_0 + B^k (B^{-1}\beta)^{T_0 \kappa^{1-i} \frac{\kappa^{i} - 1}{\kappa - 1}} x_0\\
                &= \Bigg(\left(\sum_{j=0}^{i-1} B^{js} (1 - B^{s - \frac{T_0}{\kappa^{i-j-1}}}) Z^{T_0 \kappa^{1-i}\frac{\kappa^j - 1}{\kappa - 1}} \right) + \left(\sum_{j=0}^{i-1} B^{(j+1)s} \eta (1 - \beta^{\frac{T_0}{\kappa^{i-j-1}}}) B^{-\frac{T_0}{\kappa^{i-j-1}}} Z^{T_0 \kappa^{1-i}\frac{\kappa^j - 1}{\kappa - 1}} \right) + B^k Z^{T_0 \kappa^{1-i} \frac{\kappa^{i} - 1}{\kappa - 1}}\Bigg) x_0\\
                &\quad + \left(\sum_{j=0}^{i-1} B^{(j+1)s} (1 - \eta) (1 - \beta^{\frac{T_0}{\kappa^{i-j-1}}}) B^{-\frac{T_0}{\kappa^{i-j-1}}} Z^{T_0 \kappa^{1-i}\frac{\kappa^j - 1}{\kappa - 1}} \right) u_0
            \end{align*}
    
            Using $\Gamma_{i}^{(2)}$ and $\Upsilon_{i}^{(2)}$ to show the influence of $x_0$ and $u_0$ on the agent's opinion at the start of block $i$ (in transient states) under \emph{decreasing} policy, respectively, we have:
            \[x_{i}^{(2)} = \Gamma_{i}^{(2)} x_0 + \Upsilon_{i}^{(2)} u_0 \]
            Also, we can show $\Gamma_{i}^{(2)} + \Upsilon_{i}^{(2)} = 1$.
    
            \begin{align*}
                \Gamma_{i}^{(2)} + \Upsilon_{i}^{(2)} &= \Bigg(\left(\sum_{j=0}^{i-1} B^{js} (1 - B^{s - \frac{T_0}{\kappa^{i-j-1}}}) Z^{T_0 \kappa^{1-i}\frac{\kappa^j - 1}{\kappa - 1}} \right) + \left(\sum_{j=0}^{i-1} B^{(j+1)s} \eta (1 - \beta^{\frac{T_0}{\kappa^{i-j-1}}}) B^{-\frac{T_0}{\kappa^{i-j-1}}} Z^{T_0 \kappa^{1-i}\frac{\kappa^j - 1}{\kappa - 1}} \right) + B^k Z^{T_0 \kappa^{1-i} \frac{\kappa^{i} - 1}{\kappa - 1}}\Bigg)\\
                &\quad + \left(\sum_{j=0}^{i-1} B^{(j+1)s} (1 - \eta) (1 - \beta^{\frac{T_0}{\kappa^{i-j-1}}}) B^{-\frac{T_0}{\kappa^{i-j-1}}} Z^{T_0 \kappa^{1-i}\frac{\kappa^j - 1}{\kappa - 1}} \right)\\
                &= \left(\sum_{j=0}^{i-1} B^{js} (1 - B^{s - \frac{T_0}{\kappa^{i-j-1}}}) Z^{T_0 \kappa^{1-i}\frac{\kappa^j - 1}{\kappa - 1}} \right) + B^k Z^{T_0 \kappa^{1-i} \frac{\kappa^{i} - 1}{\kappa - 1}} + \left(\sum_{j=0}^{i-1} B^{(j+1)s} (1 - \beta^{\frac{T_0}{\kappa^{i-j-1}}}) B^{-\frac{T_0}{\kappa^{i-j-1}}} Z^{T_0 \kappa^{1-i}\frac{\kappa^j - 1}{\kappa - 1}} \right)\\
                &\quad \left(\sum_{j=0}^{i-1} B^{(j+1)s} \eta (1 - \beta^{\frac{T_0}{\kappa^{i-j-1}}}) B^{-\frac{T_0}{\kappa^{i-j-1}}} Z^{T_0 \kappa^{1-i}\frac{\kappa^j - 1}{\kappa - 1}} \right) (1 - 1)\\
                &= \left(\sum_{j=0}^{i-1} B^{js} (1 - B^{s - \frac{T_0}{\kappa^{i-j-1}}}) Z^{T_0 \kappa^{1-i}\frac{\kappa^j - 1}{\kappa - 1}} \right) + B^k Z^{T_0 \kappa^{1-i} \frac{\kappa^{i} - 1}{\kappa - 1}} + \left(\sum_{j=0}^{i-1} B^{(j+1)s} (1 - \beta^{\frac{T_0}{\kappa^{i-j-1}}}) B^{-\frac{T_0}{\kappa^{i-j-1}}} Z^{T_0 \kappa^{1-i}\frac{\kappa^j - 1}{\kappa - 1}} \right)
            \end{align*}
            \begin{align*}
                \Gamma_{i}^{(2)} + \Upsilon_{i}^{(2)} &= \left(\sum_{j=0}^{i-1} B^{js} (1 - B^{s - \frac{T_0}{\kappa^{i-j-1}}}) Z^{T_0 \kappa^{1-i}\frac{\kappa^j - 1}{\kappa - 1}} \right) + B^k Z^{T_0 \kappa^{1-i} \frac{\kappa^{i} - 1}{\kappa - 1}} + \left(\sum_{j=0}^{i-1} B^{(j+1)s} B^{-\frac{T_0}{\kappa^{i-j-1}}} Z^{T_0 \kappa^{1-i}\frac{\kappa^j - 1}{\kappa - 1}} \right)\\
                &\quad - \left(\sum_{j=0}^{i-1} B^{(j+1)s} \beta^{\frac{T_0}{\kappa^{i-j-1}}} B^{-\frac{T_0}{\kappa^{i-j-1}}} Z^{T_0 \kappa^{1-i}\frac{\kappa^j - 1}{\kappa - 1}} \right)\\
                &= \left(\sum_{j=0}^{i-1} B^{js} Z^{T_0 \kappa^{1-i}\frac{\kappa^j - 1}{\kappa - 1}} \right) - \left(\sum_{j=0}^{i-1} B^{(j+1)s - \frac{T_0}{\kappa^{i-j-1}}} Z^{T_0 \kappa^{1-i}\frac{\kappa^j - 1}{\kappa - 1}} \right) + \left(\sum_{j=0}^{i-1} B^{(j+1)s-\frac{T_0}{\kappa^{i-j-1}}} Z^{T_0 \kappa^{1-i}\frac{\kappa^j - 1}{\kappa - 1}} \right)\\
                &\quad - \left(\sum_{j=0}^{i-1} B^{(j+1)s} Z^{T_0 \kappa^{1-i+j}} Z^{T_0 \kappa^{1-i}\frac{\kappa^j - 1}{\kappa - 1}} \right) + B^k Z^{T_0 \kappa^{1-i} \frac{\kappa^{i} - 1}{\kappa - 1}}\\
                &= \left(\sum_{j=0}^{i-1} B^{js} Z^{T_0 \kappa^{1-i}\frac{\kappa^j - 1}{\kappa - 1}} \right) - \left(\sum_{j=0}^{i-1} B^{(j+1)s} Z^{T_0 \kappa^{1-i}\frac{\kappa^{j+1} - 1}{\kappa - 1}} \right)  + B^k Z^{T_0 \kappa^{1-i} \frac{\kappa^{i} - 1}{\kappa - 1}}\\
                &= 1 + \left(\sum_{j=1}^{i-1} B^{js} Z^{T_0 \kappa^{1-i}\frac{\kappa^j - 1}{\kappa - 1}} \right) - \left(\sum_{j=0}^{i-2} B^{(j+1)s} Z^{T_0 \kappa^{1-i}\frac{\kappa^{j+1} - 1}{\kappa - 1}} \right) - B^{k} Z^{T_0 \kappa^{1-i}\frac{\kappa^{i} - 1}{\kappa - 1}}  + B^k Z^{T_0 \kappa^{1-i} \frac{\kappa^{i} - 1}{\kappa - 1}}\\
                &= 1 + \left(\sum_{j=1}^{i-1} B^{js} Z^{T_0 \kappa^{1-i}\frac{\kappa^j - 1}{\kappa - 1}} \right) - \left(\sum_{j=1}^{i-1} B^{js} Z^{T_0 \kappa^{1-i}\frac{\kappa^{j} - 1}{\kappa - 1}} \right) = 1
            \end{align*}
    
            Therefore, the agent's opinion at the start of block $i$ (in transient states) is a convex combination of $x_0$ and $u_0$.

            \[x_{i}^{(2)} = (1-\Upsilon_{i}^{(2)}) x_0 + \Upsilon_{i}^{(2)} u_0 \]
    
            Now, we find the agent's opinion at the start of block $i$ in steady states, where the agent follows the \emph{decreasing} policy (for any $i > m_D, i \in \mathbb{N}$). The agent stops clicking after block $m_D$, meaning there is no influence from recommendation on the opinion. As a result, the opinion at the start of block $i$ is a function of $x_0$ and $x_{m_D}^{(2)}$. Let $k=is$:

            \begin{align*}
                x_{i}^{(2)} &= (1-B)x_0 + Bx_{k-1} = (1-B)(1+B)x_0 + B^2x_{k-2} = (1-B)(1+B+B^2)x_0 + B^3x_{k-3}\\
                &= (1-B) \sum_{q=0}^{s-1} B^q x_0 + B^{s} x_{k - s} = (1-B) \sum_{q=0}^{s-1} B^q x_0 + B^{s} x_{i-1}^{(2)} \\
            \end{align*}

            The above equations shows the opinion based on a function of the innate opinion and the opinion at the start of previous block. Doing the recursive approach, we have:
    
            \begin{align*}
                x_{i}^{(2)} &= (1-B) \sum_{q=0}^{s-1} B^q x_0 + B^{s} \Bigg((1-B) \sum_{q=0}^{s-1} B^q x_0 + B^{s} x_{i-2}^{(2)}\Bigg) = (1-B) (1+B^s) \sum_{q=0}^{s-1} B^q x_0 + B^{2s} x_{i-2}^{(2)} \\
                &= (1-B) (1+B^s+B^{2s}) \sum_{q=0}^{s-1} B^q x_0 + B^{3s} x_{i-3}^{(2)} = (1-B) \Bigg(\sum_{j=0}^{i-m_D-1} B^{js}  \Bigg) \sum_{q=0}^{s-1} B^q x_0 + B^{(i-m_D)s} x_{m_D}^{(2)} \\
            \end{align*}
    
            Since $B \in [0,1]$, we can use the summation of a geometric series to demonstrate the above equation as:
    
            \begin{align*}
                x_{i}^{(2)} &= (1-B) \frac{1-B^{(i-m_D)s}}{1-B^s} \frac{1-B^{s}}{1-B} x_0 + B^{(i-m_D)s} x_{m_D}^{(2)} = (1-B^{(i-m_D)s}) x_0 + B^{(i-m_D)s} x_{m_D}^{(2)} = \Gamma_{i}^{(2)} x_0 + \Upsilon_{i}^{(2)} u_0
            \end{align*}
    
            Obviously, $\Gamma_{i}^{(2)} + \Upsilon_{i}^{(2)} = 1$, So, the opinion at the start of block $i$ at steady states (for any $i > m_D, i \in \mathbb{N}$) is again a convex combination of the innate opinion and the recommendation.

            \item To study the influence of the \emph{adaptive decreasing} policy on the agent's opinion, at the start of block $i$, where $i$ is in transient blocks, we use the same recursive approach for opinion changes (for any $i \in [1,2,\dots,m_{AD}]$). Let $m_{AD}$ be the last block in transient states, after which the agent has a fixed clicking rate. In other words, the agent decreases its clicking rate at each block from block 1 to $m_{AD}$.
            
            \begin{align*}
                x_{k} &= \frac{\alpha}{\alpha + \beta} x_0 + \frac{\beta}{\alpha + \beta} x_{k-1} = (1-B) x_0 + B x_{k-1} = (1-B) (1 + B) x_0 + B^2 x_{k-2}\\
                &=(1-B) \sum_{q=0}^{s-T_0+(i-1)\tau-1} B^q x_0 + B^{s-T_0+(i-1)\tau} x_{k-s+T_0-(i-1)\tau}\\
                &= (1-B) \sum_{q=0}^{s-T_0+(i-1)\tau-1} B^q x_0 + B^{s-T_0+(i-1)\tau} \Bigg(\alpha x_0 + \beta x_{k-s+T_0-(i-1)\tau-1} +(1-Z)u_0 \Bigg)\\
                &= (1-B) \sum_{q=0}^{s-T_0+(i-1)\tau-1} B^q x_0 + B^{s-T_0+(i-1)\tau} \Bigg(\alpha (1+\beta) x_0 + \beta^2 x_{k-s+T_0-(i-1)\tau-2} +(1-Z) (1+\beta) u_0 \Bigg)\\
                &= (1-B) \sum_{q=0}^{s-T_0+(i-1)\tau-1} B^q x_0 + B^{s-T_0+(i-1)\tau} \Bigg(\alpha (1+\beta+\beta^2) x_0 + \beta^3 x_{k-s+T_0-(i-1)\tau-3} +(1-Z) (1+\beta+\beta^2) u_0 \Bigg)\\
                &= (1-B) \sum_{q=0}^{s-T_0+(i-1)\tau-1} B^q x_0 + B^{s-T_0+(i-1)\tau} \Bigg(\alpha \sum_{q=0}^{T_0-(i-1)\tau-1} \beta^q x_0 +(1-Z) \sum_{q=0}^{T_0-(i-1)\tau-1} \beta^q u_0 \Bigg)\\
                &\quad + B^{s-T_0+(i-1)\tau} \beta^{T_0-(i-1)\tau} x_{k-s}
            \end{align*}
            The previous equation demonstrates the opinion at the start of block $i$ as a function of the innate opinion, recommendation, and the opinion at the start of the previous block. By applying the same recursive approach, we obtain:

            \begin{align*}
                x_{k} &= (1-B) \sum_{q=0}^{s-T_0+(i-1)\tau-1} B^q x_0 + B^{s-T_0+(i-1)\tau} \Bigg(\alpha \sum_{q=0}^{T_0-(i-1)\tau-1} \beta^q x_0 + (1-Z) \sum_{q=0}^{T_0-(i-1)\tau-1} \beta^q u_0 \Bigg)\\
                &\quad + B^{s-T_0+(i-1)\tau} \beta^{T_0-(i-1)\tau} \Bigg((1-B) \sum_{q=0}^{s-T_0+(i-2)\tau-1} B^q x_0\Bigg) + B^{2(s-T_0)+\tau(2i-3)} \beta^{T_0-(i-1)\tau} x_{k-2s+T_0-(i-2)\tau}\\
                &= (1-B) \sum_{q=0}^{s-T_0+(i-1)\tau-1} B^q x_0 + B^{s-T_0+(i-1)\tau} \Bigg(\alpha \sum_{q=0}^{T_0-(i-1)\tau-1} \beta^q x_0 + (1-Z) \sum_{q=0}^{T_0-(i-1)\tau-1} \beta^q u_0 \Bigg)\\
                &\quad + B^{s-T_0+(i-1)\tau} \beta^{T_0-(i-1)\tau} \Bigg((1-B) \sum_{q=0}^{s-T_0+(i-2)\tau-1} B^q x_0\Bigg) + B^{2(s-T_0)+\tau(2i-3)} \beta^{T_0-(i-1)\tau} \Bigg(\alpha \sum_{q=0}^{T_0-(i-2)\tau-1} \beta^q x_0\\
                &\quad + (1-Z) \sum_{q=0}^{T_0-(i-2)\tau-1} \beta^q u_0 \Bigg) + B^{2(s-T_0)+\tau(2i-3)} \beta^{2T_0-\tau(2i-3)} x_{k-2s}\\
                &= (1-B) \sum_{q=0}^{s-T_0+(i-1)\tau-1} B^q x_0 + B^{s-T_0+(i-1)\tau} \Bigg(\alpha \sum_{q=0}^{T_0-(i-1)\tau-1} \beta^q x_0 + (1-Z) \sum_{q=0}^{T_0-(i-1)\tau-1} \beta^q u_0 \Bigg)\\
                &\quad + B^{s-T_0+(i-1)\tau} \beta^{T_0-(i-1)\tau} \Bigg((1-B) \sum_{q=0}^{s-T_0+(i-2)\tau-1} B^q x_0\Bigg) + B^{2(s-T_0)+\tau(2i-3)} \beta^{T_0-(i-1)\tau} \Bigg(\alpha \sum_{q=0}^{T_0-(i-2)\tau-1} \beta^q x_0\\
                &\quad + (1-Z) \sum_{q=0}^{T_0-(i-2)\tau-1} \beta^q u_0 \Bigg) + B^{2(s-T_0)+\tau(2i-3)} \beta^{2T_0-\tau(2i-3)} \Bigg( (1-B) \sum_{q=0}^{s-T_0+(i-3)\tau-1} B^q x_0 + B^{s-T_0+(i-3)\tau} x_{k-3s+T_0-(i-3)\tau}\Bigg)
            \end{align*}
            \begin{align*}
                x_{k} &= (1-B) \sum_{q=0}^{s-T_0+(i-1)\tau-1} B^q x_0 + B^{s-T_0+(i-1)\tau} \Bigg(\alpha \sum_{q=0}^{T_0-(i-1)\tau-1} \beta^q x_0 + (1-Z) \sum_{q=0}^{T_0-(i-1)\tau-1} \beta^q u_0 \Bigg)\\
                &\quad + B^{s-T_0+(i-1)\tau} \beta^{T_0-(i-1)\tau} \Bigg((1-B) \sum_{q=0}^{s-T_0+(i-2)\tau-1} B^q x_0\Bigg) + B^{2(s-T_0)+\tau(2i-3)} \beta^{T_0-(i-1)\tau} \Bigg(\alpha \sum_{q=0}^{T_0-(i-2)\tau-1} \beta^q x_0\\
                &\quad + (1-Z) \sum_{q=0}^{T_0-(i-2)\tau-1} \beta^q u_0 \Bigg) + B^{2(s-T_0)+\tau(2i-3)} \beta^{2T_0-\tau(2i-3)} \Bigg( (1-B) \sum_{q=0}^{s-T_0+(i-3)\tau-1} B^q x_0\Bigg)\\
                &\quad + B^{3(s-T_0)+\tau(3i-6)} \beta^{2T_0-\tau(2i-3)} \Bigg(\alpha \sum_{q=0}^{T_0-(i-3)\tau-1} \beta^q x_0 + (1-Z) \sum_{q=0}^{T_0-(i-3)\tau-1} \beta^q u_0 \Bigg) + B^{3(s-T_0)+\tau(3i-6)} \beta^{3T_0-\tau(3i-6)} x_{k-3s}\\
                &= \Bigg[ (1-B) \left(\sum_{j=0}^{i-1} B^{js} \left(\sum_{q=0}^{s-T_0+(i-1-j)\tau-1} B^q\right) \left(\prod_{q=0}^{j-1} (B^{-1}\beta)^{T_0-(i-1-q)\tau} \right) \right)\\
                & \quad + \left(\sum_{j=0}^{i-1} B^{(j+1)s} \alpha \left(\sum_{q=0}^{T_0-(i-1-j)\tau-1}\beta^q\right) \left(\prod_{q=0}^{j} B^{-T_0+(i-1-q)\tau} \right) \left(\prod_{q=0}^{j-1} \beta^{T_0-(i-1-q)\tau} \right) \right) + B^k \left(\prod_{q=0}^{i-1} (B^{-1}\beta)^{T_0-(i-1-q)\tau} \right) \Bigg] x_0\\
                &\quad + \left(\sum_{j=0}^{i-1} B^{(j+1)s} (1 - Z) \left(\sum_{q=0}^{T_0-(i-1-j)\tau-1}\beta^q\right) \left(\prod_{q=0}^{j} B^{-T_0+(i-1-q)\tau} \right) \left(\prod_{q=0}^{j-1} \beta^{T_0-(i-1-q)\tau} \right) \right) u_0\\
                &= \Bigg[ (1-B) \left(\sum_{j=0}^{i-1} B^{js} \left(\sum_{q=0}^{s-T_0+(i-1-j)\tau-1} B^q\right) \left(\prod_{q=0}^{j-1} Z^{T_0-(i-1-q)\tau} \right) \right)\\
                & \quad + \left(\sum_{j=0}^{i-1} B^{(j+1)s} \alpha \left(\sum_{q=0}^{T_0-(i-1-j)\tau-1}\beta^q\right) \left(\prod_{q=0}^{j} B^{-T_0+(i-1-q)\tau} \right) \left(\prod_{q=0}^{j-1} \beta^{T_0-(i-1-q)\tau} \right) \right) + B^k \left(\prod_{q=0}^{i-1} Z^{T_0-(i-1-q)\tau} \right) \Bigg] x_0\\
                &\quad + \left(\sum_{j=0}^{i-1} B^{(j+1)s} (1 - Z) \left(\sum_{q=0}^{T_0-(i-1-j)\tau-1}\beta^q\right) \left(\prod_{q=0}^{j} B^{-T_0+(i-1-q)\tau} \right) \left(\prod_{q=0}^{j-1} \beta^{T_0-(i-1-q)\tau} \right) \right) u_0
            \end{align*}
            Because of the fact that both $B$ and $\beta$ are within $[0,1]$, we can use the summation of geometric series to demonstrate the above equation as:
            \begin{align*}
                x_{k} &= \Bigg[\Bigg(\sum_{j=0}^{i-1} B^{js} (1-B^{s-T_0+(i-1-j)\tau}) Z^{j(T_0-\tau i)+\tau\frac{j^2+j}{2}}\Bigg) + \eta \Bigg(\sum_{j=0}^{i-1} B^{(j+1)s-T_0+(i-1-j)\tau} (1-\beta^{T_0-(i-1-j)\tau}) Z^{j(T_0-\tau i)+\tau\frac{j^2+j}{2}}\Bigg)\\
                & \quad + B^k  Z^{iT_0+\tau\frac{i-i^2}{2}} \Bigg] x_0 + (1-\eta) \Bigg(\sum_{j=0}^{i-1} B^{(j+1)s-T_0+(i-1-j)\tau} Z^{j(T_0-\tau i)+\tau\frac{j^2+j}{2}} (1-\beta^{T_0-(i-1-j)\tau})\Bigg)u_0\\
                &= \Gamma_{i}^{(3)}x_0 + \Upsilon_{i}^{(3)}u_0
            \end{align*}

            Now, we will demonstrate that we can express the opinion as a convex combination of $x_0$ and $u_0$. This can be accomplished by showing that $\Gamma_{i}^{(3)} + \Upsilon_{i}^{(3)} = 1$.

            \begin{align*}
                \Gamma_{i}^{(3)}x_0 + \Upsilon_{i}^{(3)}u_0 &= \Bigg[\Bigg(\sum_{j=0}^{i-1} B^{js} (1-B^{s-T_0+(i-1-j)\tau}) Z^{j(T_0-\tau i)+\tau\frac{j^2+j}{2}}\Bigg)\\
                &\quad + \eta \Bigg(\sum_{j=0}^{i-1} B^{(j+1)s-T_0+(i-1-j)\tau} (1-\beta^{T_0-(i-1-j)\tau}) Z^{j(T_0-\tau i)+\tau\frac{j^2+j}{2}}\Bigg)\\
                & \quad + B^k  Z^{iT_0+\tau\frac{i-i^2}{2}} \Bigg] + (1-\eta) \Bigg(\sum_{j=0}^{i-1} B^{(j+1)s-T_0+(i-1-j)\tau} Z^{j(T_0-\tau i)+\tau\frac{j^2+j}{2}} (1-\beta^{T_0-(i-1-j)\tau})\Bigg)
            \end{align*}
            \begin{align*}
                \Gamma_{i}^{(3)}x_0 + \Upsilon_{i}^{(3)}u_0 &= \Bigg[\Bigg(\sum_{j=0}^{i-1} B^{js} (1-B^{s-T_0+(i-1-j)\tau}) Z^{j(T_0-\tau i)+\tau\frac{j^2+j}{2}}\Bigg) + B^k  Z^{iT_0+\tau\frac{i-i^2}{2}} \Bigg]\\
                & \quad + \Bigg(\sum_{j=0}^{i-1} B^{(j+1)s-T_0+(i-1-j)\tau} Z^{j(T_0-\tau i)+\tau\frac{j^2+j}{2}} (1-\beta^{T_0-(i-1-j)\tau})\Bigg)\\
                &= \Bigg[\Bigg(\sum_{j=0}^{i-1} B^{js} Z^{j(T_0-\tau i)+\tau\frac{j^2+j}{2}}\Bigg) - \Bigg(\sum_{j=0}^{i-1} B^{(j+1)s-T_0+(i-1-j)\tau} Z^{j(T_0-\tau i)+\tau\frac{j^2+j}{2}}\Bigg) + B^k  Z^{iT_0+\tau\frac{i-i^2}{2}} \Bigg]\\
                & \quad + \Bigg(\sum_{j=0}^{i-1} B^{(j+1)s-T_0+(i-1-j)\tau} Z^{j(T_0-\tau i)+\tau\frac{j^2+j}{2}}\Bigg) - \Bigg(\sum_{j=0}^{i-1} B^{(j+1)s-T_0+(i-1-j)\tau} Z^{j(T_0-\tau i)+\tau\frac{j^2+j}{2}} \beta^{T_0-(i-1-j)\tau}\Bigg)\\
                &= \Bigg(\sum_{j=0}^{i-1} B^{js} Z^{j(T_0-\tau i)+\tau\frac{j^2+j}{2}}\Bigg) + B^k  Z^{iT_0+\tau\frac{i-i^2}{2}} - \Bigg(\sum_{j=0}^{i-1} B^{(j+1)s-T_0+(i-1-j)\tau} Z^{j(T_0-\tau i)+\tau\frac{j^2+j}{2}} \beta^{T_0-(i-1-j)\tau}\Bigg)\\
                &= \Bigg(\sum_{j=0}^{i-1} B^{js} Z^{j(T_0-\tau i)+\tau\frac{j^2+j}{2}}\Bigg) + B^k  Z^{iT_0+\tau\frac{i-i^2}{2}} - \Bigg(\sum_{j=0}^{i-1} B^{(j+1)s} Z^{(j+1)(T_0-\tau i)+\tau\frac{j^2+3j+2}{2}}\Bigg)\\
                &= \Bigg(\sum_{j=0}^{i-1} B^{js} Z^{j(T_0-\tau i)+\tau\frac{j^2+j}{2}}\Bigg) + B^k  Z^{iT_0+\tau\frac{i-i^2}{2}} - B^{is} Z^{i(T_0-\tau i)+\tau\frac{i^2-2i+1+3i-3+2}{2}}\\
                &\quad - \Bigg(\sum_{j=0}^{i-2} B^{(j+1)s} Z^{(j+1)(T_0-\tau i)+\tau\frac{j^2+3j+2}{2}}\Bigg)\\
                &= \Bigg(\sum_{j=0}^{i-1} B^{js} Z^{j(T_0-\tau i)+\tau\frac{j^2+j}{2}}\Bigg) + B^k  Z^{iT_0+\tau\frac{i-i^2}{2}} - B^{k} Z^{iT_0+\tau\frac{i-i^2}{2}} - \Bigg(\sum_{j=0}^{i-2} B^{(j+1)s} Z^{(j+1)(T_0-\tau i)+\tau\frac{j^2+3j+2}{2}}\Bigg)\\
                &= 1 + \Bigg(\sum_{j=1}^{i-1} B^{js} Z^{j(T_0-\tau i)+\tau\frac{j^2+j}{2}}\Bigg) - \Bigg(\sum_{j=1}^{i-1} B^{js} Z^{j(T_0-\tau i)+\tau\frac{j^2+j}{2}}\Bigg) = 1
            \end{align*}
            So, we can write the opinion as $x_{i}^{(3)} = (1-\Upsilon_{i}^{(3)})x_0 + \Upsilon_{i}^{(3)}u_0$.
            If block $i$ is in steady blocks (for $i > m_{AD}, i \in \mathbb{N}$), then, the agent has a fixed clicking rate of $T_0 - (m_{AD} - 1)\tau$, which is similar to \emph{fixed} policy, with different initial clicking rate. For the agent under \emph{adaptive decreasing} policy in steady blocks, we have:
            
            \begin{align*}
                x_{i}^{(3)} &= \left[ \frac{1 - (B^{s} Z^{T_0 - (m_{AD} - 1)\tau})^{i-m_{AD}}}{1 - B^{s} Z^{T_0 - (m_{AD} - 1)\tau}} \left(1 - B^{s-(T_0 - (m_{AD} - 1)\tau)} + \eta B^{s-(T_0 - (m_{AD} - 1)\tau)} (1 - \beta^{T_0 - (m_{AD} - 1)\tau}) \right) \right] x_0\\
                &\quad + \left[ \frac{1 - (B^{s} Z^{T_0 - (m_{AD} - 1)\tau})^{i-m_{AD}}}{1 - B^{s} Z^{T_0 - (m_{AD} - 1)\tau}} (1 - \eta) B^{s-(T_0 - (m_{AD} - 1)\tau)} (1 - \beta^{T_0 - (m_{AD} - 1)\tau}) \right] u_0\\
                &\quad + (B^{s} Z^{T_0 - (m_{AD} - 1)\tau})^{i-m_{AD}} x_{m_{AD}}^{(3)} \\
                &= \left[ \frac{1 - (B^{s} Z^{T_0 - (m_{AD} - 1)\tau})^{i-m_{AD}}}{1 - B^{s} Z^{T_0 - (m_{AD} - 1)\tau}} \left(1 - B^{s-(T_0 - (m_{AD} - 1)\tau)} + \eta B^{s-(T_0 - (m_{AD} - 1)\tau)} (1 - \beta^{T_0 - (m_{AD} - 1)\tau}) \right) \right] x_0\\
                &\quad + \left[ \frac{1 - (B^{s} Z^{T_0 - (m_{AD} - 1)\tau})^{i-m_{AD}}}{1 - B^{s} Z^{T_0 - (m_{AD} - 1)\tau}} (1 - \eta) B^{s-(T_0 - (m_{AD} - 1)\tau)} (1 - \beta^{T_0 - (m_{AD} - 1)\tau}) \right] u_0\\
                &\quad + (B^{s} Z^{T_0 - (m_{AD} - 1)\tau})^{i-m_{AD}} \Bigg((1-\Upsilon_{m_{AD}}^{(3)})x_0 + \Upsilon_{m_{AD}}^{(3)}u_0\Bigg)\\
                &= \Bigg[ \frac{1 - (B^{s} Z^{T_0 - (m_{AD} - 1)\tau})^{i-m_{AD}}}{1 - B^{s} Z^{T_0 - (m_{AD} - 1)\tau}} \left(1 - B^{s-(T_0 - (m_{AD} - 1)\tau)} + \eta B^{s-(T_0 - (m_{AD} - 1)\tau)} (1 - \beta^{T_0 - (m_{AD} - 1)\tau}) \right)\\
                &\quad + (B^{s} Z^{T_0 - (m_{AD} - 1)\tau})^{i-m_{AD}} - (B^{s} Z^{T_0 - (m_{AD} - 1)\tau})^{i-m_{AD}} \Upsilon_{m_{AD}}^{(3)}\Bigg] x_0  + \Bigg[ \frac{1 - (B^{s} Z^{T_0 - (m_{AD} - 1)\tau})^{i-m_{AD}}}{1 - B^{s} Z^{T_0 - (m_{AD} - 1)\tau}} (1 - \eta)\\
                &\quad \times B^{s-(T_0 - (m_{AD} - 1)\tau)} (1 - \beta^{T_0 - (m_{AD} - 1)\tau}) + (B^{s} Z^{T_0 - (m_{AD} - 1)\tau})^{i-m_{AD}} \Upsilon_{m_{AD}}^{(3)}\Bigg] u_0
            \end{align*}
            
            In the previous equation, 
            \begin{align*}
                \frac{1 - (B^{s} Z^{T_0 - (m_{AD} - 1)\tau})^{i-m_{AD}}}{1 - B^{s} Z^{T_0 - (m_{AD} - 1)\tau}} &\left(1 - B^{s-(T_0 - (m_{AD} - 1)\tau)} + \eta B^{s-(T_0 - (m_{AD} - 1)\tau)} (1 - \beta^{T_0 - (m_{AD} - 1)\tau}) \right)\\
                &+ (B^{is} Z^{T_0 - (m_{AD} - 1)\tau})^{i-m_{AD}}
            \end{align*}
            exactly equals to $1 - \Upsilon_{i-m_{AD}}^{(1)}$, and $\frac{1 - (B^{s} Z^{T_0 - (m_{AD} - 1)\tau})^{i-m_{AD}}}{1 - B^{s} Z^{T_0 - (m_{AD} - 1)\tau}} (1 - \eta) B^{s-(T_0 - (m_{AD} - 1)\tau)} (1 - \beta^{T_0 - (m_{AD} - 1)\tau}) = \Upsilon_{i-m_{AD}}^{(1)}$ for the initial clicking rate of $T_0 - (m_{AD} - 1)\tau$.
            \begin{align*}
                x_{i}^{(3)} &= \Bigg(1 - \Upsilon_{i-m_{AD}}^{(1)} - (B^{s} Z^{T_0 - (m_{AD} - 1)\tau})^{i-m_{AD}} \Upsilon_{m_{AD}}^{(3)}\Bigg) x_0 + \left(\Upsilon_{i-m_{AD}}^{(1)} + (B^{s} Z^{T_0 - (m_{AD} - 1)\tau})^{i-m_{AD}} \Upsilon_{m_{AD}}^{(3)}\right)u_0\\
                &= (1 - \Upsilon_{i}^{(3)}) x_0 + \Upsilon_{i}^{(3)} u_0
            \end{align*}
            where $\Upsilon_{i}^{(3)} = \Upsilon_{i-m_{AD}}^{(1)} + (B^{s} Z^{T_0 - (m_{AD} - 1)\tau})^{i-m_{AD}} \Upsilon_{m_{AD}}^{(3)}$. Furthermore, the agent's opinion at the start of a block in steady states, under the \emph{adaptive decreasing} policy, has been demonstrated to be a convex combination of the innate opinion and the recommendation.
            
        \end{enumerate}
\end{proof}

%%%%%%%%%%%%%%%%%%%%%%%%%%%%%%%%%%%%%%%%%%%%%%%%%%%%%%%
\subsection{Proof of Proposition \ref{prop:compare-single-opinion}}\label{app-prop-compare-finite}
\begin{proof}
    Let $\alpha \geq \beta$.
    \begin{enumerate}
        \item For a \emph{fixed policy} we can express the multiplier of recommendation as \[  (1 - Z) B^{s-T_0} \Bigg(\sum_{q=0}^{T_0-1} \beta^q\Bigg) \Bigg(\sum_{j=0}^{n-1} (B^{s-T_0} \beta^{T_0})^j\Bigg) \] 
        where $Z$ increases in $\alpha$, and both $1-Z$ and $B$ decrease in $\alpha$. For a fixed non-zero $\beta \in (0,1]$, the sum $\Bigg(\sum_{q=0}^{T_0-1} \beta^q\Bigg)$ is independent of $\alpha$. Since $T_0$ and $s-T_0$ are non-negative, $(B^{s-T_0} \beta^{T_0})$ is a number in $[0,1]$, which decreases in $\alpha$. Therefore, each term in the sum from $q=0$ to $T_0 -1$, decreases as $\alpha$ increases, implying the whole term decreases as $\alpha$ increases.\\

        For transient states in a \emph{decreasing} policy, as $\alpha$ increases, $\Upsilon_{i}^{(2)}$ decreases. To demonstrate this, we show the derivative of $\Upsilon_{i}^{(2)}$ with respect to $\alpha$ is negative, meaning the whole term decreases in $\alpha$.

        \begin{align*}
            \frac{d(\Upsilon_{i}^{(2)})}{d\alpha} &= \sum_{j=0}^{i-1} (1 - \beta^{\frac{T_0}{\kappa^{i-j-1}}}) \frac{d\Bigg( 
            (1 - \eta) B^{(j+1)s-\frac{T_0}{\kappa^{i-j-1}}} Z^{\frac{T_0 \kappa^{1-i}}{\kappa - 1} (\kappa^j-1)}\Bigg)}{d\alpha}\\
            &= \sum_{j=0}^{i-1} (1 - \beta^{\frac{T_0}{\kappa^{i-j-1}}}) \Bigg[-\frac{1}{1-\beta} B^{(j+1)s-\frac{T_0}{\kappa^{i-j-1}}} Z^{\frac{T_0 \kappa^{1-i}}{\kappa - 1} (\kappa^j-1)}\\
            &\quad + (1-\eta) B^{(j+1)s-\frac{T_0}{\kappa^{i-j-1}}} Z^{\frac{T_0 \kappa^{1-i}}{\kappa - 1} (\kappa^j-1)-1} \Bigg( - ((j+1)s-\frac{T_0}{\kappa^{i-j-1}}) + (\frac{T_0 \kappa^{1-i}}{\kappa - 1} (\kappa^j-1)) \Bigg)\Bigg]
        \end{align*}
        Since $B\geq 0, Z \geq 0$ and $\beta \in (0,1]$, the first term ($-\frac{1}{1-\beta} B^{(j+1)s-\frac{T_0}{\kappa^{i-j-1}}} Z^{\frac{T_0 \kappa^{1-i}}{\kappa - 1} (\kappa^j-1)}$) is negative. For the second term if we show that $\Bigg( - ((j+1)s-\frac{T_0}{\kappa^{i-j-1}}) + (\frac{T_0 \kappa^{1-i}}{\kappa - 1} (\kappa^j-1)) \Bigg)$ is negative, then the whole derivative with respect to $\alpha$ is negative ($(1-\eta) \in [0,1]$).
        \begin{align*}
            \Bigg( - ((j+1)s-\frac{T_0}{\kappa^{i-j-1}}) + (\frac{T_0 \kappa^{1-i}}{\kappa - 1} (\kappa^j-1)) \Bigg) &= -js - (\frac{T_0 \kappa^{1-i}}{\kappa - 1} (\kappa^j-1)) -s + \frac{T_0}{\kappa^{i-j-1}}
        \end{align*}
        For any non-negative $j$, we have $i > j$, $s \geq 0$, $\kappa \geq 1$, so, $-js - (\frac{T_0 \kappa^{1-i}}{\kappa - 1} (\kappa^j-1))$ is negative. For any $s$, we know $T_0 \leq s$, then, $-s + \frac{T_0}{\kappa^{i-j-1}} < 0$. To conclude, the derivative with respect to $\alpha$ is negative.\\

        For the recommendation's multiplier in steady states under a \emph{decreasing} policy, we use \eqref{eq:single-opinion-policy-two-0}:
        \begin{align*}
            \frac{d\Upsilon^{(2)}_i}{d\alpha} = \Upsilon_{m_D}^{(2)} \frac{d(B^{(i-m_D)s})}{d\alpha} + B^{(i-m_D)s}\frac{d\Upsilon_{m_D}^{(2)}}{d\alpha}
        \end{align*}

        Because $\Upsilon_{m_D}^{(2)}$ is a sum of non-negative terms, then, it is non negative. Also, $B^{(i-m_D)s}$ is non-negative, and from previous results, we know that $\frac{d\Upsilon_{m_D}^{(2)}}{d\alpha}$ is negative. Now, to complete the proof, we show that $\frac{d(B^{(i-m_D)s})}{d\alpha}$ is negative, thus proving that $\Upsilon^{(2)}_i$ decreases in $\alpha$ in steady states, as well.

        \begin{align*}
            \frac{d(B^{(i-m_D)s})}{d\alpha} &= - (i-m_D)sB^{(i-m_D)s-1} (\frac{\beta}{Z^2})
        \end{align*}

        As this is about steady states, $i>m_D$, then the whole term is negative.\\

        Moreover, $\Upsilon_{i}^{(3)}$ decreases with increasing $\alpha$ in both transient and steady states. To prove this, it is sufficient to show that both expressions have a negative derivative with respect to $\alpha$.

        When $i$ is in transient states, using \eqref{eq:single-opinion-policy-three}:
        \begin{align*}
            \frac{d\Upsilon_{i}^{(3)}}{d\alpha} &= \sum_{j=0}^{i-1} (1-\beta^{T_0-(i-1-j)\tau}) \frac{d\Bigg(B^{(j+1)s-T_0+(i-1-j)\tau} Z^{j(T_0-\tau i)+\tau\frac{j^2+j}{2}}\Bigg)}{d\alpha}\\
            &= \sum_{j=0}^{i-1} (1-\beta^{T_0-(i-1-j)\tau}) \Bigg(B^{(j+1)s-T_0+(i-1-j)\tau} Z^{j(T_0-\tau i)+\tau\frac{j^2+j}{2}}\Bigg) \Bigg(-\frac{1}{1-\beta}\\
            &\quad + (1-\eta) Z^{-1}  \Bigg(-((j+1)s-T_0+(i-1-j)\tau) + (j(T_0-\tau i)+\tau\frac{j^2+j}{2}) \Bigg)\Bigg)
        \end{align*}

        Since $B \in [0,1], Z \in [0,1]$ and $\beta \in (0,1]$, then $(1-\beta^{T_0-(i-1-j)\tau}) \Bigg(B^{(j+1)s-T_0+(i-1-j)\tau} Z^{j(T_0-\tau i)+\tau\frac{j^2+j}{2}}\Bigg) > 0$. The first term is negative ($-\frac{1}{1-\beta}$). Also, we have :
        \[-(j+1)s+T_0-(i-1-j)\tau + (j(T_0-\tau i)+\tau\frac{j^2+j}{2}) =(j+1)(-s+T_0)-\tau(i-1-j+j(i+\frac{-j-1}{2}))\]
        We know $s \geq T_0$. As $0 \leq j \leq i-1$, then $i-1-j \geq 0$ and $i+\frac{-j-1}{2} \geq 0$. Consequently, $\Bigg(-((j+1)s-T_0+(i-1-j)\tau) + (j(T_0-\tau i)+\tau\frac{j^2+j}{2}) \Bigg)$ is non-positive. Then, the whole term is negative.

        For \emph{adaptive decreasing} policy in steady states, based on \eqref{eq:single-opinion-policy-three-fixed}, we conclude similarly that the derivative of $\Upsilon_{i}^{(3)}$ is negative.

        \begin{align*}
            \frac{d\Upsilon_{i}^{(3)}}{d\alpha} &= \frac{d\Upsilon_{i-m_{AD}}^{(1)}}{d\alpha} +  \Upsilon_{m_{AD}}^{(3)} \frac{d((B^{s}Z^{T_0-(m_{AD} - 1)\tau})^{i-m_{AD}})}{d\alpha} + (B^{s}Z^{T_0-(m_{AD} - 1)\tau})^{i-m_{AD}} \frac{d\Upsilon_{m_{AD}}^{(3)}}{d\alpha}
        \end{align*}

        From previous parts we know $\frac{d\Upsilon_{i-m_{AD}}^{(1)}}{d\alpha} < 0$ and $\frac{d\Upsilon_{m_{AD}}^{(3)}}{d\alpha} < 0$. Also, as $B \in [0,1]$, then $(B^{s}Z^{T_0-(m_{AD} - 1)\tau})^{i-m_{AD}} \geq 0$. $\Upsilon_{m_{AD}}^{(3)}$ itself is a sum of non-negative terms, so, it is non-negative. To complete the proof, we have:
        \[\frac{d((B^{s}Z^{T_0-(m_{AD} - 1)\tau})^{i-m_{AD}})}{d\alpha} = \beta^{s(i-m_{AD)}} Z^{(-s+T_0-(m_{AD} - 1)\tau)(i-m_{AD})-1}\Bigg((-s+T_0-(m_{AD} - 1)\tau)(i-m_{AD})\Bigg) \]
        We know $s \geq T_0, m_{AD} \geq 1, i > m_{AD}$, therefore, $(-s+T_0-(m_{AD} - 1)\tau)(i-m_{AD}) \leq 0$. Then the whole derivative is negative.

        \item As $i$ increases, (i.e. as we look further into the blocks), for $\Upsilon_{i}^{(1)} = \frac{1 - (B^{s} Z^{T_0})^i}{1 - B^{s} Z^{T_0}} (1 - \eta) B^{s-T_0} (1 - \beta^{T_0})$, the term $(B^{s} Z^{T_0})^i$ decreases as $B^{s} Z^{T_0} \in [0,1]$, then $(1 - \beta^{T_0})$ increases. Thus, overall, since other terms are constant, 
        $\Upsilon_{i}^{(1)}$ increases as $i$ increases. 
        
        In steady-state blocks for the \emph{decreasing} policy, $\Upsilon^{(2)}_i = B^{(i-m_D)s} \Upsilon_{m_D}^{(2)}$ decreases in $i$, as $B \in [0,1]$ and $\Upsilon_{m_D}^{(2)}$ is independent of $i$. 
        
        For transient blocks, both $\Upsilon_{i}^{(2)}$ and $\Upsilon_{i}^{(3)}$ increase as $i$ increases, because both involve sums over terms from $j=0$ to $i-1$. When $i$ increases, more positive terms are considered (non-zero $\beta$), then, both increase in $i$. However, the rate of increase decreases as $i$ increases, which can be shown by proving that the derivative with respect to $j$ is negative for each case.

        \begin{align*}
            \frac{d(\Upsilon_{i}^{(2)})}{dj} &= (1 - \eta) \sum_{j=0}^{i-1} B^{(j+1)s-T_0 \kappa^{1-i}\frac{\kappa^{j+1}-1}{\kappa-1}}\beta^{\frac{T_0 \kappa^{1-i}}{\kappa - 1} (\kappa^j-1)}\\
            &\quad \times \Bigg[(1 - \beta^{\frac{T_0}{\kappa^{i-j-1}}}) \Bigg((s-\frac{T_0 \kappa^{1-i}}{\kappa-1} \kappa^{j+1} \ln(\kappa)) \ln(B) + \frac{T_0 \kappa^{1-i}}{\kappa - 1} \kappa^j \ln(\kappa) \ln(\beta) \Bigg) - \beta^{T_0\kappa^{j+1-i}} T_0\kappa^{j+1-i} \ln(\kappa) \ln(\beta)  \Bigg] 
        \end{align*}

        \begin{align*}
            \frac{d(\Upsilon_{i}^{(3)})}{dj} &= (1-\eta) \sum_{j=0}^{i-1} B^{(j+1)(s-T_0+\tau i - \tau) - \tau\frac{j^2+j}{2}} \beta^{j(T_0-\tau i)+\tau\frac{j^2+j}{2}} \Bigg[(1-\beta^{T_0-(i-1-j)\tau})\\
            &\quad \times \Bigg((s-T_0+\tau(i - 1) - \tau\frac{2j+1}{2}) \ln(B) + ((T_0-\tau i)+\tau\frac{2j+1}{2})ln(\beta) \Bigg) - \tau \beta^{T_0-(i-1-j)\tau} \ln(\beta)  \Bigg]
        \end{align*}

        As $\ln(B)$ and $\ln(\beta)$ are negative, then the derivatives are negative for $j \in [0,i-1]$.

        \item For the \emph{fixed} policy, based on \eqref{eq:single-opinion-policy-one}, with increasing $T_0$, $\Upsilon_{i}^{(1)}$ increases, since the derivative with respect to $T_0$ is positive (non-zero $\beta$). Again, we have $\ln(\beta) < 0, \ln(B) < 0$.

        \begin{align*}
            \frac{d\Upsilon_{i}^{(1)}}{dT_0} &= (1 - \eta) B^{s-T_0}\Bigg[ \frac{(1 - \beta^{T_0}) \Bigg( -i(B^{s} Z^{T_0})^i \ln(Z)(1 - B^{s} Z^{T_0}) + B^{s} Z^{T_0} \ln(Z)(1 - (B^{s} Z^{T_0})^i)\Bigg)}{(1 - B^{s} Z^{T_0})^2}\\
            &\quad + \frac{(1 - (B^{s} Z^{T_0})^i) (1 - B^{s} Z^{T_0}) \Bigg(- \ln(B) (1-\beta^{T_0}) -\beta^{T_0}\ln(\beta) \Bigg)}{(1 - B^{s} Z^{T_0})^2}\Bigg]
        \end{align*}

        \item For the \emph{decreasing} policy in transient blocks, based on \eqref{eq:single-opinion-policy-two} and the earlier equations in Appendix \ref{app:single-opinion}, we had $\Upsilon_{i}^{(2)} = \left(\sum_{j=0}^{i-1} B^{(j+1)s} (1 - Z) \left(\sum_{q=0}^{\frac{T_0}{\kappa^{i-j-1}}-1}\beta^q\right) \left(\prod_{q=0}^{j} B^{-\frac{T_0}{\kappa^{i-q-1}}} \right) \left(\prod_{q=0}^{j-1} \beta^{\frac{T_0}{\kappa^{i-q-1}}} \right) \right)$ which increases in $\kappa$ (non-zero $\beta$). This is because $\left(\sum_{q=0}^{\frac{T_0}{\kappa^{i-j-1}}-1}\beta^q\right)$ and $\left(\prod_{q=0}^{j} B^{-\frac{T_0}{\kappa^{i-q-1}}} \right)$ decrease in $\kappa$, while $\left(\prod_{q=0}^{j-1} \beta^{\frac{T_0}{\kappa^{i-q-1}}} \right)$ increases. However, we can say that the influence of decrease in $\left(\sum_{q=0}^{\frac{T_0}{\kappa^{i-j-1}}-1}\beta^q\right)$ is more significant (because there is an exponential reduction in the number of terms) than the increase in $\left(\prod_{q=0}^{j-1} \beta^{\frac{T_0}{\kappa^{i-q-1}}} \right)$ (because of the slower growth towards 1 with bounded $\beta \leq 1$). As a result, the whole term decreases in $\kappa$.\\

        \item For the \emph{adaptive decreasing} policy in transient blocks, based on \eqref{eq:single-opinion-policy-three} and the previous equations in Appendix \ref{app:single-opinion}, we have $\Upsilon_{i}^{(3)} = \left(\sum_{j=0}^{i-1} B^{(j+1)s} (1 - Z) \left(\sum_{q=0}^{T_0-(i-1-j)\tau-1}\beta^q\right) \left(\prod_{q=0}^{j} B^{-T_0+(i-1-q)\tau} \right) \left(\prod_{q=0}^{j-1} \beta^{T_0-(i-1-q)\tau} \right) \right)$, which decreases in $\tau$. This is because $\left(\sum_{q=0}^{T_0-(i-1-j)\tau-1}\beta^q\right)$ and $\left(\prod_{q=0}^{j} B^{-T_0+(i-1-q)\tau}\right)$ decrease, when $\tau$ increases, while $\left(\prod_{q=0}^{j-1} \beta^{T_0-(i-1-q)\tau}\right)$ increases. However, we can conclude that the rate at which the summation of $\left(\sum_{q=0}^{T_0-(i-1-j)\tau-1}\beta^q\right)$ decreases is faster than the rate at which the product $\left(\prod_{q=0}^{j-1} \beta^{T_0-(i-1-q)\tau}\right)$ increases. The decrease in the summation is because of the linear reduction in the number of terms. On the other hand, the product has a gradual increase as $\beta$ is bounded by 1. Therefore, the reduction is faster than the increase. So the whole term decreases when $\tau$ increases.
        
    \end{enumerate}

\end{proof}

%%%%%%%%%%%%%%%%%%%%%%%%%%%%%%%%%%%%%%%%%%%%%%%%%%%%%%%
\subsection{Proof of Corollary \ref{cor:compare-infinite}}
    We analyze the agent's opinion at the limit case as $i \to \infty$ under three different policies (\emph{fixed}, \emph{decreasing}, and \emph{adaptive decreasing} policies).

    \begin{enumerate}
        \item \emph{Fixed} Policy: From \eqref{eq:single-opinion-policy-one}, and the fact that the opinion at the start of each block is a convex combination of $x_0$ and $u_0$, the agent's opinion at the start of block $i$ under the \emph{fixed} policy, for $s=T_0$ is:
        \[x_{i}^{(1)} = (1-\Upsilon_{i}^{(1)}) x_0 + \Upsilon_{i}^{(1)}u_0 = x_0 + (1 - (\beta^{T_0})^i) (1 - \eta) (u_0 - x_0)\]
        To find the opinion at the limit case of $i \to \infty$, we have:
        \begin{align*}
            \lim_{i\rightarrow \infty} x^{(1)}_i = \lim_{i\rightarrow \infty} \Bigg(x_0 + (1 - (\beta^{T_0})^i) (1 - \eta) (u_0 - x_0) \Bigg) =  x_0 + (1 - \eta) (u_0 - x_0) \lim_{i\rightarrow \infty}(1 - (\beta^{T_0})^i)
        \end{align*}
        Because $\beta \in [0,1]$ and $T_0 >0$, we have $(\beta^{T_0})^i \to 0$ as $i \to \infty$, then:
        \begin{align*}
            \lim_{i\rightarrow \infty} x^{(1)}_i = x_0 + (1 - \eta) (u_0 - x_0) =  \eta x_0 + (1 - \eta) u_0 
        \end{align*}
        \item Similarly, according to \eqref{eq:single-opinion-policy-two-0}, the agent's opinion under \emph{decreasing policy} at the start of block $i$ (when $i$ is in steady states) is:
        \[x_{i}^{(2)} = (1-\Upsilon_{i}^{(2)}) x_0 + \Upsilon_{i}^{(2)}u_0 = x_0 + B^{(i-m_D)s} \Upsilon_{m_D}^{(2)} (u_0 - x_0) \]
        Then, to find the limit of the opinion when $i \to \infty$, we have:
        \[\lim_{i\rightarrow \infty} x^{(2)}_i = \lim_{i\rightarrow \infty} \Bigg(x_0 + B^{(i-m_D)s} \Upsilon_{m_D}^{(2)} (u_0 - x_0)\Bigg) = x_0 +  (u_0 - x_0) \Upsilon_{m_D}^{(2)} \lim_{i\rightarrow \infty} \Bigg(B^{(i-m_D)s}\Bigg)\]
        Since $B \in [0,1]$ and $i > m_D$, when $i \to \infty$, $B^{(i-m_D)s} \to 0$ and $\Upsilon_{m_D}^{(2)}$ is independent of $i$. Therefore:
        \[\lim_{i\rightarrow \infty} x^{(2)}_i = x_0 + 0 = x_0\]

        \item Following \eqref{eq:single-opinion-policy-three-fixed}, the agent's opinion at the start of block $i$ (where $i$ is in steady states) under \emph{adaptive decreasing} policy is:
        \[x_{i}^{(3)} = (1-\Upsilon_{i}^{(3)}) x_0 + \Upsilon_{i}^{(3)}u_0 = x_0 + \Bigg(\Upsilon_{i-m_{AD}}^{(1)} +  (B^{s}Z^{T_0-(m_{AD} - 1)\tau})^{i-m_{AD}} \Upsilon_{m_{AD}}^{(3)}\Bigg) (u_0 - x_0) \]
        Then, as $i \to \infty$, the limit of the opinion is:
        \begin{align*}
            \lim_{i\rightarrow \infty} x^{(3)}_i &= \lim_{i\rightarrow \infty} \Bigg(x_0 + \Bigg(\Upsilon_{i-m_{AD}}^{(1)} +  (B^{s}Z^{T_0-(m_{AD} - 1)\tau})^{i-m_{AD}} \Upsilon_{m_{AD}}^{(3)}\Bigg) (u_0 - x_0)\Bigg)\\
            & = x_0 + (u_0 - x_0) \lim_{i\rightarrow \infty} \Upsilon_{i-m_{AD}}^{(1)} +(u_0 - x_0) \lim_{i\rightarrow \infty} \Bigg((B^{s}Z^{T_0-(m_{AD} - 1)\tau})^{i-m_{AD}} \Upsilon_{m_{AD}}^{(3)}\Bigg)
        \end{align*}
        Because $B, Z \in [0,1]$, $s \geq 0$, $T_0-(m_{AD} - 1)\tau \geq 0$ and $i > m_D$, then, $\lim_{i\rightarrow \infty} (B^{s}Z^{T_0-(m_{AD} - 1)\tau})^{i-m_{AD}} \to 0$ and $\Upsilon_{m_D}^{(3)}$ is independent of $i$. Therefore:
        \[\lim_{i\rightarrow \infty} x^{(3)}_i = x_0 + (u_0 - x_0) \lim_{i\rightarrow \infty} \Upsilon_{i-m_{AD}}^{(1)}\]
        
        Substituting $\Upsilon_{i-m_{AD}}^{(1)}$ from \eqref{eq:single-opinion-policy-one} with initial clicking rate of $T_0 -(m_{AD} - 1)\tau$, we have:
        \begin{align*}
            \lim_{i\rightarrow \infty} x^{(3)}_i &= x_0 + (u_0 - x_0) \lim_{i\rightarrow \infty} \frac{1 - (B^{s} Z^{T_0 -(m_{AD} - 1)\tau})^{i-m_{AD}}}{1 - B^{s} Z^{T_0 -(m_{AD} - 1)\tau}} (1 - \eta) B^{s-(T_0 -(m_{AD} - 1)\tau)} (1 - \beta^{T_0 -(m_{AD} - 1)\tau})\\
            &= x_0 + (u_0 - x_0) \frac{(1 - \eta) B^{s-(T_0 -(m_{AD} - 1)\tau)} (1 - \beta^{T_0 -(m_{AD} - 1)\tau})}{1 - B^{s} Z^{T_0 -(m_{AD} - 1)\tau}}
        \end{align*}
        The agent under \emph{adaptive decreasing} policy at the start of each block has the opinion drift less that $x_\text{drift}$ during the steady states. As a result, the opinion drift at the start of block $i \to \infty$ is less than a given tolerance $x_\text{drift}$. So we have:
        \[|\lim_{i\rightarrow \infty} x^{(3)}_i - x_0| = |u_0 - x_0| \frac{(1 - \eta) B^{s-(T_0 -(m_{AD} - 1)\tau)} (1 - \beta^{T_0 -(m_{AD} - 1)\tau})}{1 - B^{s} Z^{T_0 -(m_{AD} - 1)\tau}} < x_\text{drift}\]
        We know $\frac{(1 - \eta) B^{s-(T_0 -(m_{AD} - 1)\tau)} (1 - \beta^{T_0 -(m_{AD} - 1)\tau})}{1 - B^{s} Z^{T_0 -(m_{AD} - 1)\tau}} \geq 0$. If $u_0 > x_0$:
        \[\lim_{i\rightarrow \infty} x^{(3)}_i - x_0 = (u_0 - x_0) \frac{(1 - \eta) B^{s-(T_0 -(m_{AD} - 1)\tau)} (1 - \beta^{T_0 -(m_{AD} - 1)\tau})}{1 - B^{s} Z^{T_0 -(m_{AD} - 1)\tau}} < x_\text{drift}\]
        \[\lim_{i\rightarrow \infty} x^{(3)}_i < x_\text{drift} + x_0\]
        Since the opinion cannot exceed 1, we conclude: \[\lim_{i\rightarrow \infty} x^{(3)}_i \leq \min\{1 , x_0 + x_\text{drift}\}\]
        For the lower bound on the opinion at the limit case, we analyze the scenario where $\tau$ is sufficiently large (for example $\tau = T_0$), and $x_{\text{drift}} = \epsilon$, where $0 < \epsilon < 0.001$ is a small positive number close to zero. Under this scenario, the agent's opinion drifts back to its innate opinion, at the limit case of $i \to \infty$, as the clicking rate converges to zero after a few blocks. In other words, after a certain number of blocks, the recommendation does not have any effects on the opinion (meaning $\lim_{i\rightarrow \infty} \Upsilon_{i-m_{AD}}^{(1)} \to 0$), and we have \[\lim_{i\rightarrow \infty} x^{(3)}_i  = x_0\]
        Hence, the agent's opinion (under \emph{adaptive decreasing} policy), at the limit case of $i \to \infty$, is bounded as:
        \[x_0 \leq \lim_{i\rightarrow \infty} x^{(3)}_i \leq \min\{1 , x_0 + x_\text{drift}\}\]

        Similarly, if $u_0 \leq x_0$, we have:
        \[x_0 - \lim_{i\rightarrow \infty} x^{(3)}_i = (x_0 - u_0) \frac{(1 - \eta) B^{s-(T_0 -(m_{AD} - 1)\tau)} (1 - \beta^{T_0 -(m_{AD} - 1)\tau})}{1 - B^{s} Z^{T_0 -(m_{AD} - 1)\tau}} < x_\text{drift}\]
        \[x_0 - x_\text{drift} <\lim_{i\rightarrow \infty} x^{(3)}_i\]
        The upper bound on the opinion is when the agent's opinion drifts back to $x_0$ and the lower bound is when the opinion has its maximum drift which is $x_0 - x_\text{drift}$. As the agent's opinion cannot be less than $-1$, the lower bound would be $\max\{-1, x_0 - x_\text{drift}\}$. Therefore:
        \[\max\{-1 , x_0 - x_\text{drift}\} \leq \lim_{i\rightarrow \infty} x^{(3)}_i \leq x_0\]
    \end{enumerate}

%%%%%%%%%%%%%%%%%%%%%%%%%%%%%%%%%%%%%%%%%%%%%%%%%%%%%%%
\subsection{Proof of Proposition \ref{prop:comapre-utility}}

    The agent's reward function is defined as $R^A (|x_i - u_i|) = 1$. For all policies where $T_0 = s$, by using \eqref{eq:agent-utility}, and the results of Corollary \ref{cor:compare-infinite}, we can derive the following:

    \begin{enumerate}
        \item An agent following a \emph{fixed} policy clicks exactly $s = T_0$ time steps in each block. To determine the limit of the utility for this policy as the horizon $K=ns$ grows, where $n \to \infty$, we analyze:
        \[\lim_{n\rightarrow \infty} U^{(1)}(h_K ^A) = \lim_{n\rightarrow \infty} \Bigg(\lambda \frac {1} {K} \sum_{i=0}^{K-1} \text{clk}_i -  (1 - \lambda) |x_K - x_0|\Bigg) = \lambda \lim_{n\rightarrow \infty} \frac {1} {nT_0} nT_0 -  (1 - \lambda) \lim_{n\rightarrow \infty} |x_K - x_0|\]
        From Corollary \ref{cor:compare-infinite}, we know that $\lim_{n\rightarrow \infty} |x_K - x_0| = |\eta x_0 +  (1 - \eta) u_0 - x_0| = |-(1-\eta) x_0 +  (1 - \eta) u_0|$. Therefore, the limit of the utility is:
        \[\lim_{n\rightarrow \infty} U^{(1)}(h_K ^A) = \lambda -  (1 - \lambda) (1-\eta)| u_0 - x_0|\]

       \item For an again following a \emph{decreasing} policy, the number of clicks converge to zero in the long term. The limit of the opinion is given by Corollary \ref{cor:compare-infinite}. Then the limit of the utility as the horizon $K=ns$ grows, where $n \to \infty$, is:
       \[\lim_{n\rightarrow \infty} U^{(2)}(h_K ^A) = \lim_{n\rightarrow \infty} \Bigg(\lambda \frac {1} {K} \sum_{i=0}^{K-1} \text{clk}_i -  (1 - \lambda) |x_K - x_0|\Bigg)\]
        As $n \to \infty$, the clicking rate converges to zero and the opinion converges to innate opinion $x_0$:
       \[\lim_{n\rightarrow \infty} U^{(2)}(h_K ^A) = \lambda \lim_{n\rightarrow \infty} \frac {1} {nT_0} \sum_{j=0}^{m_D} \frac{T_0}{\kappa^{j}} -  (1 - \lambda) \lim_{n\rightarrow \infty} |x_K - x_0| = 0 + 0 = 0\]

       \item For an agent following a \emph{adaptive decreasing} policy, the clicking rate decreases during transient blocks but stabilizes at a fixed rate in steady states. The limit of the opinion follows from Corollary \ref{cor:compare-infinite}. So, the limit of the utility as the horizon $K=ns$ grows, where $n \to \infty$, is given by:
       \begin{align*}
           \lim_{n\rightarrow \infty} U^{(3)}(h_K ^A) &= \lim_{n\rightarrow \infty} \Bigg(\lambda \frac {1} {K} \sum_{i=0}^{K-1} \text{clk}_i -  (1 - \lambda) |x_K - x_0|\Bigg)\\
           &= \lambda \lim_{n\rightarrow \infty} \frac {1} {nT_0} \Bigg(\sum_{j=1}^{m_{AD} - 1} (T_0 - (j-1) \tau) + \sum_{j=m_{AD}}^{n} (T_0 - (m_{AD}-1) \tau)\Bigg) -  (1 - \lambda) \lim_{n\rightarrow \infty} |x_K - x_0|\\
           &= \lambda \lim_{n\rightarrow \infty} \frac {1} {nT_0} \Bigg( \frac{(m_{AD} - 1)(2T_0 - (m_{AD}-2) \tau)}{2} + (n-m_{AD}+1)(T_0 - (m_{AD}-1) \tau)\Bigg)\\
           &\quad -  (1 - \lambda) \lim_{n\rightarrow \infty} |x_K - x_0|\\
       \end{align*}

       For the lower bound on the utility, the lowest average cumulative clicking is achieved when the user stops clicking in long term, which corresponds to the \emph{decreasing policy}. Under this policy, the agent can adjust the tolerance $x_{\text{drift}}$ to be $x_{\text{drift}} =\epsilon$, where $0 < \epsilon < 0.001$ is a small positive value close to 0, and $\tau$ to be sufficiently large (for example $\tau = T_0$). These adjustments prevent negative utility and force the clicking rate to converge to zero. Therefore, the lower bound on the limit of the utility, as the horizon $K=ns$ grows, where $n \to \infty$, is: \[0\leq \lim_{n\rightarrow \infty} U^{(3)}(h_K ^A)\]

        For the upper bound, consider scenarios (such as when $x_0 = u_0$) where the user sticks to its initial clicking rate in the first block, achieving the highest possible average cumulative clicking rate (as $m_{AD} = 1$) and the lowest possible opinion drift which is zero, as the agent's opinion does not deviate from the innate value. Then, the upper bound on the limit of the utility is $\lambda$.

        Thus, the limit of the utility as the horizon $K=ns$ grows, where $n \to \infty$, is bounded by:
        \[0\leq \lim_{n\rightarrow \infty} U^{(3)}(h_K ^A) \leq \lambda\]

       \item If $\lim_{n\rightarrow \infty} U^{(1)}(h_K ^A) \geq 0$, under the \emph{adaptive decreasing} policy, where $x_{\text{drift}} = (1 - \eta)| u_0 - x_0|$, we have $\lim_{n\rightarrow \infty} U^{(3)}(h_K ^A) = \lim_{n\rightarrow \infty} U^{(1)}(h_K ^A)$. On the other hand, under this same policy (the \emph{adaptive decreasing} policy), there exist parameters $\tau$ and $x_{\text{drift}}$ such that $\lim_{n\rightarrow \infty} U^{(3)}(h_K ^A) > \lim_{n\rightarrow \infty} U^{(1)}(h_K ^A)$. To determine the conditions under which this occurs, consider for $\tau=1$, there exist $\epsilon_1$ and $\epsilon_2$, such that $x_{\text{drift}} = (1 - \eta)| u_0 - x_0| \epsilon_1 + \epsilon_2$, where $\epsilon_1 = \frac{B(1-\beta^{s-1})}{1-B\beta^{s-1}}$ and $0 < \epsilon_2 \leq 0.001$, if  $\epsilon_1 < 1 - \frac{\lambda}{s} \frac{1}{(1-\lambda)(1 - \eta)| u_0 - x_0|}$, we can guarantee $\lim_{n\rightarrow \infty} U^{(3)}(h_K ^A) > \lim_{n\rightarrow \infty} U^{(1)}(h_K ^A)$. Otherwise, if $x_{\text{drift}} = (1 - \eta)| u_0 - x_0|$, we have $\lim_{n\rightarrow \infty} U^{(3)}(h_K ^A) = \lim_{n\rightarrow \infty} U^{(1)}(h_K ^A)$. So, by this approach, the \emph{adaptive} decreasing policy can achieve the utility of the \emph{fixed} policy, or for some cases improves it.

        The new $x_{\text{drift}}$ is defined as follows: we know that the \emph{adaptive decreasing} policy can imitate \emph{fixed} policy. Now, consider a scenario where the agent does not click for just one time step within a block. Then, there is an opinion shift due to the lack of clicking compared to always clicking case. Let $K=ns$ and $n\to \infty$, the reduced clicking rate, remains unchanged in the long term. So, the opinion at time step $K$ is $x^{(3)}_n = x_0 + \Bigg(\Upsilon_{n-m_{AD}}^{(1)} +  (B^{s}Z^{T_0-(m_{AD} - 1)\tau})^{n-m_{AD}} \Upsilon_{m_{AD}}^{(3)}\Bigg) (u_0 - x_0)$, where does not matter what exactly $m_{AD}$ is. As $n \to \infty$, $(B^{s}Z^{T_0-(m_{AD} - 1)\tau})^{n-m_{AD}}$ tends to 0, since $(B^{s}Z^{T_0-(m_{AD} - 1)\tau}) \in [0,1]$. Therefore, we have $x^{(3)}_n = x_0 +\Upsilon_{n-m_{AD}}^{(1)} (u_0 - x_0) = x_0 + \Bigg(\tfrac{1 - (B^{s} Z^{s-1})^{n-m_{AD}}}{1 - B^{s} Z^{s-1}} (1 - \eta) B^{s-(s-1)} (1 - \beta^{s-1})\Bigg)(u_0 - x_0)$. $(B^{s} Z^{s-1})^{n-m_{AD}} \to 0$ also converges to zero, by $n \to \infty$, as $B^{s} Z^{s-1} \in [0,1]$. Then, we conclude $x^{(3)}_n = x_0 + \Bigg(\tfrac{1}{1 - B \beta^{s-1}} (1 - \eta) B (1 - \beta^{s-1})\Bigg)(u_0 - x_0) = x_0 + (1 - \eta)(u_0 - x_0)\epsilon_1 $. Consequently, the deviation from $x_0$ is $|x^{(3)}_n - x_0| = (1 - \eta)|u_0 - x_0|\epsilon_1$ which is less than $x_{\text{drift}} = (1 - \eta)| u_0 - x_0| \epsilon_1 + \epsilon_2$, where $\epsilon_2$ is a small positive number. Thus, the utility is $\lim_{n\rightarrow \infty} U^{(3)}(h_K ^A) = \lambda - \frac{\lambda}{s} - (1-\lambda) |x^{(3)}_n - x_0| = \lambda - \frac{\lambda}{s} - (1-\lambda)  (1 - \eta) |u_0 - x_0| \epsilon_1 $. We can guarantee that if $\epsilon_1 < 1 - \frac{\lambda}{s} \frac{1}{(1-\lambda)(1 - \eta)| u_0 - x_0|}$, then $\lim_{n\rightarrow \infty} U^{(3)}(h_K ^A) > \lim_{n\rightarrow \infty} U^{(1)}(h_K ^A)$.
    \end{enumerate}

%%%%%%%%%%%%%%%%%%%%%%%%%%%%%%%%%%%%%%%%%%%%%%%%%%%%%%%
\subsection{Additional Experiments}

The following figures illustrate the verification of the generality of the results obtained from both theoretical and numerical analyses. The experiments were conducted over $N=10$ blocks, each consisting of $s=8$ time steps, and under fixed recommendation policy for the platform. All reported values refer to measurements taken at the final time step. We focused on the utility at the last time step because under the adaptive decreasing policy, the agent only considers its final opinion at the end of each block. This implies that there may be deviations from $x_{drift}$ within a block, but the agent adjusts the clicking rate only after completing all the steps within that block.

\begin{figure}[htbp]
    \centering
    \begin{subfigure}[b]{0.3\textwidth}
        \includegraphics[width=\textwidth]{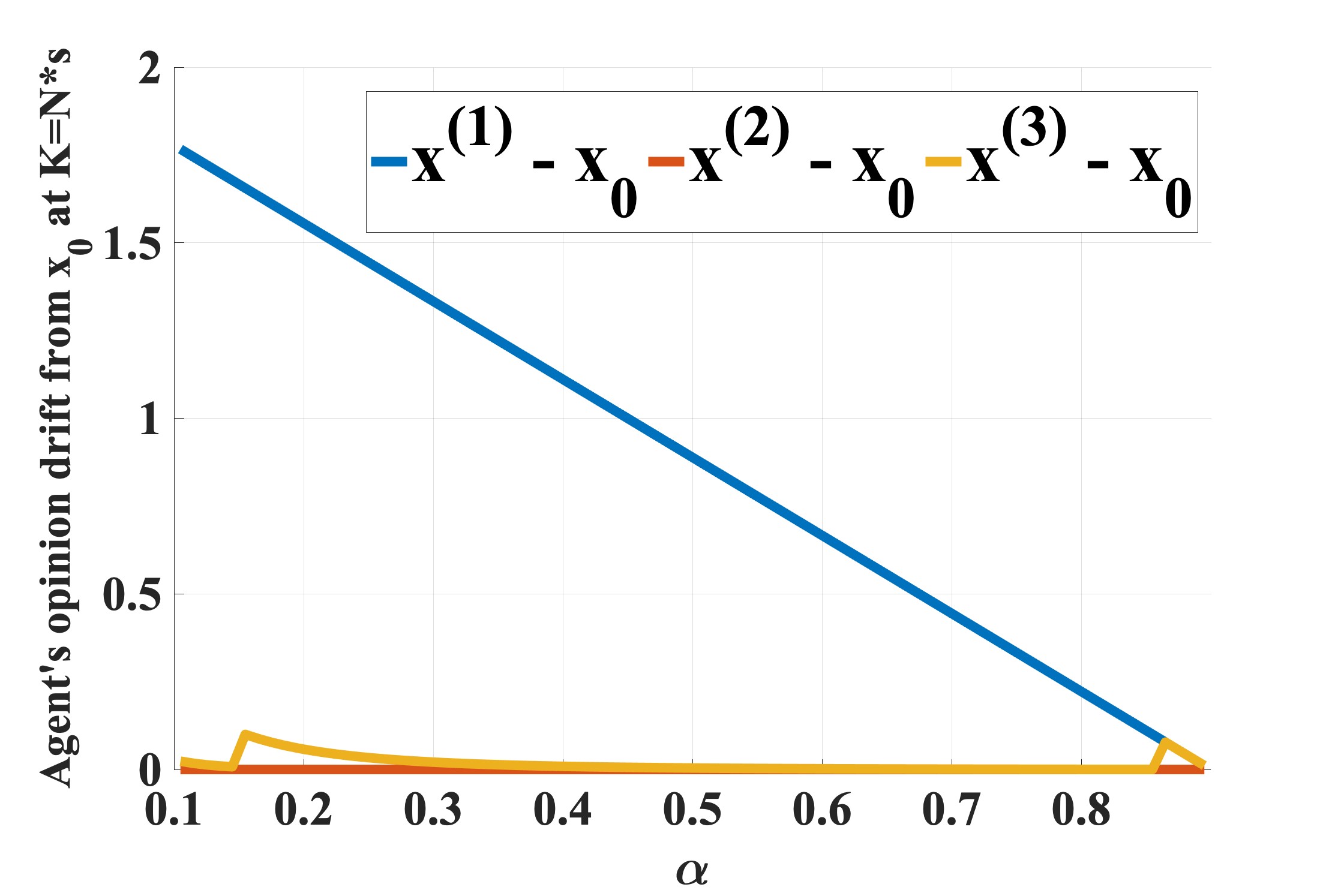}
        \caption{Deviation of agent's final opinion from innate opinion}
        \label{fig:drift_alpha}
    \end{subfigure}
    \hfill
    \begin{subfigure}[b]{0.3\textwidth}
        \includegraphics[width=\textwidth]{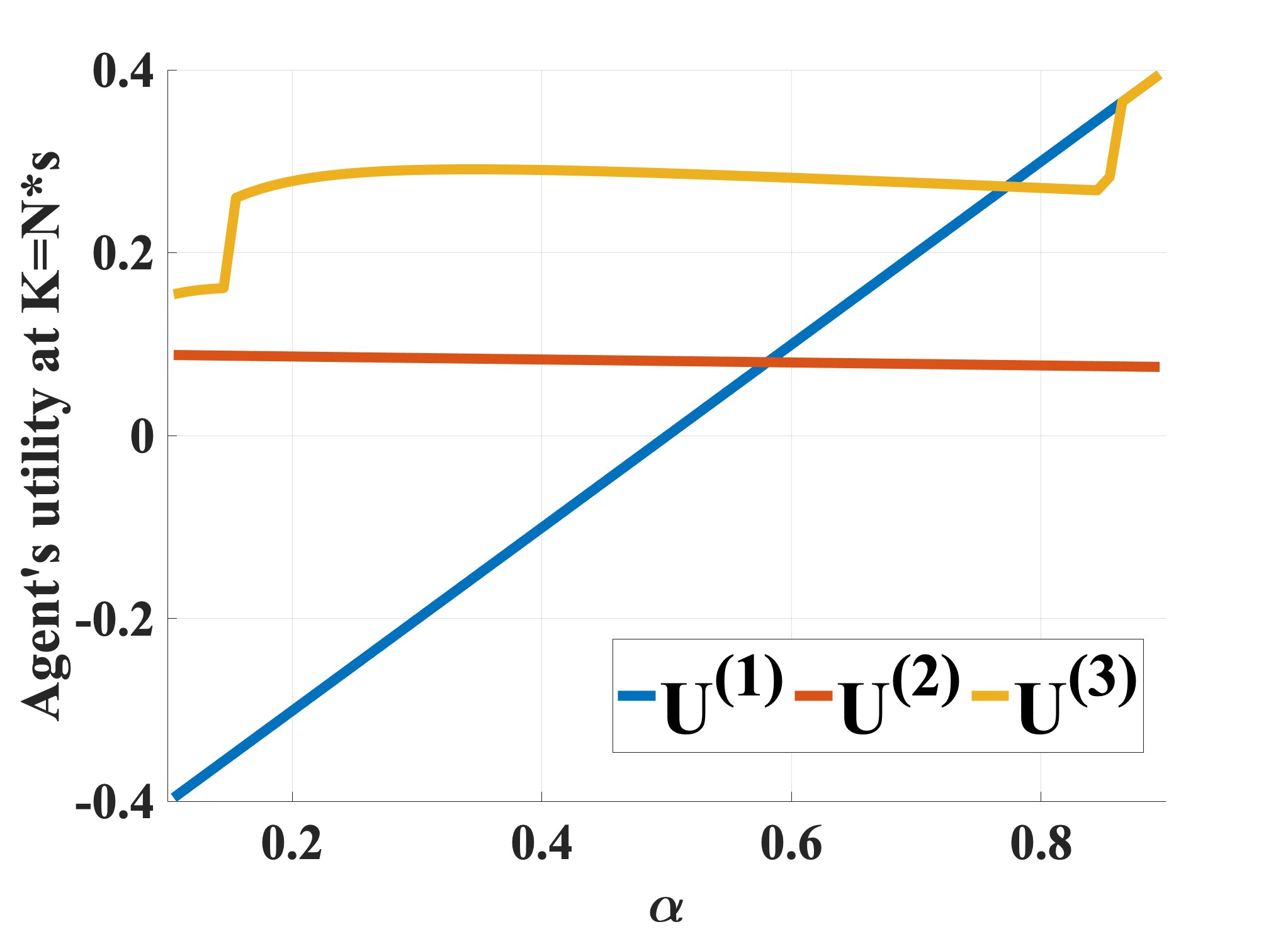}
        \caption{Final agent's utility}
        \label{fig:agent_utility_alpha}
    \end{subfigure}
    \hfill
    \begin{subfigure}[b]{0.3\textwidth}
        \includegraphics[width=\textwidth]{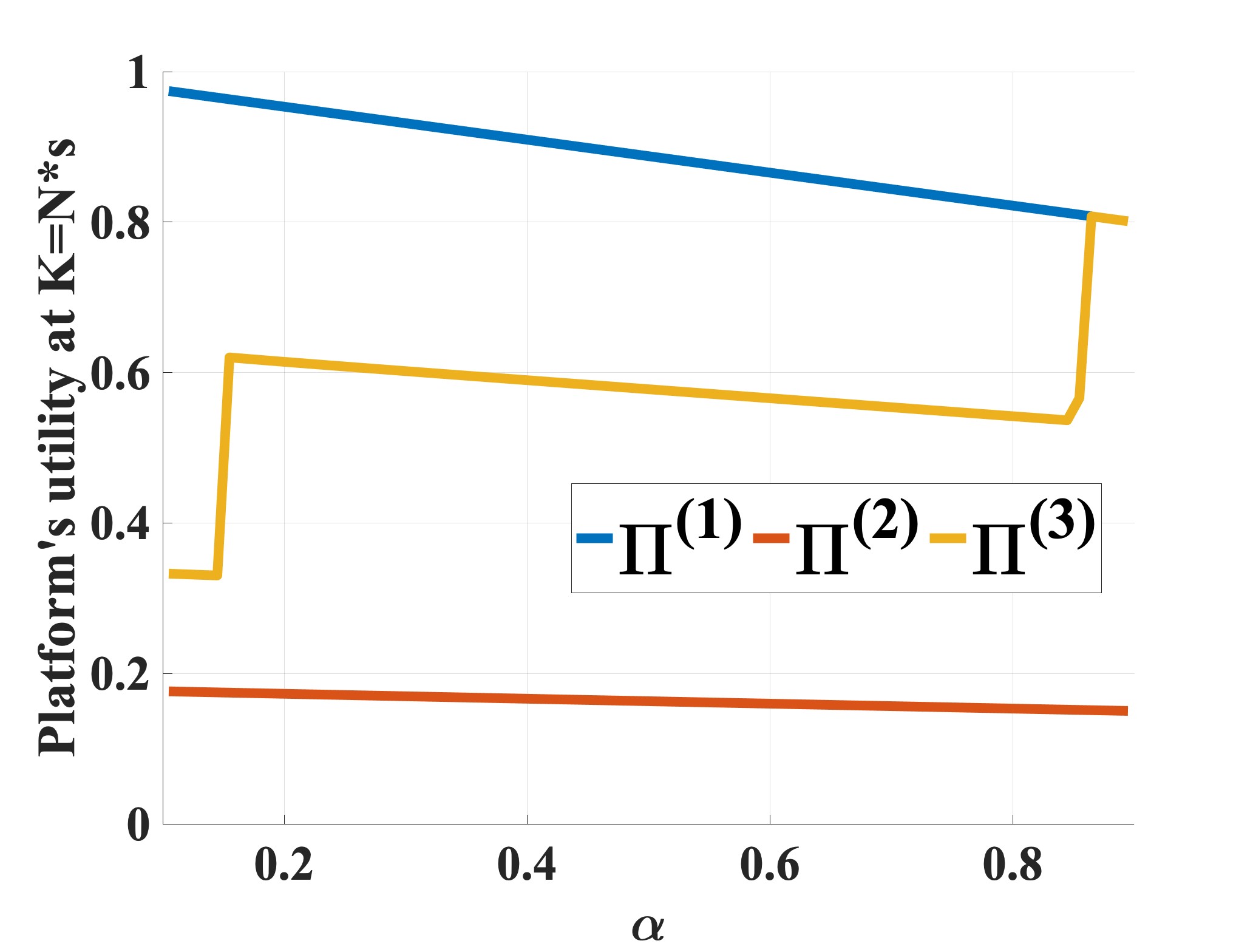}
        \caption{Final platform's utility}
        \label{fig:platform_utility_alpha}
    \end{subfigure}

    \caption{Impact of varying $\alpha$ on the agent's final opinion, agent's final utility, and platform's final utility}
    \label{fig:vary_alpha}
\end{figure}

Figure \ref{fig:vary_alpha} is related to experiments in which the value of $\alpha$ changes from $0.105$ to $0.895$ in increments of $0.01$, when the other parameters are fixed at: $\beta = 0.1, \lambda = 0.5, T_0 = s = 8, \kappa = 2, c = 0.1, \tau=3, x_0 =-1, u_0=1, x_{drift}=0.1$. Figure \ref{fig:drift_alpha} shows that for different values of $\alpha$, the agent under policy~\ref{alg:policy-one} (fixed clicking policy) always has the maximum deviation from the innate opinion, under policy~\ref{alg:policy-two} (decreasing clicking policy) always comes back to the innate opinion, while the diviation for policy~\ref{alg:policy-three} (adaptive decreasing policy) is less than $x_{drift}$.

According to Figure~\ref{fig:agent_utility_alpha}, for low values of $\alpha$, the fixed clicking policy results in a considerable deviation, reducing the agent's utility (negative value), while the decreasing clicking policy consistently maintains the agent's utility within a stable range (above zero), due to no deviation and no clicks after sufficient time steps. However, as $\alpha$ increases, the influence of recommendations decreases, reducing the deviation even when clicks continue, and the utilities under adaptive decreasing and fixed clicking policies become comparable. For most of the values of $\alpha$, adaptive decreasing policy leads to higher agent's utility.

Regarding platform utility (Figure~\ref{fig:platform_utility_alpha}), because the higher clicking rate leads to the higher utility, the fixed clicking policy yields the highest platform utility across varying $\alpha$ values. Conversely, the decreasing clicking policy produces the lowest utility for the platform due to a lack of clicks, while the adaptive decreasing policy results in intermediate platform utility. At higher values of $\alpha$, the adaptive decreasing and fixed clicking policies result in similar platform utilities.

%fig1:varying alpha

\begin{figure}[htbp]
    \centering
    \begin{subfigure}[b]{0.3\textwidth}
        \includegraphics[width=\textwidth]{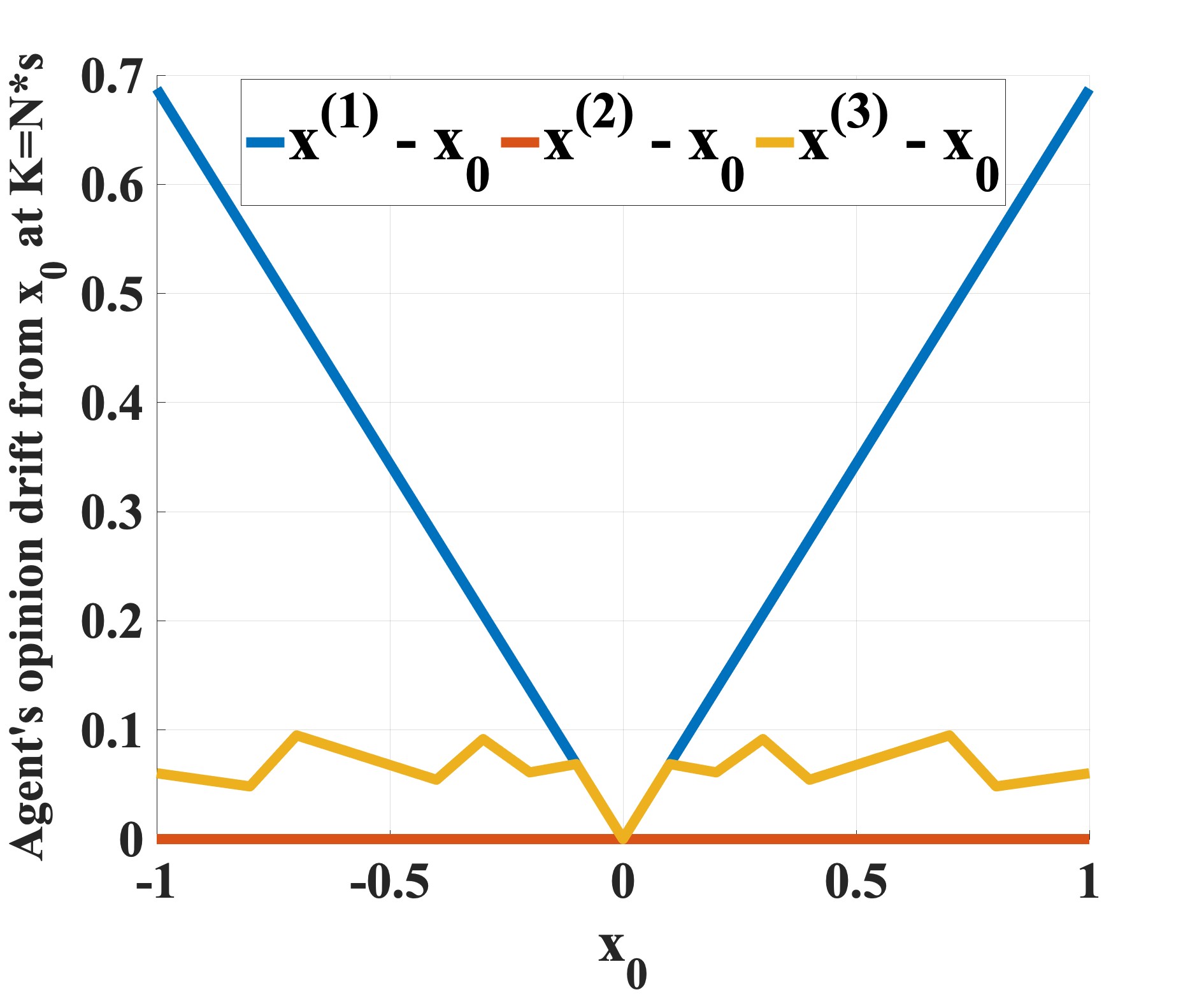}
        \caption{Deviation of agent's final opinion from innate opinion}
        \label{fig:drift_x0}
    \end{subfigure}
    \hfill
    \begin{subfigure}[b]{0.3\textwidth}
        \includegraphics[width=\textwidth]{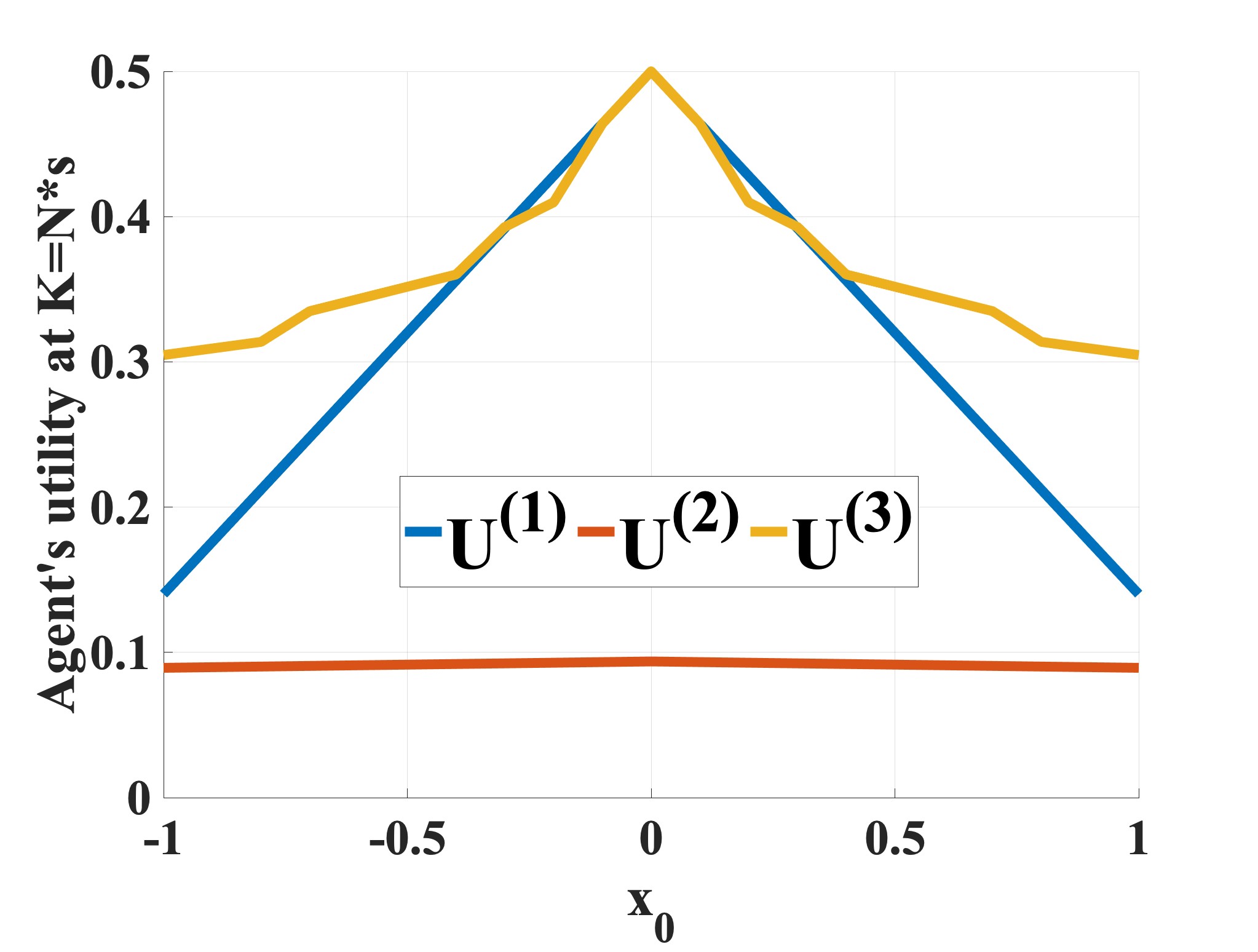}
        \caption{Final agent's utility}
        \label{fig:agent_utility_x0}
    \end{subfigure}
    \hfill 
    \begin{subfigure}[b]{0.3\textwidth}
        \includegraphics[width=\textwidth]{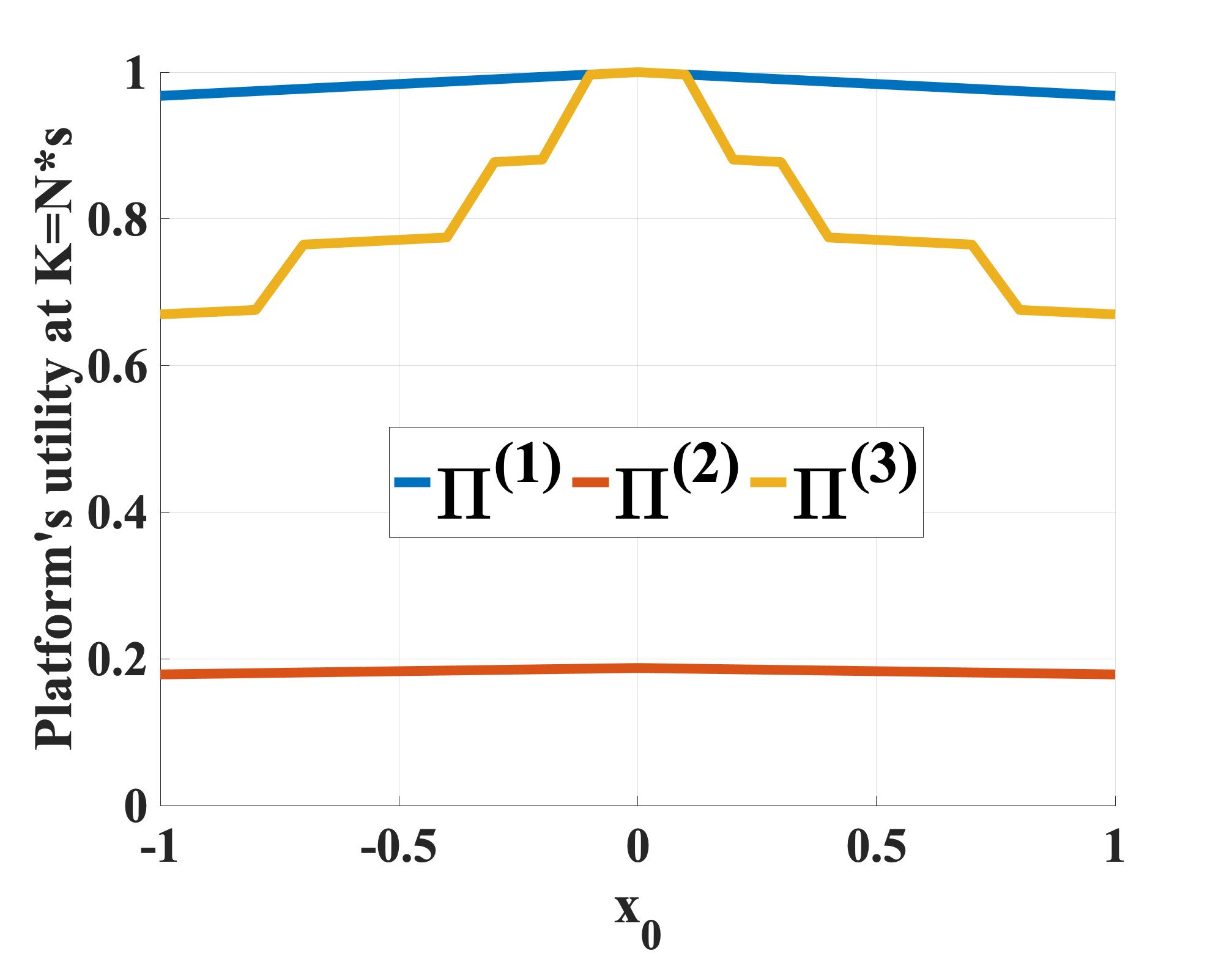}
        \caption{Final platform's utility}
        \label{fig:platform_utility_x0}
    \end{subfigure}

    \caption{Impact of varying innate opinion $x_0$ on the agent's final opinion, agent's final utility, and platform's final utility}
    \label{fig:vary_x0}
\end{figure}

Figure \ref{fig:vary_x0} corresponds to experiments in which the value of $x_0$ changes from $-1$ to $1$ in increments of $0.1$, while the other parameters are fixed at: $\alpha=0.25, \beta = 0.2, \lambda = 0.5, T_0 = s = 8, \kappa = 2, c = 0.1, \tau=3, u_0=0, x_{drift}=0.1$. Each figure from this experiment consists of two stages: the first stage is when the innate opinion is lower than the recommendation, and the second stage is where the innate opinion is higher than the recommendation.
Figure \ref{fig:drift_x0} demonstrates that the agent under the fixed clicking policy always has the maximum deviation, except when the recommendation is identical to the innate opinion (in which case no deviation occurs under any policy, even if clicks occur). The least deviation is related to the decreasing clicking policy, while the adaptive decreasing policy keeps deviation below $x_{drift}$. Figure \ref{fig:agent_utility_x0} illustrates that the adaptive decreasing policy generally leads to the highest agent utility, while the decreasing clicking policy results in the lowest agent's utility (still not negative). Finally, Figure \ref{fig:platform_utility_x0} shows that the platform reaches the highest utility under the fixed clicking policy for the agent (when the number of clicks is highest), the lowest utility under the decreasing clicking policy for the agent (no clicks), and intermediate utility under the case of an agent with the adaptive decreasing policy.

%fig2:varying x_0

\begin{figure}[htbp]
    \centering
    \begin{subfigure}[b]{0.3\textwidth}
        \includegraphics[width=\textwidth]{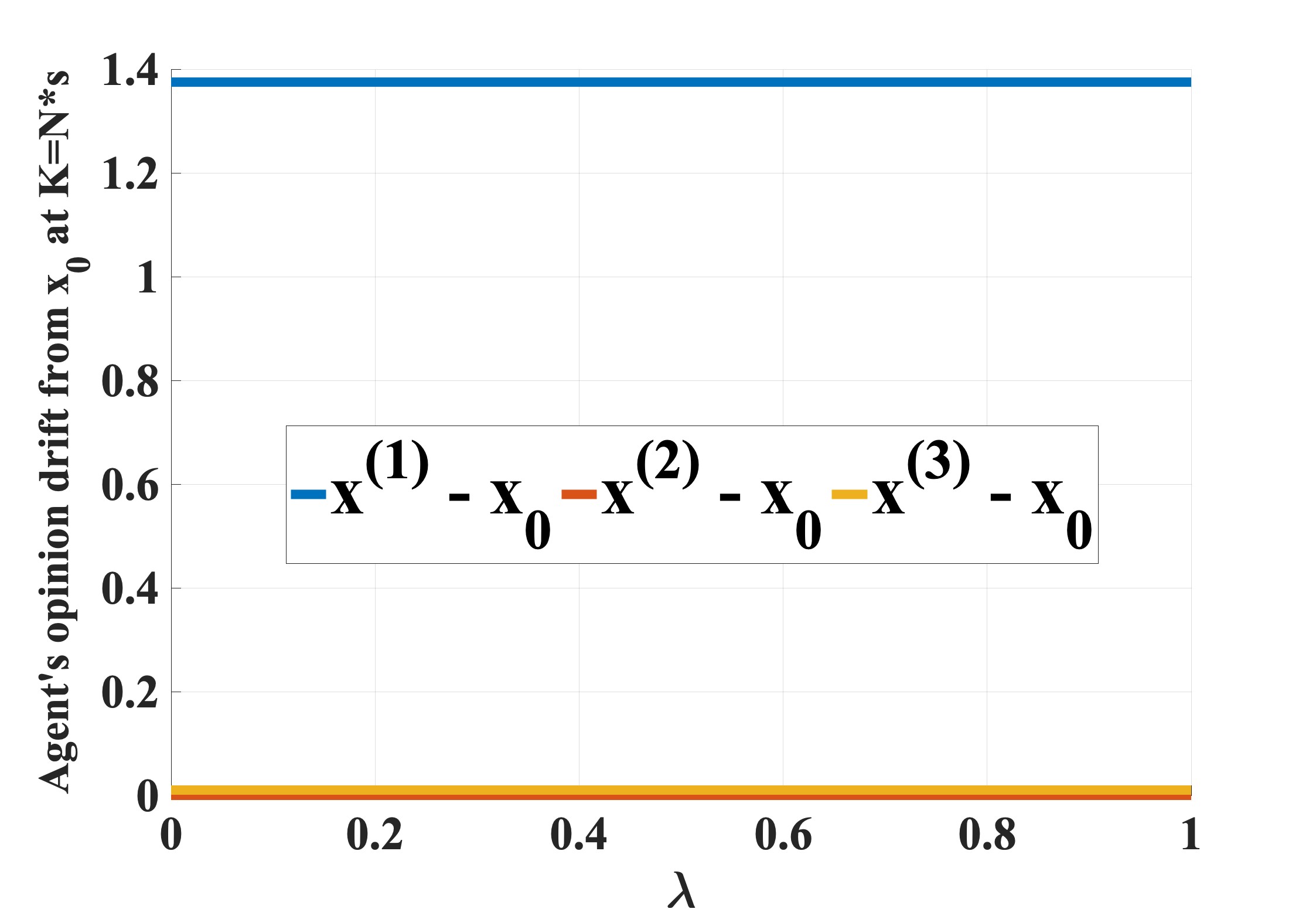}
        \caption{Deviation of agent's final opinion from innate opinion}
        \label{fig:drift_lambda}
    \end{subfigure}
    \hfill 
    \begin{subfigure}[b]{0.3\textwidth}
        \includegraphics[width=\textwidth]{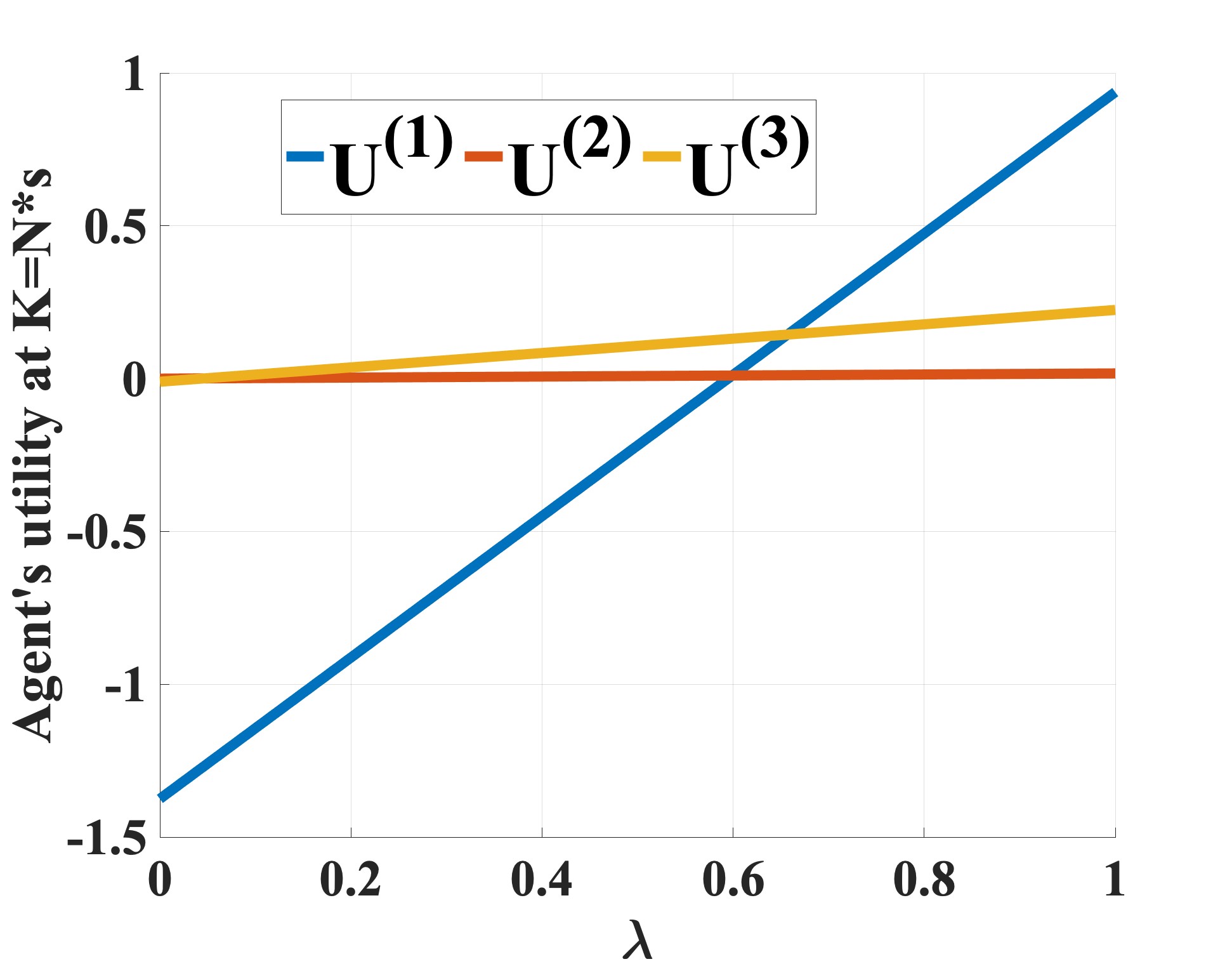}
        \caption{Final agent's utility}
        \label{fig:agent_utility_lambda}
    \end{subfigure}
    \hfill 
    \begin{subfigure}[b]{0.3\textwidth}
        \includegraphics[width=\textwidth]{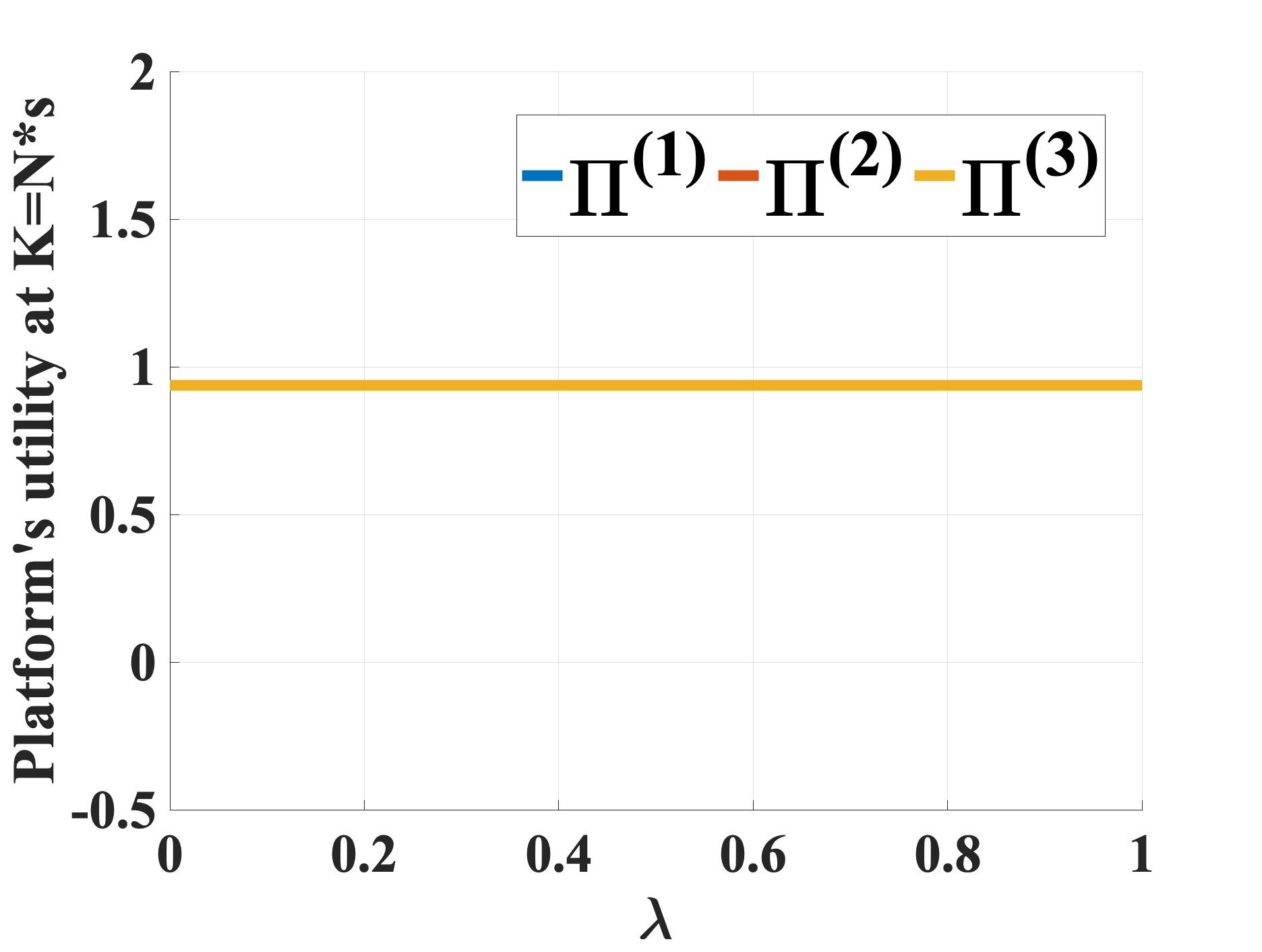}
        \caption{Final platform's utility}
        \label{fig:platform_utility_lambda}
    \end{subfigure}

    \caption{Impact of varying $\lambda$ on the agent's final opinion, agent's final utility, and platform's final utility}
    \label{fig:vary_lambda}
\end{figure}

Figure \ref{fig:vary_lambda} shows experiments with varying $\lambda$ from $0$ to $1$ in increments of $0.05$. Other parameters are fixed at $\alpha=0.25, \beta = 0.2, T_0 = s = 8, \kappa = 2, c = 0.1, \tau=3, x_0=-1, u_0=1, x_{drift}=0.1$. Figures \ref{fig:drift_lambda} and \ref{fig:platform_utility_lambda} remain unchanged, as $\lambda$ is a parameter in the agent's utility, representing the relative importance of the average number of clicks and the deviation from the innate opinion. Considering Figure \ref{fig:agent_utility_lambda}, it can be seen that for lower values of $\lambda$, where less deviation from the innate opinion is more important, the agent's utility is negative under the fixed clicking policy (as it has the highest clicking rate among the three policies), and highest under the adaptive decreasing policy. This trend reverses for higher values of $\lambda$, where clicking carries greater weight, leading to the highest agent's utility under those values of $\lambda$.

%fig3:varying lambda

\begin{figure}[htbp]
    \centering
    \begin{subfigure}[b]{0.3\textwidth}
        \includegraphics[width=\textwidth]{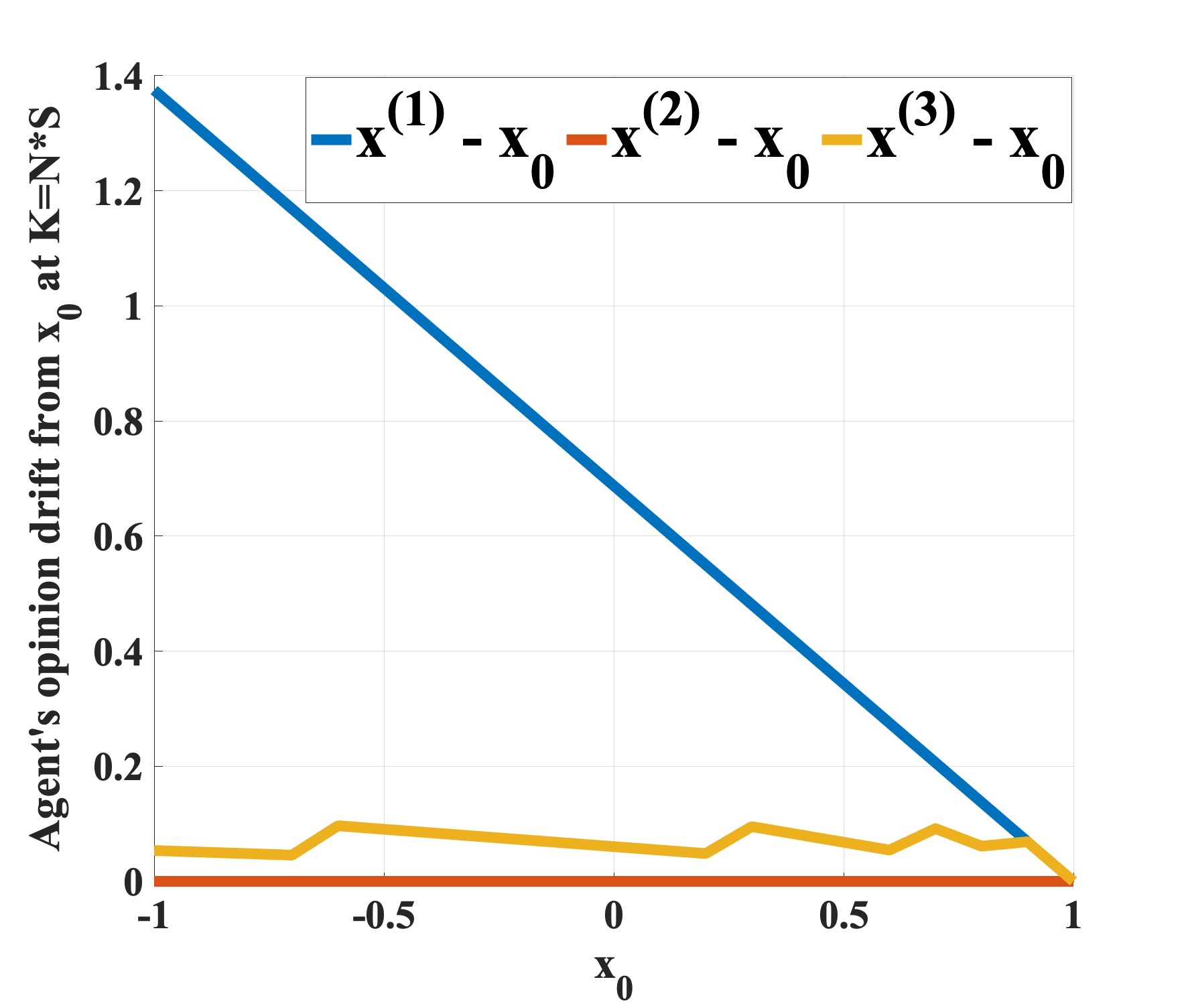}
        \caption{Deviation of agent's final opinion from innate opinion}
        \label{fig:drift_u_0}
    \end{subfigure}
    \hfill 
    \begin{subfigure}[b]{0.3\textwidth}
        \includegraphics[width=\textwidth]{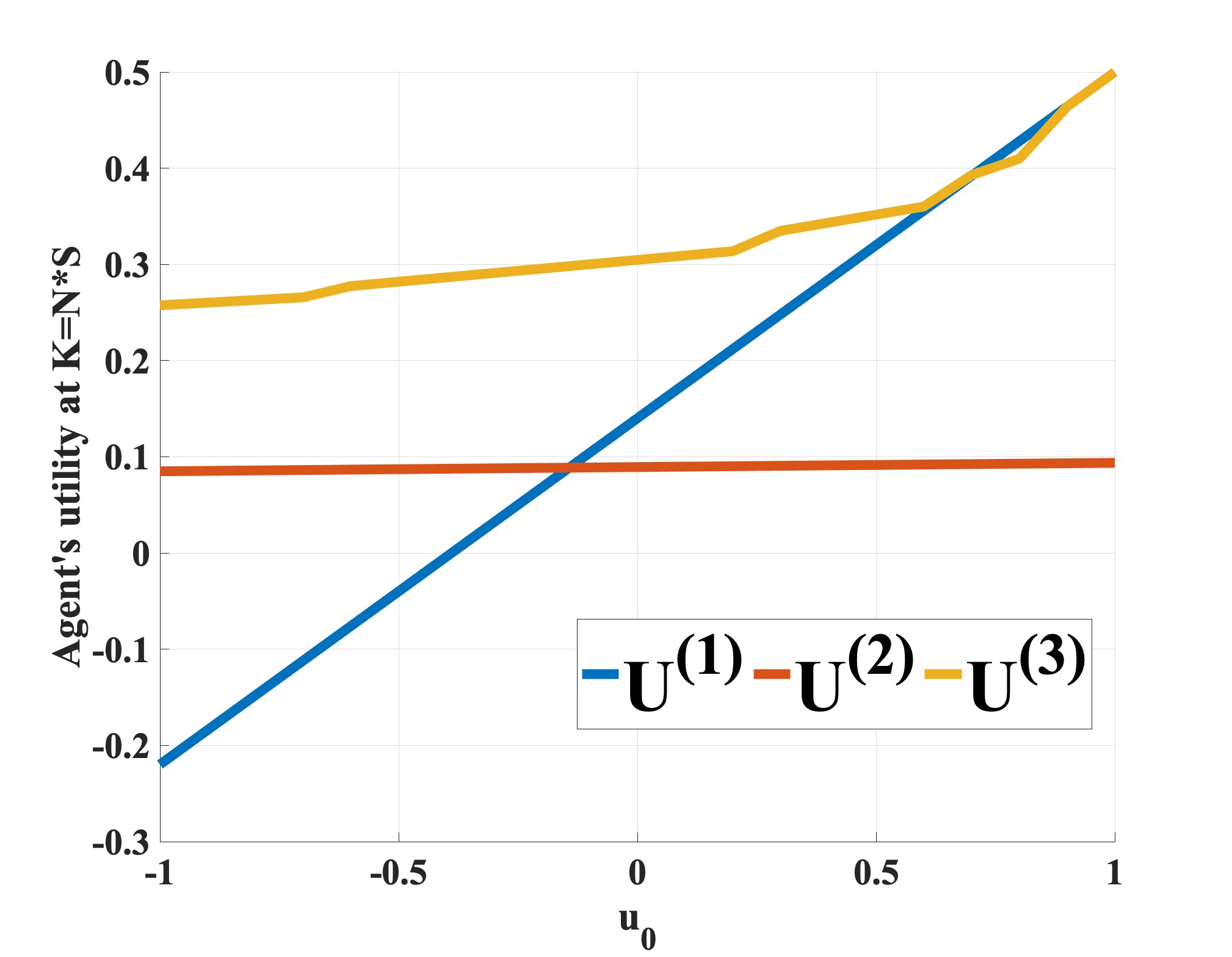}
        \caption{Final agent's utility}
        \label{fig:agent_utility_u_0}
    \end{subfigure}
    \hfill
    \begin{subfigure}[b]{0.3\textwidth}
        \includegraphics[width=\textwidth]{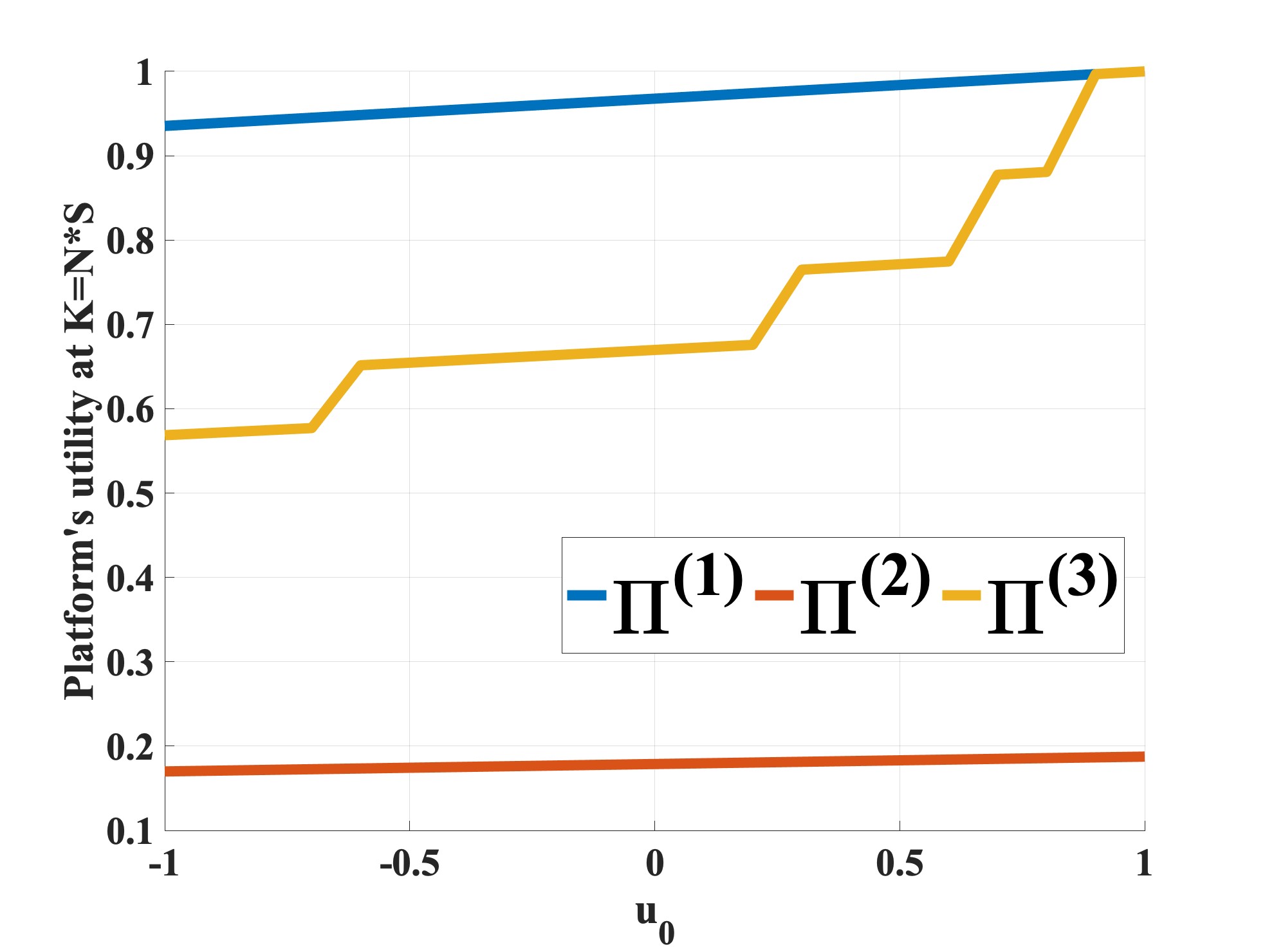}
        \caption{Final platform's utility}
        \label{fig:platform_utility_u_0}
    \end{subfigure}

    \caption{Impact of varying $u_0$ on the agent's final opinion, agent's final utility, and platform's final utility}
    \label{fig:vary_u_0}
\end{figure}

Figure \ref{fig:vary_u_0} illustrates experiments with varying $u_0$ from $-1$ to $1$ in increments of $0.1$. Other parameters are fixed at $\alpha=0.25, \beta = 0.2, T_0 = s = 8, \kappa = 2, c = 0.1, \tau=3, x_0=1, x_{drift}=0.1$. Similar to before, based on Figure \ref{fig:drift_u_0}, the deviation of the agent's final opinion from the innate opinion is maximum under the fixed clicking policy, minimum under the decreasing clicking policy, and less than $x_{drift}$ under the adaptive decreasing policy. By increasing $u_0$, as the recommendation value gets closer to the innate opinion, the deviation from the innate opinion decreases under the fixed clicking policy while clicks occur. Considering Figure \ref{fig:agent_utility_u_0}, the agent's final utility under the decreasing clicking policy does not change significantly, while for lower values of $u_0$, the agent under the fixed clicking policy receives negative utility. This utility increases with higher values of $u_0$, yet remains lower than the agent's final utility under the adaptive decreasing policy (which is not negative) for most recommendation values. The agent's utility values under the fixed and adaptive decreasing policies converge as $u_0$ approaches $x_0$. Figure \ref{fig:platform_utility_u_0} shows that the platform gains the highest utility when the agent follows the fixed clicking policy and the lowest utility when the agent follows the decreasing clicking policy. An agent following the adaptive decreasing policy can lead to improvements in the platform's utility as the recommendation becomes more similar to the innate opinion (more clicks can occur without deviation exceeding $x_{drift}$).

%fig4:varying u_0
\end{appendices}

\addtolength{\textheight}{-12cm}   % This command serves to balance the column lengths

\end{document}